\documentclass[useAMS,usenatbib]{mn2e}

\usepackage								{amsmath}
\usepackage								{natbib}

\usepackage{graphicx}
\usepackage{color}
\usepackage{tabularx}
\usepackage{aas_macros}

\usepackage {hyperref}

\title[\texttt{Fervent}: Chemistry-coupled radiative transfer]{\texttt{Fervent}: Chemistry-coupled, ionising and non-ionising radiative feedback in magnetohydrodynamical simulations}
\author[C. Baczynski, S. C. O. Glover, R. S. Klessen]{C. Baczynski$^{1}$\thanks{E-mail:
baczynski@uni-hd.de}, S. C. O. Glover$^{1}$, R. S. Klessen$^{1,2,3}$   \\
$^{1}$Zentrum f\"ur Astronomie der Universit\"at Heidelberg, Institut f\"ur Theoretische Astrophysik, Albert-Ueberle-Str. 2, 69120 Heidelberg, Germany\\
$^{2}$Department of Astronomy and Astrophysics, University of California, 1156 High Street, Santa Cruz, CA 95064, USA \\  
$^{3}$Kavli Institute for Particle Astrophysics and Cosmology, Stanford University, SLAC National Accelerator Laboratory, Menlo Park, CA 94025, USA}

\begin{document}

\date{draft version -- submitted to MNRAS March 31.}


\maketitle


\begin{abstract}
We introduce a radiative transfer code module for the magnetohydrodynamical adaptive mesh refinement code \texttt{FLASH 4}. It is coupled to an efficient chemical network which explicitly tracks the three hydrogen species $\mathrm{H, H_2, H^+}$ as well as $\mathrm{C^+}$ and $\mathrm{CO}$. The module is geared towards modeling all relevant thermal feedback processes of massive stars, 
and is able to follow the non-equilibrium time-dependent thermal and chemical state of the present-day interstellar medium as well as that of dense molecular clouds. We describe in detail the implementation of all relevant thermal stellar feedback mechanisms, i.e.\ photoelectric, photoionization and $\mathrm{H_2}$ dissociation heating as well as pumping of molecular hydrogen by UV photons. All included radiative feedback processes are extensively tested. We also compare our module to dedicated photon-dominated region (PDR) codes and find good agreement in our modeled hydrogen species once our radiative transfer solution reaches equilibrium. In addition, we show that the implemented radiative feedback physics is insensitive to the spatial resolution of the code and show under which conditions it is possible to obtain well-converged  evolution in time. Finally, we briefly explore the robustness of our scheme for treating combined ionizing and non-ionizing radiation.
\end{abstract}


\setcitestyle{comma}
\section{Introduction}
Massive stars continuously shape their environments by feedback in the form of radiation, winds and supernovae. Not only do they regulate the thermal and velocity structure of the interstellar medium (ISM) (see e.g.\ \citealt{lopez11,kt12,hopkins13,walch14}, or \citealt{kg14} for a general overview), they also determine the total fraction of their parent molecular cloud  that gets converted into stars \citep{murray11,Dale2012,fg13,hopkins14}.

However, the impact of the deposited radiative energy and momentum on the star formation efficiency, cloud lifetime and turbulent velocity structure is still subject to debate (compare e.g.\ the results of \citealt{grit09} and \citealt{walch12} with those of \citealt{Dale2012}). A powerful approach to explore the interaction of stellar feedback with its surroundings are numerical simulations in which we control the implemented physics and can choose the boundary conditions and initial conditions. Simulations of this kind are technically challenging and important physical processes are often omitted or treated in a simplified fashion in order to make the calculations computationally tractable. Unfortunately, it appears that if too much of the real physics is omitted, then this can significantly influence the effectiveness of the feedback in the simulations \citep[see e.g.][]{iliev09,rkuiper12}. It is therefore important to continue to search for ways in which to model the physics of stellar feedback accurately and efficiently.

For radiative feedback from massive stars, many different approaches have been proposed in the literature, ranging from simple approximations \citep[e.g.][]{Dale2007,krumholzfld,poormanrt}, to moment-based schemes \citep[e.g.][]{hayes03,ramsesrt,athenart}, ray-tracing techniques \citep[e.g][]{Rijkhorst,peters10,moray} and finally Monte Carlo methods \citep[e.g.][]{torus} as well as hybrid schemes \citep{JPpaardekooper2010,Kuiper2010}. These different approaches have different degrees of sophistication and often wildly different computational costs. In general, the higher the accuracy of the method, the more costly it is, limiting the applicability of the most accurate approaches to situations with only one or a few radiation sources, and simulation durations that cover only a small fraction of the total stellar lifetime.


One important aspect of any model of radiative feedback from massive stars is the way in which the radiative output of the star is coupled to the thermal and chemical evolution of the gas. Although highly sophisticated models of the effect of stellar radiation on the chemical composition and temperature structure of interstellar gas are available \citep[e.g.][]{ercolano03,ferland13}, they do not account for the hydrodynamical evolution of the gas, and hence cannot be used to study the effect of radiative feedback on star formation. At the other extreme, many hydrodynamical models of radiative feedback account for the chemical state of the gas only to the extent that they distinguish between ionized and neutral regions. In addition, they often adopt a highly simplified treatment of the thermal state of the gas, for instance by assuming that all ionized gas is isothermal with a temperature $T_{\rm ion} = 10^{4}$~K, and all neutral gas is isothermal with a temperature $T_{\rm neut} = 10$~K \citep[see e.g.][]{Dale2007,grit09}. Those hydrodynamical models that do attempt to move beyond this simple approximation and to properly account for the coupling between the radiation, the chemical and thermal state of the gas and its hydrodynamical evolution are typically designed to model radiative feedback in high-redshift, metal-free gas \citep[see e.g.\ many of the codes compared in][]{iliev06,iliev09}. This gas generally has a low molecular fraction, no dust, and relatively simple microphysics in comparison to that making up the present-day ISM.

In this paper we present \texttt{Fervent}, a radiative transfer module designed to model the coupled chemical, thermal and dynamical effects of radiative feedback on the present-day ISM. \texttt{Fervent} consists of an adaptive ray-tracing scheme, inspired by the \texttt{Moray} algorithm of \citet{moray}, coupled to a simplified chemical network and a detailed cooling function. It allows us to self-consistently calculate the chemical state of interstellar gas affected by radiation from massive stars, as well as to address the question of how the thermal state of the gas is influenced. This is vital for an accurate modelling of H{\sc ii} region dynamics, as it is the thermal structure of the gas that determines the strength of the hydrodynamical response to the deposited radiative energy, which in the end constrains the effectiveness of the feedback. In other words, our scheme has the advantage that it connects the underlying chemical processes directly to the hydrodynamics without an intermediate model for the thermal structure of the gas. Ultimately, we aim to quantify the range and strength of stellar radiative feedback from the ground up, by including all of the important chemical processes that influence the behaviour of the gas.   


Our paper is structured in the following way. In Section~\ref{model},  we present the physical details of the model and an outline of our ray-tracing scheme. Tests of our method are described in section \ref{tests} and we summarize our findings in section \ref{summary}. Additional technical details regarding the ray-tracing scheme are given in 
Appendix~\ref{AppendixA}.

\section{Radiative transfer physics and numerics}\label{model}
The rate of radiative energy transfer $\mathrm{d}E / \mathrm{d} t$ in $\mathrm{erg \ s^{-1}} $ at every position in space $\mathbf{x}$ and time $t$ is given by
\begin{equation}
\begin{aligned}
\mathrm{d}E = I(\nu, \mathbf{n},\mathbf{x},t) \ \mathrm{cos} \ \theta_{\mathbf{n}} \  \mathrm{d} \Omega_{\mathbf{n}}  \ \mathrm{d} A_{\mathbf{n}} \ \mathrm{d} \nu \ \mathrm{d} t 
\end{aligned}
\end{equation}
where $I(\nu,\mathbf{n},\mathbf{x},t)$ is the specific intensity in $\mathrm{erg\ s^{-1}\ Hz^{-1} \ cm^{-2} \ sr^{-1}}$ of the radiation field seen in the direction of $\mathbf{n}$ in the frequency range $ \nu$ to $\nu + \mathrm{d} \nu$. In addition, the amount of locally transferred energy depends on how much of the emitting source area $A_{\mathbf{n}}$ is seen in the solid angle $\Omega_{\mathbf{n}}$. 

The evolution of $I$ is modeled by a transport equation of the form:
\begin{equation}\label{radTransferI}
\begin{aligned}
\frac{ \partial{I} }{ \partial{t} } +  \mathbf{n} \cdot \nabla{I} = -\kappa(\nu,\mathbf{x},t) \ I + j(\nu,\mathbf{x},t)
\end{aligned}
\end{equation}
The right hand side consists of local source and sink terms, i.e.~the attenuation of the field by some medium with absorption coefficient $\kappa$ and spontaneous emission coefficient $j$. On the left hand side, the propagation of the field in time and its geometric dilution along the direction $\mathbf{n}$ is described. 
It is convenient to reduce equation \eqref{radTransferI} to one dimension and express the specific intensity in terms of the photon flux $P$ from a single radiation source. By integrating $I$ over the solid angle and a closed surface containing the emitting source, the photon flux $P_\nu$ in the frequency range $ \nu$ to $\nu + \mathrm{d} \nu$ is obtained,
\begin{equation}\label{Pflux}
\begin{aligned}
\frac{1}{c}\frac{ \partial{P_\nu} }{ \partial{t}} + \frac{ \partial{P_\nu} }{ \partial{r}}  = -\kappa \ P_\nu,
\end{aligned}
\end{equation}
where we have neglected the local emission $j$ due to e.g.~scattering processes.
In practice one cannot always stay in the ray-frame, as we will show later for processes other than hydrogen-ionizing radiation. 

To allow us to solve for the evolution of the photon flux as a function of position and time within our computational volume, we implement a ray-tracing approach similar to that proposed by \cite{moray} in the magneto-hydrodynamical simulation code \texttt{FLASH} 4 \citep{flash, dubey2008}. We briefly describe the method in the next section. For implementation details concerning the general code structure, parallelization, box-ray intersection calculation and ray propagation, see Appendix~\ref{AppendixA}. Coupling to a chemical network explicitly tracking the species $\mathrm{CO}$, $\mathrm{C^+}$, $\mathrm{H^+}$, $\mathrm{H}$ and $\mathrm{H_2}$ is described in section \ref{chemistry} and an overview of missing physics in the current implementation is given in section \ref{miss}.
\subsection{Ray-tracing}\label{raytrace}
We follow a split approach to calculate the effects of the radiation field on the gas in our simulation domain. \\
First, by tracing rays from all radiation sources outward, we gather information on the chemical composition and radiation intensity and use this to calculate the radiative heating rate, the atomic hydrogen photoionization rate and the H$_{2}$ photodissociation rate. Each cell of the mesh has to know the gas column and photon flux to and from each source illuminating it in order to calculate all quantities necessary for input into the chemistry module. In a subsequent step, the gas composition and temperature is updated locally on a cell-by-cell basis, with no further input from its surroundings. 

Ideally, after a single chemistry update another ray-tracing step should be performed to check whether the new state is converged. 
However, multiple ray-tracing and chemistry updates during a single simulation time step are currently computationally too expensive. Instead we impose a limiter on the size of the hydrodynamical timestep $\mathrm \Delta t$ by requiring that the change in the fractional abundance of the most radiatively affected species, atomic hydrogen,
\begin{equation}
\begin{aligned}
\left| \mathrm{\Delta} x_{\mathrm{H,new}} - \mathrm{\Delta} x_{\mathrm{H,old}} \right| = \mathrm{\Delta x_{H}}
\end{aligned}
\end{equation}
is less than some factor $f_{\mathrm{H}}$. This implies that the radiation field is assumed static during the chemistry step. We use the H abundance in our limiter, because the main photochemical processes that we are interested in modelling -- the photoionization of atomic hydrogen and the photodissociation of H$_{2}$ -- both change the H abundance significantly.  Ensuring that the global evolution time step satisfies
\begin{equation}
\begin{aligned}
\mathrm{\Delta} t_{\mathrm{new}} \leq \frac{f_{\mathrm{H}}} { \mathrm{\Delta x_{H}}}  \ \mathrm{\Delta} t_{\mathrm{old}},
\end{aligned}
\end{equation}
where the smallest $\mathrm{\Delta} t_{\mathrm{new}}$ of all cells is taken, should therefore accurately capture the chemical evolution of the gas for small enough $f_{\mathrm{H}}$.  Other processes that produce atomic hydrogen (e.g.\ the collisional dissociation of H$_{2}$) generally do not lead to changes in the H abundance that occur rapidly enough to trigger the timestep limiter, provided that we choose a reasonable value for  $f_{\mathrm{H}}$. We discuss appropriate values for $f_{\mathrm{H}}$ in Section~\ref{rtype}. After the new thermal and chemical state of all cells has been calculated, the hydrodynamical equations are advanced. 

During each simulation timestep, rays originating from all radiation sources are generated at the same time. The rays are spaced evenly on a spherical surface centered on the source locations by utilizing the \texttt{HEALPix} software package \citep{HEALPix}. \texttt{HEALPix} is an equal area decomposition of a unit sphere with a nested grid structure. The base level of refinement consists of twelve pixels, where each pixel can be split independently into four nested child pixels. All pixels are numbered consecutively which leads to a limit of thirteen levels of refinement, where the pixel number just fits inside a long integer. This corresponds to a maximum angular resolution of $3.2 \times 10^{-3} \ \mathrm{arcseconds}$. The independent splitting makes it inherently applicable to an adaptive mesh refinement (AMR) code such as \texttt{FLASH} 4. 

After generating the initial sphere of rays, intersections between them and the cubic cells of the simulation mesh are calculated during the ray-tracing step. Rays are represented by an equation of a line in vector form $\mathbf{p}(r) = \mathbf{s} + r \mathbf{n}$, with the unit direction vector $\mathbf{n}$, source location $\mathbf{s}$ and total traversed distance $r$. Each time a ray enters a new cell the intersection distance $\mathrm{d}r$ from the entry to the exit point is calculated. To compute the radiative heating rate and the various photochemical rates, we need to know for each ray and each cell the value of the ray segment $\mathrm{d}r$ in that cell, as well as the total column densities of atomic and molecular hydrogen, $N_{\rm H}$ and $N_{\rm H_{2}}$, traversed by the ray between the source and the cell. These column densities are given by
\begin{equation}
\begin{aligned}
N_{\mathrm{H},\mathrm{H_2}} = \sum_{i=0}^k x_{\mathrm{H},\mathrm{H_2}}(i) \ n_i \mathrm{d}r_i,
\end{aligned}
\end{equation}
where $i = 0$ represents the cell containing the point source, $k$ is the current cell, $\mathrm{d}r_i$ is the ray segment in cell $i$, $x_{\rm H, H_{2}}(i)$ are the fractional abundances of H and H$_{2}$, respectively, in cell $i$, and $n_{i}$ is the number density of H nuclei in cell $i$. These quantities are computed as we move along the rays and are
saved in the mesh data structure in the same way as any other hydrodynamical state variable. For the propagation of the radiation field, the gas columns are carried to the next neighboring cell as a ray property (see Appendix~\ref{AppendixA} for details). 

For a full sampling of the radiation field, each mesh cell has to be intersected by at least one ray. At this point, we make use of the \texttt{HEALPix} software's ability to split pixels by assigning each ray a \texttt{HEALpix} level $l$ and index $h$. We use the splitting criterion introduced in \cite{moray}, which calculates the ratio of the mesh cell size and the pixel size at a distance $r$,
\begin{equation}
\begin{aligned}
f_{\Omega} = \frac{A}{r^2 \Omega} = \frac{d^2} {r^2  \frac{4\pi}{N_{\mathrm{pixel}}} },
\end{aligned}
\end{equation}
where $A$ is the face area of a cubic cell with side lengths $d$, $\Omega$ is the solid angle associated with a single pixel at the current refinement level and $N_{\mathrm{pixel}} = 12 \times 2^l \times 2^l $ is the total number of pixels on refinement level $l$. Provided that $f_{\Omega} > 1$, this criterion guarantees that each cell is sampled by at least one ray. Conveniently, this criterion also accounts for regions refined during the course of the simulation, where instead of a global uniform grid with constant area $A$, local regions are resolved by a finer mesh. This results in smaller cell sizes and rays are split to adapt to this difference. Multiple rays per cell are treated by summing up all individual contributions to the photochemical rates and heating rates, i.e.\
\begin{equation}
\begin{aligned}
k_{\mathrm{tot}} = \sum^{N_{\mathrm{rays}}}_{i = 0} k_i,
\end{aligned}
\end{equation} 
without any additional weighting. In the same vein, multiple sources are taken into account. We do not distinguish between rays from different sources, and so the order in which the individual contributions are calculated is not controlled. 

This ray-tracing approach is intrinsically serial as new rays have to be generated during the traversal of the simulation domain. The total number of rays is therefore unpredictable at the beginning of a simulation step. In contrast, the most efficient parallelizing schemes expect that all ray information is available from the beginning of the ray-tracing step and parallelize ray-cell intersection calculations, see e.g.~\cite{Rijkhorst}. However, this has the disadvantage that for a large number of sources, all rays from all sources have to be generated at the same time, consuming large amounts of memory. In addition, this introduces an overhead in communication as rays have to be known globally everywhere to account for possible local contributions.  
A simple way to reduce the required amount of communication is a pruning step where only the rays that actually enter the local domain are to be communicated. Another way to prepare the rays could be to pre-generate all intersections in a tree-walk step. 

In our scheme each domain containing a source starts to locally ray-trace. If all sources are distributed evenly over all domains this approach is efficient, but in most astrophysical applications sources are expected to cluster. However, as rays are traced from the source outward, once the optical depth $\tau$ becomes too large they can be terminated, essentially introducing an on-the-fly pruning. This greatly reduces the cost of the ray-tracing step.


\subsection{Photochemistry and radiative heating}\label{chemistry}
In this section, we describe how we account for the effects of the radiation field on the chemistry and thermal balance of the gas. The processes that we currently treat are the photoelectric heating of the gas by the dust, the photodissociation of the molecules $\mathrm{H_2}$ and $\mathrm{CO}$, the pumping of excited rotational and vibrational states of $\mathrm{H_2}$ by ultraviolet (UV) photons, and the photoionization of atomic and molecular hydrogen. To do this, we couple the reaction and heating rates gathered during ray-tracing to the NL97 chemical network described in \citet{glover2012}; note that details of the implementation of this network in \texttt{FLASH} can be found in \citet{walch14}.

Our main concern is radiation from stars, which for simplicity we approximate by a blackbody spectrum $B_{\nu}$, i.e.\ we neglect any absorption or emission line features. We divide this spectrum into four energy bins: $5.6 \ \mathrm{eV} < E_{5.6} < 11.2 \ \mathrm{eV}$, $11.2 \ \mathrm{eV} < E_{11.2} < 13.6 \ \mathrm{eV}$, $13.6 \ \mathrm{eV} < E_{13.6} < 15.2 \ \mathrm{eV}$ and $ E_{15.2{+}} > 15.2 \ \mathrm{eV}$. Using a larger number of energy bins would allow us to account in more detail for the frequency-dependent opacity of the gas, but at the cost of slowing down the entire calculation. We have found that using four bins appears to give us the best trade-off between speed and the accuracy with which we can model the thermal balance of the gas and the details of the hydrogen chemistry.  For the tracked species we explicitly calculate the rates of the reactions:
\begin{equation}
\begin{aligned}
&\mathrm{i)}   &&\ \mathrm{H_2}  & +& \ \gamma_{11.2 } && \rightarrow \mathrm{H}     &+& \  \mathrm{H}   \\
&\mathrm{ii)}  &&\ \mathrm{H}    & +& \ \gamma_{13.6} && \rightarrow \mathrm{H^+}    &+& \  \mathrm{e^-} \\
&\mathrm{iii)} &&\ \mathrm{H}    & +& \ \gamma_{15.2+} && \rightarrow \mathrm{H^+}   &+& \  \mathrm{e^-} \\
&\mathrm{iv)}  &&\ \mathrm{H_2}  & +& \ \gamma_{15.2+} && \rightarrow \mathrm{H_2^+} &+& \  \mathrm{e^-}, \\
\end{aligned}
\end{equation}
where the subscript number denotes the lower end of the photon energy bin and the plus sign that the bin extends to infinity. Reaction (i) describes the dissociation of $\mathrm{H_2}$ through excitation by far ultraviolet photons. We assume that the only photons responsible for this process are those in the energy range $11.2 < h\nu < 13.6$~eV. Lower energy photons cannot dissociate H$_{2}$ molecules that are in the vibrational ground state and hence become significant only if we are concerned with the photodissociation of H$_{2}$ in hot, high density gas ($T > 2000$~K, $n > 10^{4} \: {\rm cm^{-3}}$; see \citealt{glover15}), conditions in which collisional dissociation of H$_{2}$ is in any event likely to dominate. Photons with energies $h \nu > 13.6$~eV can dissociate H$_{2}$, but have a much larger probability of being absorbed by atomic hydrogen resulting in photoionization.

Reactions (ii) and (iii) account for the photoionization of atomic hydrogen by the radiation field. For reasons which will become clear below, we distinguish between photons with energies $13.6 < h\nu < 15.2$~eV that can ionize atomic hydrogen but not molecular hydrogen, and photons with energies $h \nu > 15.2$~eV that can ionize both H and H$_{2}$. Finally, reaction (iv) accounts for the photoionization of H$_{2}$. This produces H$_{2}^{+}$ ions, but this chemical species is not directly tracked in the NL97 chemical network. Instead, we assume that all of the H$_{2}^{+}$ ions are destroyed by dissociative recombination
\begin{equation}
{\rm H_{2}^{+} + e^{-}} \rightarrow {\rm H + H},
\end{equation}
so that a single photoionization of H$_{2}$ ultimately results in the production of two hydrogen atoms. Since the rate coefficient for this reaction is large, and we expect there to be a high electron density close to the ionization front, it is a good approximation to treat this process as effectively instantaneous (i.e.\ we do not expect to build up a significant abundance of H$_{2}^{+}$). 

We note that if one does not account for the effects of photoionizing radiation on the H$_{2}$ then it is easy to obtain unphysical results in highly molecular gas, since in this case the bulk of photons with energies larger than $13.6 \ \mathrm{eV}$ can freely escape from a fully molecular medium, where the abundance of atomic hydrogen and thus the effective optical depth are zero. In that sense, our treatment here also forms a numerical closure. We considered several alternative ways of dealing with this problem, such as slowly ramping up the 
strength of the radiation sources (allowing a photodissociation region (PDR) time to form before the gas starts to become ionized), or artificially creating an initial PDR, but we found that the behaviour of these alternatives depended too strongly on the spatial resolution of the simulation and was hard to parameterize. 
\subsubsection{Ionizing radiation: chemical effects}
Photons with an energy $E_{\nu} = h \nu$ greater than $13.6 \ \mathrm{eV}$ can ionize atomic hydrogen. A star with a blackbody spectrum and effective temperature of $T_{\mathrm{eff}}$ emits
\begin{equation}
\begin{aligned}
P_\mathrm{ion} =  \int^{\infty}_{\nu_{13.6}} \! \! \! \! \! \! \! 4 \pi R^2_{*} \ {\frac{\pi  B_\nu (T_{\mathrm{eff}})} { E_{\nu} } }\  \mathrm{d} \nu = \int^{\infty}_{\nu_{13.6}} \! \! \! \! \! \! \! P_\nu \ \mathrm{d} \nu 
\end{aligned}
\end{equation}
photons capable of ionization per second, where $h$ is the Planck constant and $R_{*}$ is the radius of the star. We assume the photon flux to be constant over one timestep, which is reasonable since on the scales that we are interested in modelling, the hydrodynamical timestep is orders of magnitude smaller than the lifetime of even the most massive O-type star.
In this case, equation~\eqref{Pflux} reduces to
\begin{equation}\label{DP}
\begin{aligned}
\frac{ \mathrm{d}{P_\mathrm{ion}} }{ \mathrm{d}{r}}  = -\kappa \ P_\mathrm{ion}
\end{aligned}
\end{equation}
where we insert the ionizing photon flux $P_\mathrm{ion}$ for $P_\nu$. At this point two approaches are possible to calculate the attenuated flux in a given cell. One option is to use the total distance $r$ from the source location at $r=0$ to the current position
\begin{equation}\label{nonconservative}
\int^{P}_{P_0} \frac{1}{P'_\mathrm{ion}} \  \mathrm{d} P'_\mathrm{ion} = \int^r_0 {-\kappa  \ \mathrm{d}r'},
\end{equation}
from which it follows that
\begin{equation}
P(r) = P_0  \ \mathrm{e}^{-\tau},
\end{equation}
where $P_0 = P_\mathrm{ion}(r=0) $, $P = P_\mathrm{ion}(r)$, and $\tau = \int^r_0 \kappa  \ \mathrm{d}r'$ is the total optical depth along the ray.
Alternatively, we can use the ray segments $\mathrm{d}r$, 
\begin{equation}
\int^{P+\mathrm{d}P}_{P} \! \! \frac{1}{P'_\mathrm{ion}} \  \mathrm{d} P'_\mathrm{ion} = \int^{r+\mathrm{d}r}_{r} { \! \! \! \! \! \! \! -\kappa  \ \mathrm{d}r'}
\end{equation}
to arrive at
\begin{equation}
\mathrm{d} P = P \ ( 1 - \mathrm{e}^{-\mathrm{d}\tau} ),
\end{equation}
which gives the change in $P$ over a single cell. Each treatment of the photon flux requires different properties to be carried along the ray, either $P_0$ or the incrementally attenuated flux $P_{i+1} = P_{i} - \mathrm{d}P_i$, where $i$ denotes the current cell index. Depending on how photons of the radiation field are absorbed by the medium, the first or second formalism is more advantageous. If there is a reaction that maps a single photon to a transition from one species to another and all attenuated photons lead to this transition, $\mathrm{d}P$ provides an easy prescription for equating all absorbed photons to the change in the appropriate species. In this case the scheme is photon conservative and the incremental change in optical depth $\mathrm{d}\tau = \sigma n_s \mathrm{d}r$ can be expressed in terms of a reaction cross-section $\sigma$, the number density of the affected species $n_s$ and the ray segment $\mathrm{d}r$. 

In comparison, if a transition to a new species only occurs for a fraction of the absorbed photons, which is determined by some non-local quantity such as the total column of gas to the current cell position, $\mathrm{d}P$ cannot be photon conservative. In that case the transition rate has to be determined from the flux at the source $P_0$. 

To model ionizing radiation, we follow  \citet{moray}, who use the second approach and attenuate the photon flux based on ray segments $\mathrm{d}r$. We start by converting $P_0$ to the absolute number of ionizing photons emitted in a timestep $\Delta t$ at the source position
\begin{equation}
\begin{aligned}
N_\mathrm{ion}(r=0) = P_0 \ \Delta t.
\end{aligned}
\end{equation}
This number of photons is then divided evenly amongst $N_{\mathrm{pixel}}$ initial rays, so that each ray starts with $P_{0} \ \Delta t / N_{\rm pix}$ ionizing photons associated with it. This initial set of rays is set up so that each ray has a direction vector centered on (and normal to) the associated  \texttt{HEALPix} pixel, pointing away from the source. 
We then walk along each ray. As we enter each new cell along a given ray, we reduce the number of ionizing photons currently associated with that ray, $N_{\rm ion}(r)$, by an amount $\mathrm{d}N_{\mathrm{ion}}$ that is given by
\begin{equation}\label{dion}
\mathrm{d} N_{\rm ion} = N_{\rm ion}(r) \ ( 1 - \mathrm{e}^{-\mathrm{d}\tau} ).
\end{equation}
We next
exploit the fact that $\mathrm{d}N_{\mathrm{ion}}$ is a dimensionless quantity to allow us to formulate the ionization rate in a given cell in a way that is independent of spatial resolution. 
This is done by calculating the total number of neutral hydrogen atoms in a cell, using the local number density of H atoms, $n_{\rm H}$, and the volume of the cell,  $V_{\mathrm{cell}} = d^m$, where $d$ is the length of one side of the cell and $m$ is the dimensionality of the mesh. The ionization rate then follows as
\begin{equation}\label{kion}
\begin{aligned}
k_{\mathrm{ion}} = \frac{\mathrm{d}N_{\mathrm{ion}}}{n_{\mathrm{H}} V_{\mathrm{cell}} \ \Delta t}.
\end{aligned}
\end{equation}
This formulation of the ionization rate ensures, by construction, that the number of ionizing photons absorbed is equal to the ionization rate per unit volume integrated over the volume of the cell, i.e.\ it is photon conservative. 

\begin{figure*}
\centering
\includegraphics[scale=0.6]{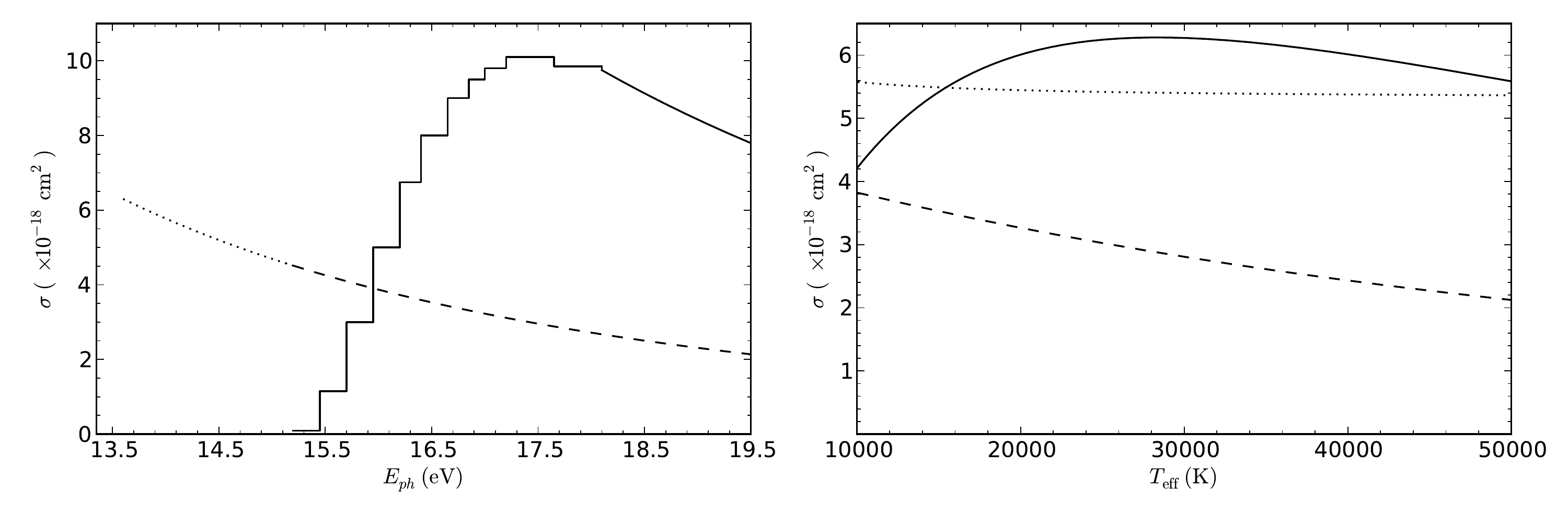}
\caption{{\it Left}: frequency dependent cross-sections $\sigma$ for the ionization of $\mathrm{H}$ and $\mathrm{H_2}$  by photons with energy $E_{\rm ph}$. The hydrogen photoionization cross-section is denoted by a dotted line up to an energy of $15.2 \ \mathrm{eV}$, the H$_{2}$ photoionization threshold, and by a dashed line at higher energies. 
The solid line denotes the H$_{2}$ photoionization cross-section 
{\it Right}: intensity-weighted mean H and H$_{2}$ photoionization cross-sections for a range of different effective temperatures. The dotted line shows $\langle \sigma_{\rm H}^{13.6} \rangle$,
the intensity-weighted mean photoionization cross-section for H in the $E_{13.6}$ energy bin, and the dashed line shows $\langle \sigma_{\rm H}^{15.2} \rangle$, the same quantity in the $E_{15.2+}$ energy bin. The solid line shows $\langle \sigma_{\mathrm{H_2}} \rangle$, the intensity-weighted mean photoionization cross-section for H$_{2}$ in the $E_{15.2+}$ energy bin; the corresponding value in the lower energy bin is zero.
\label{fig:crossSections}}
\end{figure*}

For photons in the $E_{13.6}$ energy bin, this is all we need to worry about, at least as far as the photoionization rate is concerned. For photons in the $E_{15.2+}$ energy bin, however, things are more complicated, since these photons can ionize either atomic or molecular hydrogen. We therefore have to determine how many photons take part in each process. One approach would be to simply divide the photons into two groups by using the relative abundances of $\mathrm{H_2}$ and $\mathrm{H}$. However, this approach fails to account for the fact that the photoionization cross-section of H$_{2}$, $\sigma_{\mathrm{H_2}} $, is different from that of H, $\sigma_{\mathrm{H}}$. We therefore divide up the photons in a way that accounts for both the relative abundances of H$_{2}$ and H and the relative size of their ionization cross-sections. We denote the number of photons in the $E_{15.2+}$ energy bin that are removed from the ray in a given cell by H$_{2}$ photoionization as ${\rm d} N^{15.2}_{\rm dis}$ and the number that are removed by H photoionization as ${\rm d} N^{15.2}_{\rm ion}$. These are given by
\begin{equation}
\begin{aligned}
{\rm d}N^{15.2}_{{\mathrm{dis}}} &= {\rm d}N_{\mathrm{15.2+}}  \frac{  f_{\rm r} x_{\mathrm{H_2}} }{(x_{\mathrm{H_2}} f_{\rm r} + x_{\mathrm{H}})}, \\
{\rm d}N_{{\mathrm{ion}}}^{15.2} &= {\rm d}N_{\mathrm{15.2+}} \frac{x_{\mathrm{H}} }{(x_{\mathrm{H_2}} f_{\rm r} + x_{\mathrm{H}})}.
\end{aligned}
\end{equation}
Here, ${\rm d}N_{\mathrm{15.2+}}$ is the total number of photons in the $E_{15.2+}$ energy bin that are absorbed in the cell in question and
\begin{equation}
f_{\rm r} = \frac{\langle \sigma_{\mathrm{H_2}} \rangle}{\left \langle \sigma^{\mathrm{15.2+}}_{\mathrm{H}}\right \rangle},
\end{equation}
where $\langle \sigma_{\mathrm{H_2}} \rangle$ and $\langle \sigma^{\mathrm{15.2+}}_{\mathrm{H}} \rangle$ are intensity-weighted mean cross-sections for H and H$_{2}$ ionization by photons with $E > 15.2$~eV. These are given by the expressions
\begin{equation}
\begin{aligned}
\langle \sigma_{\mathrm{H_2}}(T_{\mathrm{eff}}) \rangle &= \int^{\infty}_{\nu_{15.2}}\! \! \! \frac{\sigma_{\mathrm{H_2}}(\nu) B_{\nu}}{h \nu} \ \mathrm{d} \nu \Big/ \int^{\infty}_{\nu_{15.2}}  \frac{B_{\nu}}{h \nu}  \ \mathrm{d} \nu, \\
\langle \sigma^{\mathrm{15.2+}}_{\mathrm{H}}(T_{\mathrm{eff}}) \rangle &= \int^{\infty}_{\nu_{15.2}} \! \! \! \frac{\sigma_{\mathrm{H}}(\nu) B_{\nu}}{h \nu} \ \mathrm{d} \nu \Big/ \int^{\infty}_{\nu_{15.2}}  \frac{B_{\nu} }{h \nu}  \ \mathrm{d} \nu, \\
\end{aligned}
\end{equation}
where $T_{\rm eff}$ is the effective temperature of the stellar source, $B_{\nu} = B_{\nu}(T_{\rm eff})$ is the corresponding Planck function, $\sigma_{\rm H}(\nu)$ and $\sigma_{\rm H_{2}}(\nu)$ are the frequency-dependent ionization cross-sections for H and H$_{2}$ and $h \nu_{15.2} = 15.2$~eV. For $\sigma_{\rm H}(\nu)$, we use the simple approximation
\citep{osterbrock}
\begin{equation} 
\label{simpleHcross}
\sigma_{\mathrm{H}}(\nu) = \sigma_{\mathrm{H}, 0} \left(\frac{\nu_{13.6}}{\nu} \right)^{3},
\end{equation}
as this is sufficiently accurate for the photon energies of interest here. For $\sigma_{\rm H_{2}}(\nu)$, we fit a piecewise constant cross-section to the analytical results of \cite{liu}
for energies in the range $15.2 < E < 18.1$~eV, obtaining the values shown in Table~\ref{table1}. At higher energies, we assume that the H$_{2}$ ionization cross-section falls off
as $\sigma_{\rm H_{2}}(\nu) = \sigma_{\rm H_{2}, high} (\nu_{18.10} / \nu)^{3}$, where $h \nu_{18.10} = 18.10$~eV and $\sigma_{\rm H_2, high} = 9.75  \times 10^{-18} \ \mathrm{cm^2}$. At these energies, our adopted values differ from those suggested by experiment by no more than 20\% \citep{Chung1993}. The behavior of $\sigma_{\rm H}(\nu)$
and $\sigma_{\rm H_{2}}(\nu)$ as a function of energy is illustrated in the left-hand panel of Figure~\ref{fig:crossSections}.

\begin{table}
\caption{Piecewise Cross-Section for Ionization of $\mathrm{H_2}$}
\centering
\begin{tabularx}{\columnwidth}{l@{\hskip 0.5in}l@{\hskip 0.5in}l@{\hskip 0.5in}}\label{table1}
$\sigma \ (\mathrm{Mb})$ & $E_{\mathrm{min}} \  (\mathrm{eV})$ & $E_{\mathrm{max}} \ (\mathrm{eV})$ \\
\hline
$0.00$ & 13.60 & 15.20  \\
$0.09$ & 15.20 & 15.45  \\
$1.15$ & 15.70 & 15.95  \\
$3.00$ & 15.95 & 16.20  \\
$5.00$ & 16.20 & 16.40  \\
$6.75$ & 16.40 & 16.65  \\
$8.00$ & 16.65 & 16.85  \\
$9.00$ & 16.85 & 17.00  \\
$9.50$ & 17.00 & 17.20  \\
$9.80$ & 17.20 & 17.65  \\
$10.10$ & 17.65 & 18.10 \\
$\sigma_{\mathrm{H_2}} (\nu_{18.10}/\nu)^3$ \footnote{$\nu_{\mathrm{18.10}}$ is the frequency corresponding to photons with an energy of $18.10 \ \mathrm{eV}$ } & 18.10 & $\infty$ \\
\hline
\end{tabularx}
\end{table}

Once we have computed ${\rm d}N^{15.2}_{\rm dis}$ and ${\rm d}N_{\rm ion}^{15.2}$, it is straightforward to then calculate the rate at which H$_{2}$ is destroyed by photons in the $E_{15.2+}$ energy bin:
\begin{equation}
k^{15.2}_{\mathrm{dis}} = \frac{\mathrm{d}N^{15.2}_{\mathrm{dis}}}{n_{\mathrm{H_{2}}} V_{\mathrm{cell}} \ \Delta t}.
\end{equation}
Similarly, the rate at which atomic hydrogen is photoionized by photons in the $E_{15.2+}$ energy bin is simply
\begin{equation}
k^{15.2}_{\mathrm{ion}} = \frac{\mathrm{d}N^{15.2}_{\mathrm{ion}}}{n_{\mathrm{H}} V_{\mathrm{cell}} \ \Delta t}.
\end{equation}
Note that with this formulation of the H$_{2}$ and H photoionization rates, we guarantee that the total number of photoionizations balances the total number of photons absorbed in the cell, regardless of the relative sizes of $n_{\rm H}$ and $n_{\rm H_{2}}$. 

In order to compute the total photoionization rate of atomic hydrogen, we need to account for the ionizing photons in the $E_{13.6}$ energy bin (reaction ii) as well as those in the $E_{15.2+}$ energy bin (reaction iii). To compute the rate due to the softer photons, which do not couple to H$_{2}$, we simply compute 
\begin{equation}
k^{13.6}_{\mathrm{ion}} = \frac{\mathrm{d}N_{13.6}}{n_{\mathrm{H}} V_{\mathrm{cell}} \ \Delta t},
\end{equation}
where ${\rm d}N_{13.6}$ is the total number of photons in the $E_{13.6}$ energy bin that are absorbed in the cell. The total photoionization rate then follows trivially:
\begin{equation}
k_{\rm ion} = k_{\rm ion}^{13.6} + k_{\rm ion}^{15.2+}.
\end{equation}

Another aspect of the problem that needs to be treated with care is the fact that if there is no atomic hydrogen in the cell, then none of the photons in the $E_{13.6}$ energy bin can be absorbed. If we therefore compute the number of photoionizations from this bin first and from the $E_{15.2+}$ bin second, we arrive at an unphysical scenario: the lower energy photons merely propagate through the cell without any being absorbed (because at that point in the calculation there is no H present) and the higher energy photons then destroy some of the H$_2$, creating some atomic hydrogen, but too late for this to affect the lower energy ones. To avoid this problem, we simply ensure that we account for the effects of H$_{2}$ dissociation before dealing with the $E_{13.6}$ energy bin. If the H$_{2}$ formation time is long compared to the H$_{2}$ dissociation time (which is almost always a good approximation in the clouds that we model), then the change in the atomic hydrogen number density due to the destruction of H$_{2}$ by energetic photons can be written as
\begin{equation}
{\rm d}n_{\rm H} = 2 k_{\rm dis} n_{\rm H_{2}} \Delta t = 2 \frac{{\rm d}N_{\rm dis}^{15.2+}}{V_{\rm cell}}.
\end{equation}
When computing $k_{\rm ion}^{13.6}$, we therefore do not use $n_{\rm H}$, the number density of atomic hydrogen at the start of the timestep, but instead use the improved estimate $n_{\rm H}^{\prime} = n_{\rm H} + {\rm d}n_{\rm H}$ that accounts for the destruction of H$_{2}$ by energetic photons during the timestep. If $n_{\rm H}$ is very small or zero, then this procedure allows us to avoid the coupling problem described above. On the other hand, if $n_{\rm H}$ is larger (i.e.\ if most of the H$_2$ in the cell has already been destroyed), then it becomes only a minor correction.


To complete our specification of how we determine $k_{\rm ion}$, we need to describe how we calculate ${\rm d}N_{15.2+}$ and ${\rm d}N_{\rm ion}^{13.6}$. We start by writing them as
\begin{equation}
{\rm d}N_{13.6}  =  N_{13.6} \left[1 - \exp(-{\rm d}\tau_{13.6}) \right], 
\end{equation}
and
\begin{equation}
{\rm d}N_{15.2+}  = N_{15.2+} \left[1 - \exp(-{\rm d}\tau_{15.2+}) \right],
\end{equation}
where $N_{13.6}$ and $N_{\rm 15.2+}$ are the numbers of photons in the $E_{13.6}$ and $E_{15.2+}$ energy bins at the point where the ray enters the grid cell, and ${\rm d}\tau_{13.6}$ and ${\rm d}\tau_{15.2+}$ are the change in the optical depth of the two energy bins taking place across the cell. These are given by
\begin{equation}
{\rm d}\tau_{13.6} = \langle \sigma_{\rm H}^{13.6} \rangle n_{\rm H} {\rm d}r,
\end{equation}
and
\begin{equation}
{\rm d}\tau_{15.2+} = \left[ \langle \sigma_{\rm H}^{15.2+} \rangle n_{\rm H} + \langle \sigma_{\rm H_{2}} \rangle n_{\rm H_{2}} \right] {\rm d}r,
\end{equation}
where 
\begin{equation}
\langle \sigma^{\mathrm{13.6}}_{\mathrm{H}}(T_{\mathrm{eff}}) \rangle = \int^{\infty}_{\nu_{13.6}} \! \! \! \frac{\sigma_{\mathrm{H}}(\nu) B_{\nu}}{h \nu} \ \mathrm{d} \nu \Big/ \int^{\infty}_{\nu_{13.6}}  \frac{B_{\nu} }{h \nu}  \ \mathrm{d} \nu
\end{equation}
is the intensity-weighted mean photoionization cross-section for H in the $E_{13.6}$ energy bin, and where $\langle \sigma_{\rm H}^{15.2+} \rangle$ and $\langle \sigma_{\rm H_{2}} \rangle$ have the same values as before. When computing $\langle \sigma_{\rm H}^{13.6} \rangle$, we again use the simple approximation for the frequency-dependent atomic hydrogen photonionization cross-section given in equation~\eqref{simpleHcross}.

Finally, once we have computed $k_{\rm ion}$ for each ray passing through the grid cell, we sum the contributions from all of the rays to get the total photoionization rate:
\begin{equation}
k_{\rm ion, tot} = \sum_{\rm rays} k_{\rm ion}.
\end{equation}
This is then passed to the chemistry module, which solves the following rate equation for the H$^{+}$ abundance:
\begin{equation}\label{prate}
\begin{aligned}
\frac{\mathrm{d}x_{\mathrm{H^+}}}{\mathrm{d}t} = & \  C_{\rm cl} n_{\rm e} x_{\mathrm{H}} + (C_{\rm xr} + C_{\rm cr} + k_\mathrm{ion, tot}) \ x_{\mathrm{H}} + C^{*}_{\rm cr} \ x_{\mathrm{H_2}}   \\
                                                 & - \alpha_\mathrm{B} \ n_{\rm e} x_{\mathrm{H^+}} - h_{\rm gr} x_{\mathrm{H^+}}.
\end{aligned}
\end{equation}
Here, in addition to photoionization, we also account for the ionization of atomic hydrogen by cosmic rays ($C_{\rm cr}$), X-rays ($C_{\rm xr}$), and its collisional ionization by electrons ($C_{\rm cl}$).  Additional H$^{+}$ ions are also produced by the dissociative ionization of H$_{2}$ by cosmic rays ($C^*_{\rm cr}$), while H$^{+}$ ions are removed from the gas by case B radiative recombination ($\alpha_\mathrm{B}$) and recombination on the surfaces of dust grains ($h_{\rm gr}$). Full details of our implementation of all of these processes can be found in \citet{glover2010} and \citet{walch14}, and so for brevity we do not repeat them here. Finally, we note that when computing the electron abundance, we do not simply account for the electrons coming from ionized hydrogen, but also those from atomic carbon, i.e.\ $x_{\rm e} = x_{\rm H^+} + x_{\rm C^+}$.
\subsubsection{Ionizing radiation: thermal effects}
\begin{figure}
\centering
\includegraphics[scale=0.6]{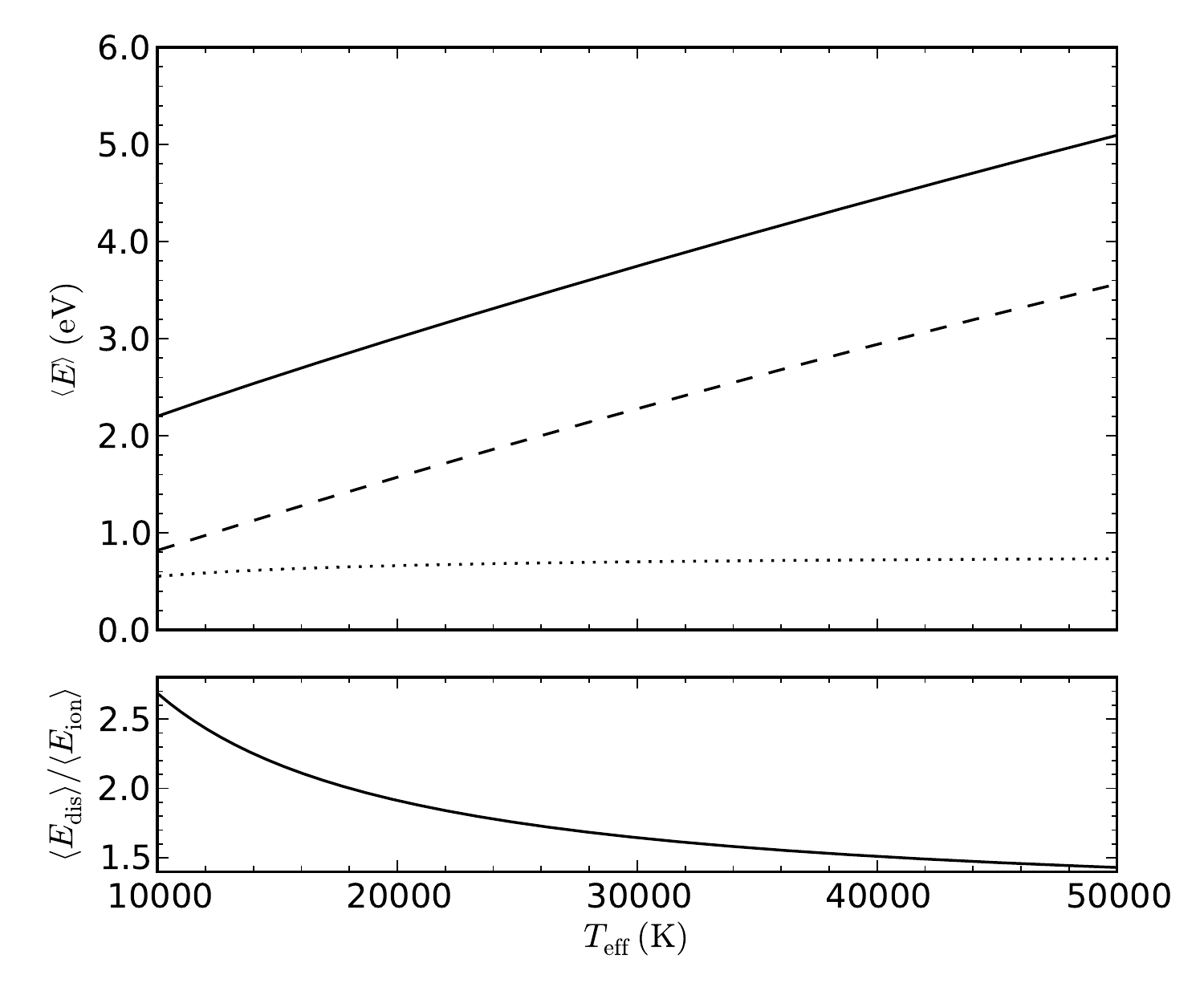}
\caption{{\em Top}: average excess energy per H ionization, for ionizing photons in the energy bins $E_\mathrm{13.6}$ (dotted) and $E_\mathrm{15.2+}$ (dashed), plotted as a function of the effective temperature of the stellar source. The solid line shows the average excess energy per H$_{2}$ ionization. {\em Bottom}:  Ratio of the heating rate per H ionization to that for H$_{2}$ ionization, considering only photons in the $E_\mathrm{15.2+}$ energy bin.}
\label{fig:avgEnergy}
\end{figure}

Each time a photon unbinds an $\mathrm{H_2}$ molecule or ionizes a hydrogen atom, its excess energy is deposited in the surrounding gas in the form of heat. The average deposited energy $ \langle E \rangle$ is given by the expression
\begin{equation}
\begin{aligned}
\left< E \right> = &\int^{\infty}_{\nu_\mathrm{0}} B_{\nu} \Big(1 - \frac{E}{E_{\nu}}\Big) \ \sigma_{\nu} \ \mathrm{d} \nu \  \Big/  \\
&\int^{\infty}_{\nu_{0}} \frac{B_{\nu}}{E_{\nu}} \ \sigma_{\nu} \ \mathrm{d} \nu
\end{aligned}
\end{equation}
where the appropriate frequency-dependent cross-sections $\sigma^{\mathrm{15.2+}}_{\mathrm{H},\nu}$, $\sigma^{\mathrm{13.6}}_{\mathrm{H},\nu}$, $\sigma_{\mathrm{H_2}, \nu}$ and lower bounds should be inserted to yield the mean energies deposited by photoionization of H due to photons in the $E_{15.2+}$ energy bin, $\langle E^\mathrm{15.2}_{\mathrm{ion}} \rangle$,
photoionization of H due to photons in the $E_{13.6}$ energy bin, $\langle E^\mathrm{13.6}_{\mathrm{ion}} \rangle$, and the dissociation of H$_{2}$ due to photoionization, $\langle E_{\mathrm{dis}} \rangle$. Figure \ref{fig:avgEnergy} shows how these quantities vary as a function of the effective temperature of the source.

Given the average energy per ionization, it is then straightforward to compute the corresponding heating rate per unit volume. We have
\begin{equation}
\begin{aligned}
&\Gamma_{\mathrm{ion}} &= &\ \left[k^\mathrm{13.6}_{\mathrm{ion}} \langle E^\mathrm{13.6}_{\mathrm{ion}} \rangle  + k^\mathrm{15.2}_{\mathrm{ion}} \langle E^\mathrm{15.2}_{\mathrm{ion}} \rangle \right] n_{\rm H},   \\
&\Gamma_{\mathrm{dis}} &= &\ \left[k_{\mathrm{dis}} \langle E_{\mathrm{dis}} \rangle \right] n_{\rm H_{2}}.
\end{aligned}
\end{equation}
These heating rates are accounted for when we determine the change in the internal energy of the gas during the timestep, as we describe in Section~\ref{energy_eq} below.

\subsubsection{Non-ionizing radiation: chemical effects}\label{non-ion}
Even though a minimum photon energy of about $15.2 \ \mathrm{eV}$ is needed to ionize molecular hydrogen \citep{liu}, less energetic photons are able to dissociate $\mathrm{H_2}$ by a two-step process \citep{stecher}. A photon with energy larger than $11.2 \ \mathrm{eV}$ is able to electronically excite $\mathrm{H_2}$ to the $B^{1}\Sigma_{u}^{+}$ or $C\ ^{1}\Pi_{u}$ states (also known as the Lyman and Werner states). As there are a large number of different bound rotational and vibrational levels in both the electronic ground state and the Lyman and Werner excited states, these electronic transitions occur via a series of discrete lines that together make up the Lyman and Werner band systems.\footnote{Since these sets of lines overlap in terms of frequency, it is common to refer to the range of energies corresponding to the combined set of lines as the Lyman-Werner band.} In most cases, the excited H$_{2}$ molecule decays back into a bound rotational and  vibrational level in the electronic ground state. However, roughly 15\% of the time, the decay occurs instead into the vibrational continuum, resulting in the dissociation of the molecule \citep{DBT1996}. 

In contrast to photoionization, which is a continuum process, H$_{2}$ photodissociation is inherently a line-driven process. This means that it is not possible to treat it in the same way that we treat photoionization, as it is important to account for the effects of H$_{2}$ self-shielding. Consider a beam of radiation propagating along a ray that passes through some molecular hydrogen. As our beam of radiation propagates through the gas, the photons that are most likely to be absorbed by the H$_2$ are those with frequencies corresponding to the center of one of the strong Lyman-Werner absorption lines. As these photons are removed from the beam, the probability of additional photons being absorbed by H$_{2}$ decreases and so the photodissociation rate declines. This effect is known as self-shielding, and it is frequently the dominant process responsible for shielding molecular gas from the effects of photodissociation. Importantly, the rate at which the Lyman-Werner absorption lines become optically thick depends on the strength of the lines: we can easily find ourselves in a situation in which the strongest lines are optically thick, while the weaker lines remain optically thin.  In addition, even once the strongest lines become optically thick at their centers, the wings of the line often remain optically thin. As a result, the photodissociation rate at a given point along the ray depends in a non-trivial way on the H$_{2}$ column density and the velocity structure between that point and the source of the radiation. Consequently, we cannot easily use the ray segment method to treat the effects of photodissociation, as the reduction in the photodissociation rate that occurs as our ray crosses any given grid cell does not depend simply on the column density of H$_{2}$ within that cell, but also on the total column density of H$_{2}$ between the cell and the source.


We therefore employ a modified version of equation~\eqref{nonconservative} to treat the H$_2$ photodissociation rate:
\begin{equation}\label{uvrate}
\begin{aligned}
k_{\mathrm{UV}} = N_{\mathrm{11.2}}(r=0) \mathrm{e}^{-\tau_{\rm d}}  \frac{ \sigma_{\mathrm{UV}} \ \mathrm{d}r}{V_{\mathrm{cell}} \ \Delta t}.
\end{aligned}
\end{equation}
Here $N_{\mathrm{11.2}}(r=0) = P_{\mathrm{11.2}}(r=0) \Delta t$ is the absolute number of UV photons with energies 
$11.2 < E < 13.6$~eV emitted by our source in a timestep $\Delta t$, $\sigma_{\rm UV}$ is the effective photodissociation cross-section, 
and $\tau_{\rm d}$ is the mean optical depth in the Lyman-Werner band due to dust absorption.

The effective photodissociation cross-section, $\sigma_{\rm UV}$, is defined as the ratio of the photodissociation rate $D$ to the photon flux in the Lyman-Werner bands, $F$, i.e.\ $\sigma_{\rm UV} = D / F$. We evaluate this using the photodissociation rate in the optically thin limit, $D = 5.18 \times 10^{-11}  \ \mathrm{s^{-1}}$ \citep{roellig2007}, which was derived for the \citet{draine1978} interstellar radiation field (ISRF). The ISRF approximates the average spectrum found in the interstellar medium (ISM) originating from reprocessed starlight. With the photon flux in the Lyman-Werner bands that one has for the same ISRF, namely $F = 2.1 \times 10^7 \ \mathrm{s^{-1} cm^{-2}}$ \citep{DBT1996}, we find $\sigma_{\rm UV, thin} = 2.47 \times 10^{-18} \ \mathrm{cm^{2}}$ in optically thin gas. As the H$_{2}$ column density increases, however, the gas starts to become optically thick in the Lyman-Werner lines. This reduces $D$ but has little effect on $F$ until we reach extremely large H$_{2}$ column densities. To account for this, we write the effective photodissociation cross-section as
\begin{equation}
\sigma_{\rm UV} = \sigma_{\rm UV, thin} f_{\rm shd},
\end{equation}
where $f_{\rm shd}$ is a self-shielding function that parameterizes how $D$ decreases as the H$_{2}$ column density increases.  
For this shielding function, we adopt the expression given in \citet{DBT1996}:
\begin{equation}
\begin{aligned}
f_{\mathrm{shd}} = \frac{0.965}{(1+x/b_5)^2} + \frac{0.035}{(1+x)^{1/2}} \times \\
           \exp \left[{-8.5 \times 10^{-4} (1+x)^{1/2}}\right].
\end{aligned}
\end{equation}
Here, $x = N_{\mathrm{H_2}}/(5 \times 10^{14} \ \mathrm{cm}^{-2})$, where $N_{\rm H_{2}}$ is the column density of H$_{2}$ along the ray between the cell and the source, and $b_{5} = b / 10^{5} \: {\rm cm \: s^{-1}}$, and $b$ parameterizes the change in shielding behavior for lines broadened due to the Doppler effect. For simplicity, in the current version of \texttt{Fervent} we do not relate $b$ to the temperature or velocity distribution of the H$_{2}$ along the ray, but instead simply set $b_{5} = 1.0$. We plan to relax this assumption in future work, but do not expect that it will make a large difference to the outcome of our models.

We assume that the dust has properties characteristic of the diffuse ISM. In this case, $\tau_{\rm d} = \gamma A_{\rm V}$ with $\gamma = 3.5$, and the visual extinction $A_{\rm V}$ is related to the total hydrogen column density via \citep{DBT1996}
\begin{equation}
A_{\rm V} = \frac{N_{\rm H} + 2 N_{\rm H_{2}} + N_{\rm H^{+}}}{1.87 \times 10^{21} \ \mathrm{cm^{-2}}} f_{\rm d/g}, 
\end{equation}
where $f_{\rm d/g}$ is the dust-to-gas ratio, normalized to the value in the local ISM. The effective dust absorption cross-section per hydrogen nucleus is therefore $\sigma_{\rm d} = \tau_{\rm d}/( 1.87 \times 10^{21} ) \ \mathrm{cm^{2}} = 1.34 \times 10^{-21}  f_{\rm d/g} \ \mathrm{cm^2}$. This is much smaller than $\sigma_{\rm UV, thin}$, and dust attenuation only becomes important when the H$_{2}$ fraction is low or when there is a very large column density of H$_{2}$ between the cell and the source. We are therefore justified in treating the effects of H$_{2}$ self-shielding prior to those of dust absorption when accounting for the photons absorbed in the grid cell. Although not important in the present context, this becomes important once we consider photoelectric heating, as we have to take care to avoid the double-counting of photons: those absorbed by H$_{2}$ are not available to be absorbed by dust and hence cannot contribute to the photoelectric heating rate. We return to this point in Section~\ref{pe} below.

Putting everything together, our final rate equation for the molecular hydrogen abundance reads:
\begin{equation}
\begin{aligned}
\frac{\mathrm{d} x_\mathrm{H_2}}{\mathrm{d}t} = \ & - (C_\mathrm{cl,\mathrm{H}} \ n_\mathrm{H}+ C_\mathrm{cl,\mathrm{e}} \ n_\mathrm{e} + C_\mathrm{cl,\mathrm{H_2}} \  n_\mathrm{H_2} ) \ x_\mathrm{H_2}  \\
&- ( k_{\mathrm{cr}} + k_{\mathrm{ISRF}} +  k_{\mathrm{UV}} + k_{\mathrm{dis}}) \ x_\mathrm{H_2} + R_{\rm d} x_\mathrm{H},
\end{aligned}
\end{equation}
where  $C_\mathrm{cl,\mathrm{H}}$, $C_\mathrm{cl,\mathrm{e}}$ and $C_\mathrm{cl,\mathrm{H_2}}$ are the rate coefficients for the collisional dissociation of H$_{2}$ by H atoms, electrons and H$_{2}$ molecules, respectively, $k_{\mathrm{cr}}$ is the rate at which H$_{2}$ is destroyed by cosmic rays, $k_{\mathrm{ISRF}}$ is the rate at which it is photodissociated by the ISRF, and $R_{\rm d}$ is the rate at which it forms on dust grains. For more details on all of these processes and their implementation in \texttt{FLASH}, we refer the reader to \citet{glover2010} and \citet{walch14}.

Finally, we should also briefly consider the other main chemical effect of the non-ionizing photons in our model, namely the photodissociation of CO. The CO photodissociation threshold is 11.09~eV, similar to that of H$_2$, and so in principle both CO and H$_{2}$ compete for the same set of photons in the $E_{11.2}$ energy bin. However, the details of the CO self-shielding are quite different from those of the H$_{2}$ self-shielding, and in principle one should also account for the shielding of CO by H$_{2}$ and vice versa \citep[see e.g.][]{visser09}. Properly accounting for these effects would add significant complexity to the code, but is of questionable utility given our highly simplified treatment of CO chemistry, which is known to overproduce CO by a factor of a few \citep{glover2012}. Therefore, for the time being we have made the major simplification of assuming that the CO photodissociation rate scales directly with the total H$_{2}$ dissociation rate, i.e.\
\begin{equation}
\begin{aligned}
k_{\mathrm{CO}} = f_{\mathrm{conv}} (k_{\mathrm{UV}} + k_{\mathrm{dis}}).
\end{aligned}
\end{equation}
The conversion factor is taken to be $f_{\mathrm{conv}} = 3.86$ \citep{roellig2007}, the ratio of the $\mathrm{CO}$ to $\mathrm{H_2}$ photodissociation rates in optically thin gas. We note also that it is important to include the $k_{\rm dis}$ term to ensure that CO cannot survive in highly ionized regions, as this behavior would be unphysical.
This approximation produces reasonable results in highly irradiated gas, but predicts that $k_{\rm CO}$ falls off too rapidly in the $A_{\rm V} \sim 0.1$--1 regime. As our tests in Section~\ref{PDRtest} demonstrate, this can result in an order of magnitude overestimation of the total CO column density in optically thick gas. Our current scheme is therefore suitable for predicting whether a given molecular cloud or clump is likely to be CO-bright or CO-faint, but not for making detailed predictions of e.g.\ CO or C$^{+}$ emissivities. Note, however, that our overestimation of the CO abundance is unlikely to significantly affect the thermal structure of the gas. CO is a more effective coolant than C$^{+}$ only in very cold gas \citep[see e.g.][]{gc12b}, but the region where we overestimate its abundance is warmed by the same radiation responsible for destroying the H$_{2}$ and CO (see e.g.\ Section~\ref{PDRtest}).

\subsubsection{Non-ionizing radiation: thermal effects}\label{non-ion-heat}
Each Lyman-Werner band photodissociation of an H$_{2}$ molecule deposits $\langle E_{\mathrm{UV}} \rangle$ of energy as heat. Typically, we find that $\langle E_{\mathrm{UV}} \rangle
\simeq 0.4$~eV \citep{black1977}. In addition to this, UV photons can also heat the gas by indirectly exciting the vibrational levels of the H$_2$ molecule \citep{burton1990}. As we have already mentioned, the absorption of a Lyman-Werner band photon results in photodissociation only around 15\% of the time. The rest of the time, the electronically excited H$_{2}$ molecule decays back to a bound ro-vibrational level in the electronic ground state. A small fraction of these decays put the molecule back in the $v = 0$ vibrational ground state, but in most cases, the $\mathrm{H_2}$ molecule is left with a considerable residual internal energy in the form of vibrational excitation. In low density gas, this energy is simply radiated away as the H$_{2}$ molecule undergoes a series of radiative transitions that eventually place it back in the vibrational ground state. In dense gas, however, collisional de-excitation can be more effective than radiative de-excitation, and in this case most of this energy is redistributed as heat. 

The rate at which vibrationally-excited H$_{2}$ is produced -- often referred to as the UV pumping rate -- is related to the H$_{2}$ photodissociation rate by 
\begin{equation}
k_{\mathrm{pump}} = f_{\mathrm{pump}} k_{\mathrm{UV}}.
\end{equation}
Following the notation of \cite{DBT1996}, the conversion factor between the two rates is given by
\begin{equation}
f_{\mathrm{pump}} = \frac{1.0 -  \langle p_\mathrm{diss} \rangle - \langle p_\mathrm{ret} \rangle}{\langle p_\mathrm{diss} \rangle},
\end{equation}
where $\langle p_\mathrm{diss} \rangle$ is the mean photodissociation probability and $\langle p_\mathrm{ret} \rangle$ is the mean probability that the molecule decays directly back into the $v=0$ level of the electronic ground state. These values vary a little depending on the shape of the incident spectrum, the degree of H$_{2}$ self-shielding that is occurring, and the density and temperature of the gas, but in typical ISM conditions, $f_{\rm pump} = 6.94$ \citep{DBT1996}.


The mean energy that is converted to heat per UV pumping event has been calculated by \citet{burton1990} and can be written as
\begin{equation}
\langle E_{\rm pump} \rangle = 2 \: {\rm eV}  \times \frac{C_\mathrm{dex}}{C_\mathrm{dex}+C_\mathrm{rad}}, 
\end{equation}
where $C_\mathrm{dex}$ is a representative value for the collisional de-excitation rate, given by \citet{burton1990} as 
\begin{equation}
\begin{aligned}
C_\mathrm{dex} = &10^{-12}(1.4 \ \mathrm{e}^{-18100 /(T+1200)} x_{\mathrm{H_2}} + \\
                 &1.0 \ \mathrm{e}^{-1000/T} x_{\mathrm{H}}) \sqrt{T} \ n,
\end{aligned}
\end{equation}
where $T$ is the temperature in Kelvin, n is the gas number density in $\rm{cm^{-3}}$, and where $C_{\rm rad} = 2 \times 10^{-7} \ \mathrm{s^{-1}}$ is a representative value for the radiative de-excitation rate. In warm atomic gas, $C_{\rm dex} \sim C_{\rm rad}$ for number densities around $n \sim 10^{4} \: {\rm cm^{-3}}$, while in cold molecular gas an even higher density is needed. It is therefore clear that this process is important only in relatively dense PDRs.

Finally, putting these two contributions together, we can write the total heating rate per unit volume due to the absorption of Lyman-Werner band photons as
\begin{equation}
\begin{aligned}
\Gamma_\mathrm{UV} = & \left( k_{\mathrm{pump}} \langle E_{\mathrm{pump}} \rangle + k_{\rm UV} \langle E_{\rm UV} \rangle \right) n_{\rm H_{2}}.
\end{aligned}
\end{equation}

\subsubsection{Photoelectric heating}\label{pe}
The final important effect that we need to account for is photoelectric heating.
Photons with energies greater than $5.6 \ \mathrm{eV}$ are able to dislodge electrons from dust grains, and these electrons go on to collisionally heat the surrounding gas. We assume that the only photons that contribute to the photoelectric heating rate are those in the $E_{5.6}$ and $E_{11.2}$ energy bins, since all of the photons in the two higher energy bins will be consumed in the ionization of H or (if energetic enough) H$_{2}$.
The lower end of the photo-electric heating energy range is not well defined and we adopt for simplicity the cut-off used in \cite{DBT1996} for the ISRF. 

We follow the prescription by \cite{bakestielens1994}, who assume a dust grain size distribution that extends down to structures as small as polycyclic aromatic hydrocarbons (PAHs) and calculate the heating rate $\Gamma_\mathrm{pe}$ in $\mathrm{erg \ cm^{-3} \ s^{-1} }$, which takes into account electron and ion recombination cooling
\begin{equation}\label{elheat}
\begin{aligned}
\Gamma_\mathrm{pe} = 1.3 \times 10^{-24} \ \epsilon G_0 f_{\rm d/g} n.
\end{aligned}
\end{equation}
The photoelectric heating efficiency is given by
\begin{equation}\label{peeff}
\begin{aligned}
\epsilon  = &\frac{4.9 \times 10^{-2}}{1+4 \times 10^{-3} (G_0 \sqrt{T}/n_\mathrm{e}\phi_{\mathrm{pah}})^{0.73} }  \\ 
            &+\frac{3.7 \times 10^{-2} (T/10^4)^{0.7} }{1+2 \times 10^{-4}(G_0 \sqrt{T}/ n_\mathrm{e}\phi_{\mathrm{pah}} ) },
\end{aligned}
\end{equation}
where $G_0$ is the strength of the local, incident interstellar radiation field normalized to the integrated Habing field \citep{habing1968}, $T$ is again the temperature in Kelvin and $n_{\rm e}$ is the electron number density in units of $\rm{cm^{-3}}$. 
The version of $\epsilon$ shown here is taken from \citet{wolfire2003}, and includes a dimensionless scaling parameter $\phi_\mathrm{pah}$ that was not included in the original \citet{bakestielens1994} prescription, as well as a $30\%$ larger pre-factor in equation \eqref{elheat}. These modifications were introduced by \citeauthor{wolfire2003} to account for the fact that observations of the ISM by the Infrared Space Observatory imply the presence of a larger abundance of PAHs than was assumed in \citeauthor{bakestielens1994}'s original treatment. In general, we follow \citeauthor{wolfire2003} and set $\phi_\mathrm{pah} = 0.5$.
\begin{figure*}
\centering
\includegraphics[scale=0.6]{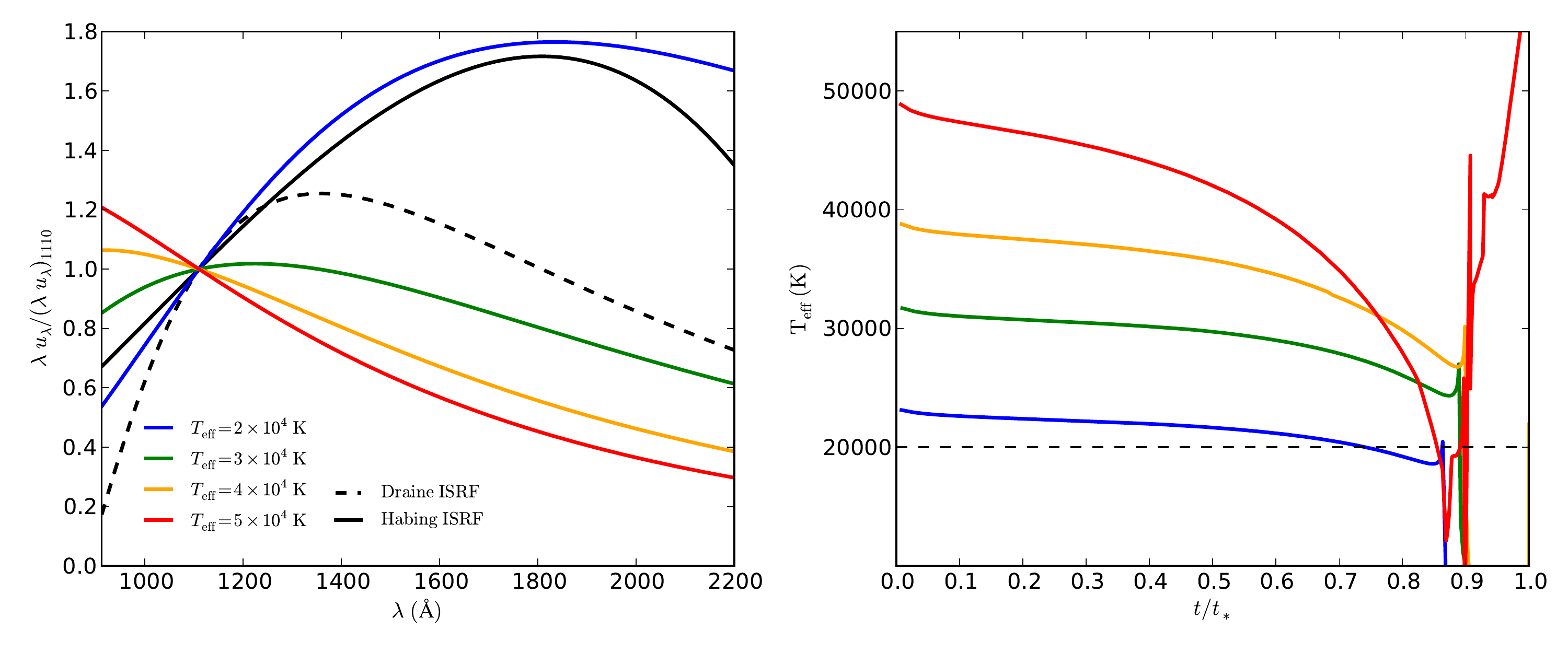}
\caption{\label{fig:ISRFcomparison} {\em Left panel}: blackbody spectra for different effective temperatures as well as the Habing and Draine ISRFs. All spectra are normalized to their energy density at $1110 \ \mathrm{\AA} $. The black dashed line is the Habing ISRF and the black solid line is the Draine ISRF. Colored solid lines correspond to effective temperatures of $2 \times 10^4$ (blue), $3 \times 10^4$ (green), $4 \times 10^4$ (orange) and $5 \times 10^4$ (red)  Kelvin. {\em Right panel}: the evolution of the effective temperatures over the lifetime $t_*$ of typical massive stars with tracks taken from \citet{ekstromA} and \citet{georgy}. Here the colors denote the mass of the star, where the more massive stars have a higher blackbody temperature. The masses are, from top to bottom, $60$, $25$, $15$ and $8 \ \mathrm{M_{\odot}}$.}
\end{figure*}

One obvious question that arises at this point is whether it is valid to take a prescription for the photoelectric heating rate that was designed primarily to model the behavior of the diffuse ISM and to apply it to a dense PDR. After all, the interstellar radiation field seen by a representative patch of the diffuse ISM has a spectral shape which differs significantly from that of a blackbody. Therefore,  before we can apply the \citet{bakestielens1994} prescription for photoelectric heating, we have to make sure that the assumptions that they made in the derivation of $\Gamma_\mathrm{pe}$ are not too strongly violated for our radiative transfer method. There are two main concerns: the spectral shape of the ISRF is not that of a blackbody, and absorption by dust is not the only attenuating process in the relevant energy range. 

We quantify the deviation from the ISRF by taking a reference wavelength of $1110 \ \mathrm{\AA}$ (i.e.\ 11.17~eV) and normalize blackbody spectra at different effective temperatures so that they all have the same energy density at this reference wavelength (see Figure~\ref{fig:ISRFcomparison}). For effective temperatures $T_{\rm eff} = 20000$--50000~K that are relevant for massive stars, we find that the energy densities of our normalized spectra at the wavelengths of interest typically differ by no more than a factor of three from the \citet{draine1978} model for the ISRF. 
\cite{bakestielens1994} calculate the heating rate for a radiation field with the shape of a $T_\mathrm{eff} = 3 \times 10^4 \ \mathrm{K}$ blackbody and compare this to the results they obtain for the Draine ISRF, normalized so that $G_{0}$ is the same for both spectra. They  find that in this case, the difference in spectral shape makes around a 25\% difference to the photoelectric heating rate. A spectrum with effective temperature of $T_\mathrm{eff} = 2 \times 10^4 \ \mathrm{K}$ nearly reproduces the Draine field and so in this case, we would expect the error to be much smaller, around a few percent. For higher effective temperatures of  $4 \times 10^4 \ \mathrm{K}$ and $5 \times 10^4 \ \mathrm{K}$,  the shape of the spectrum does not change drastically from the reference $3 \times 10^4 \ \mathrm{K}$ blackbody, and so although the error will be larger, we would still expect it to be less than 50\%. We are therefore justified in using the same efficiency parameter  $\epsilon$ as in \citet{wolfire2003}, as any errors introduced by the difference in spectra shape are relatively small and are probably dwarfed by the uncertainties stemming from other features of the dust physics, such as the PAH abundance in dense clouds or the exact geometric shape of the dust grains.

As a star evolves on the main sequence and beyond, its spectrum changes. We use non-rotating, non-magnetized solar metallicity tracks for typical massive stars taken from \citet{ekstromA} and \citet{georgy} to illustrate that these stars stay inside the explored range of $T_\mathrm{eff}$ (see the right hand side of Figure~\ref{fig:ISRFcomparison}).
A massive star enters the Wolf-Rayet phase at the end of its lifetime $t_*$, at which point its effective temperature varies rapidly. For this stage, our modeling for the photoelectric heating efficiency  breaks down as equation~\eqref{elheat} is no longer applicable.  Fortunately, only a short period of about $10$ to $15 \%$ of $t_*$ is spent in this phase.

The other issue that we need to address is how to properly account for the fact that photons in the $E_{11.2}$ energy bin can be absorbed by either dust or H$_{2}$. In principle, we should treat the photons in this energy bin in the same careful way that we treat the photons in the $E_{15.2}$ energy bin, by first calculating the total number that are absorbed and then partitioning them between H$_{2}$ and dust. However, in practice this proves to be unnecessary. The reason is that in the regime where both H$_{2}$ photodissociation and photoelectric heating are important, the effective H$_{2}$ photodissociation cross-section $\sigma_{\rm H_{2}}$ is orders of magnitude larger than the effective dust absorption cross-section, which for typical ISM dust peaks at around 700~\AA~with a value of $\sigma_\mathrm{d} \approx 3 \times 10^{-21} \ \mathrm{cm^2}/\mathrm{H} $ and then drops off rapidly at both larger and smaller wavelengths \mbox{\citep{draine2003,weingartner2001a}}.  Therefore, in these conditions H$_{2}$ absorption dominates. This means that rather than working out the number of photons absorbed by the combined effects of H$_{2}$ absorption and dust absorption and then partitioning these photons up between the two absorbers, we can instead safely split up the calculation into two steps: we first compute how many photons are absorbed by H$_{2}$ and remove these from the $E_{11.2}$ energy bin, and only then use the remaining flux to estimate $G_{0}$ and $\Gamma_{\rm pe}$.


In practice, it is convenient to use the incident energy flux, $F_{\mathrm{pe}}$ in $\rm{erg \: s^{-1}}$, in place of the photon flux when we estimate the local stellar intensity $G_{0}$ in units of the Habing field. This consists of two contributions. The first term treats photons with energies in the range of $5.6 \ \mathrm{eV}$ to $11.2 \ \mathrm{eV}$. These photons are unable to electronically excite H$_{2}$ from the ground state and so are only absorbed by dust. The second term adds the contribution from photons in the energy range $11.2 \ \mathrm{eV}$ to $13.6 \ \mathrm{eV}$ that have not already been absorbed by molecular hydrogen. We therefore have:
\begin{equation}
\begin{aligned}
F_{\mathrm{pe}} = & \  F_{\mathrm{5.6}}(r=0) \ \mathrm{e}^{- \gamma A_{\rm V}} +\\
                   & ( f_\mathrm{shd} N_{11.2} \mathrm{e}^{-\tau_{\rm d}} / \Delta t - (1 + f_\mathrm{pump}) \  k_\mathrm{UV} ) \langle E_{\mathrm{11.2}} \rangle
\end{aligned}
\end{equation}
where $F_{\mathrm{5.6}}(r=0)$ is the energy flux at the stellar surface in the energy range from $5.6 \ \mathrm{eV}$ to $11.2 \ \mathrm{eV}$, which we evenly distribute over the initial rays, and $\gamma = 2.5$, taken from \citet{bergin2004}. To determine the FUV energy flux, we first determine the FUV photon flux entering the cell and then subtract from this the photon flux absorbed by H$_{2}$ in the cell. Finally, to convert the result from a photon flux to an energy flux, we multiply by the average energy $ \langle E_{\mathrm{11.2}} \rangle$ of the photons in the $E_{11.2}$ energy bin.

The energy flux $F_{\mathrm{5.6}}(r=0)$ is carried along the ray as a ray property and attenuation is always calculated from this value. The process is not photon-conservative as we do not equate the extinction of the photo-electric heating photons with the eventual number of photons that heat the surrounding gas, but this is unlikely to introduce a significant error into our calculation of the heating rate.

After the ray-tracing step is complete and the energy flux of all contributing rays in each cell is added up, $F_{\mathrm{pe}}$ has to be expressed in units of the integrated Habing field as input for equation \eqref{elheat}. We convert $F_{\rm pe}$ to an approximate flux per unit area by dividing it by the area of one face of the cell, $d^{2}$. We then divide the resulting 
energy flux per unit area by the fiducial value of $G_\mathrm{conv} = 1.6 \times 10^{-3} \ \mathrm{erg \ s^{-1} \ cm^{-2}}$ \citep{habing1968} to obtain $G_{0}$, i.e.\
\begin{equation}\label{pefinal}
\begin{aligned}
G_0 = \frac{F_{\mathrm{pe}}}{d^2 G_\mathrm{conv}}. 
\end{aligned}
\end{equation}
Note that this conversion to $G_0$ naturally accounts for adaptively refined cells with different physical sizes $d$.  

\subsubsection{Solving the energy equation}
\label{energy_eq}
In order to solve for the evolution of the internal energy, we need to know the heating rates due to photoionization, H$_{2}$ photodissociation, the UV pumping of H$_{2}$ and photoelectric heating, which are computed as described in the previous sections. In addition, we need to know the radiative cooling rate of the gas, and must also account for any other chemical heating or cooling terms (e.g.\ H$_{2}$ formation heating, H$^{+}$ recombination cooling). To compute these additional terms, we use a modified version of the heating and cooling function introduced in \citet{glover2010} and updated in \citet{glover2012}. This accounts for all of the important chemical heating and cooling terms, as well as radiative cooling from the fine structure lines of C$^{+}$ and O, the rotational and vibrational lines of H$_{2}$ and CO, the electronic transitions of atomic hydrogen (``Lyman-$\alpha$'' cooling) and thermal emission from dust. We have supplemented this by including the additional forbidden and semi-forbidden transitions of C$^{+}$, O, O$^{+}$ and N$^{+}$ that are summarized in Table~9 of \citet{hm89}, as these become important coolants close to $T \sim 10^{4}$~K and hence make an appreciable difference to the equilibrium temperature of our H{\sc ii} regions. As we do not track O$^{+}$ and N$^{+}$ explicitly in our chemical network, we make the simplifying assumption that $x_{\rm O^{+}} / x_{\rm O} = x_{\rm N^{+}} / x_{\rm N}
= x_{\rm H^{+}} / x_{\rm H}$. Finally, we also include cooling from the electronic transitions of atomic helium and from metals, using the rates tabulated by \citet{gnat}. These rates assume that the gas is in collisional ionization equilibrium, which is not the case within our H{\sc ii} regions, since the latter are dominated by photoionization. However, these processes become important in comparison to Lyman-$\alpha$ cooling only at $T \gg 10^{4} \: {\rm K}$, and hence do not strongly affect the thermal evolution of the gas in our simulations. Nevertheless, these high temperature cooling processes are needed for a complete treatment of the ISM temperature structure and will be neccessary for future extensions of the simulation code.

Putting all of these pieces together, we can write the final net heating/cooling rate as
\begin{equation}\label{hrate}
\begin{aligned}
\Gamma_{\mathrm{tot}} = \ &( \Gamma - \Lambda )_{\rm G10} - \Lambda_{\mathrm{HM89}} - \Lambda_{\mathrm{GF12}} \  \\
                        &+ \Gamma_{\mathrm{UV}} + \Gamma_{\mathrm{pe}} + \Gamma_{\mathrm{ion} }  + \Gamma_{\mathrm{dis}},
\end{aligned}
\end{equation}
where $( \Gamma - \Lambda )_{\rm G10}$ denotes the heating and cooling function from \citet{glover2010}, $\Lambda_{\mathrm{HM89}}$ corresponds to the forbidden and semi-forbidden line cooling from \citet{hm89} and $\Lambda_{\mathrm{GF12}}$ corresponds to the high-temperature helium and metal-line cooling from \citet{gnat}.

\subsubsection{Missing physics}\label{miss}
In our radiative transfer scheme we do not treat any scattering processes or diffuse radiation. This leads to shadows, cast by optically thick gas, that are too sharp. \\ 
In addition, radiation pressure in the form of momentum imparted on $\mathrm{H}$ nuclei is also not taken into account. It could be included in a similar fashion as in \cite{moray}, who modeled  dust-free primordial gas, but as we aim for a dusty present-day ISM, an isolated treatment acting only on hydrogen is incomplete. Ideally, an effective radiation pressure based on the complete spectrum, dust composition and size distribution should be implemented. Furthermore, in harsh environments such as in close proximity to massive stars the assumption that dust is tightly coupled to gas would have to be reevaluated.

We do not include any ionization of other elements besides hydrogen. Energetically, only helium has a small effect on the overall energy deposition as it makes up about $10 \%$ of the ISM in abundance. Other elements, such as oxygen or nitrogen, only exist in trace amounts unable to capture any significant amount of radiation. We can roughly estimate the upper limit of the thermal energy deposition from ionization of helium for the most extreme blackbody temperature of $T_\mathrm{eff} = 50000 \ \mathrm{K}$ we consider here, by considering an ISM that consists only of helium and one that is pure hydrogen. Integration of the blackbody spectrum with an He ionization cross-section taken from \cite{Verner96} yields a total possible energy deposition ten times less in the case of the helium ISM in comparison to the hydrogen one. Taking into account that only around one-tenth of the interstellar medium is made up of helium, 
this roughly amounts to a percent effect in additional energy converted from the radiation field.

\section{Tests}\label{tests}
In this section we study the influence of spatial and temporal resolution on the expansion of ionization and photo-dissociation fronts. Our testing strategy follows the standard approach of using simplified setups to check the included physics piece by piece and then in combination. The first three tests only include atomic hydrogen ionization, in a static and dynamic density field, with known analytical solutions for the expansion of the ionization front in its initial (R-type) and later stage (D-type). 

The fourth test checks the non-ionizing radiation coupling to the chemical network in a test designed for photodissociation region (PDR) codes. It first probes the chemical state of the gas at fixed temperature, then the thermal state of the gas is allowed to change. The last two test cases are of a qualitative nature, where we include all modeled physics. Here, we check for any apparent numerical artifacts or unexpected behavior in the radiation-gas coupling.

\subsection{R-type ionization front expansion}\label{rtype}
The ionization front (I-front) expansion velocity during its R-type phase is much greater than the sound speed of the ambient neutral gas.  Hence, its supersonic expansion leaves no time for the photo-heated, over-pressured gas to compress the surrounding medium. Numerically, simulating the R-type phase tests the ray-tracing and rate calculation algorithm as well as the chemical network, although for simplicity we ignore H$_{2}$ in this test and simply consider the ionization of atomic hydrogen. 

\cite{stromgren} derived the size of an ionized (H{\sc ii}) region by considering the equilibrium between photoionization and recombination. In a uniform density gas of pure hydrogen, the ionized region is spherical (a Str\"omgren sphere) with radius
\begin{equation}\label{ström}
\begin{aligned}
R_{\rm s} = \left( \frac{3}{4 \pi} \frac{Q}{n^2 \alpha_\mathrm{B}} \right)^{1/3},
\end{aligned}
\end{equation}
where $n$ is the hydrogen nuclei number density of the ambient medium, $Q = \mathrm{d} N_\mathrm{ion} / \mathrm{d} t$ the rate at which ionizing photons are produced by the source, and $\alpha_\mathrm{B}$ is the case B hydrogen recombination rate, where recombination to the ground state is excluded, because the emitted photon is assumed to be quickly absorbed again by a hydrogen atom in the vicinity with no net change in the ionization fraction.
The time evolution of the radius of the ionization front as it approaches this equilibrium solution can be recovered by considering the possible fates for the ionizing photons. In a dust-free gas of pure hydrogen, all of these photons are either absorbed within the H{\sc ii} region, where the ionization state has to be maintained against  continuous  recombination, or propagate through it and reach the shell-like ionization front at position $r_\mathrm{i}$, where they ionize the gas, enlarging the H{\sc ii} region. If ${\rm d}N_{\rm ion}$ photons reach the I-front, then they ionize a thin shell with thickness ${\rm d}r_{\rm i}$, where ${\rm d}N_{\rm ion} = 4 \pi n r^2_{\rm i} {\rm d}r_{\rm i}$. It therefore follows that
\begin{equation}\label{eq:ionization}
\begin{aligned}
\frac{\mathrm{d} N_\mathrm{ion}}{\mathrm{d}t} = 4 \pi r_\mathrm{i}^2 n \frac{\mathrm{d} r_\mathrm{i}}{\mathrm{d}t} + \frac{4}{3} \pi r_\mathrm{i}^3 n^2 \alpha_\mathrm{B} 
\end{aligned}
\end{equation}
where the second term on the right-hand side accounts for recombinations within the H{\sc ii} region, which we have assumed is almost completely ionized. Rearranging this expression and solving for $r_{\rm i}(t)$ then yields
\begin{equation}
\begin{aligned}
r_\mathrm{i}(t) = R_{\rm s} (1-\mathrm{e}^{-t/t_\mathrm{rec}})^{1/3},
\end{aligned}
\end{equation}
where $t_\mathrm{rec} = (\alpha_\mathrm{B} n)^{-1}$ is the recombination time. 

For our test, we set up a simulation in which we fix $\alpha_{\rm B}$ to the constant value $\alpha_\mathrm{B} = 2.59 \times 10^{-13} \ \mathrm{cm^3 \ s^{-1}} $ and consider gas that has a quasi-isothermal equation of state, with $\gamma = 1.0001$. The atomic hydrogen number density is set to $n = 10^{-3} \ \mathrm{cm^{-3}}$, and we consider a 
simulation domain that extends from 0 to $6.4 \ \mathrm{kpc} $ in the $x$-, $y$- and $z$ directions. The source emits $Q = 10^{49}$ photons per second, and we assume that these photons are monochromatic, with energy $E = 13.6 \ \mathrm{eV}$, the ionization threshold of hydrogen. This choice means that there is essentially no photo-heating, and also allows us to consider a fixed photoionization cross section, $\sigma_\mathrm{H} = 6.3 \times 10^{-18} \ \mathrm{cm^2}$. 
We capture an octant of the spherical I-front expansion by positioning the source at the origin. For the purposes of this simple test, we neglect the effects of dust and consider no chemical processes other than photoionization and radiative recombination.

\begin{figure}
\centering
\includegraphics[scale=0.7]{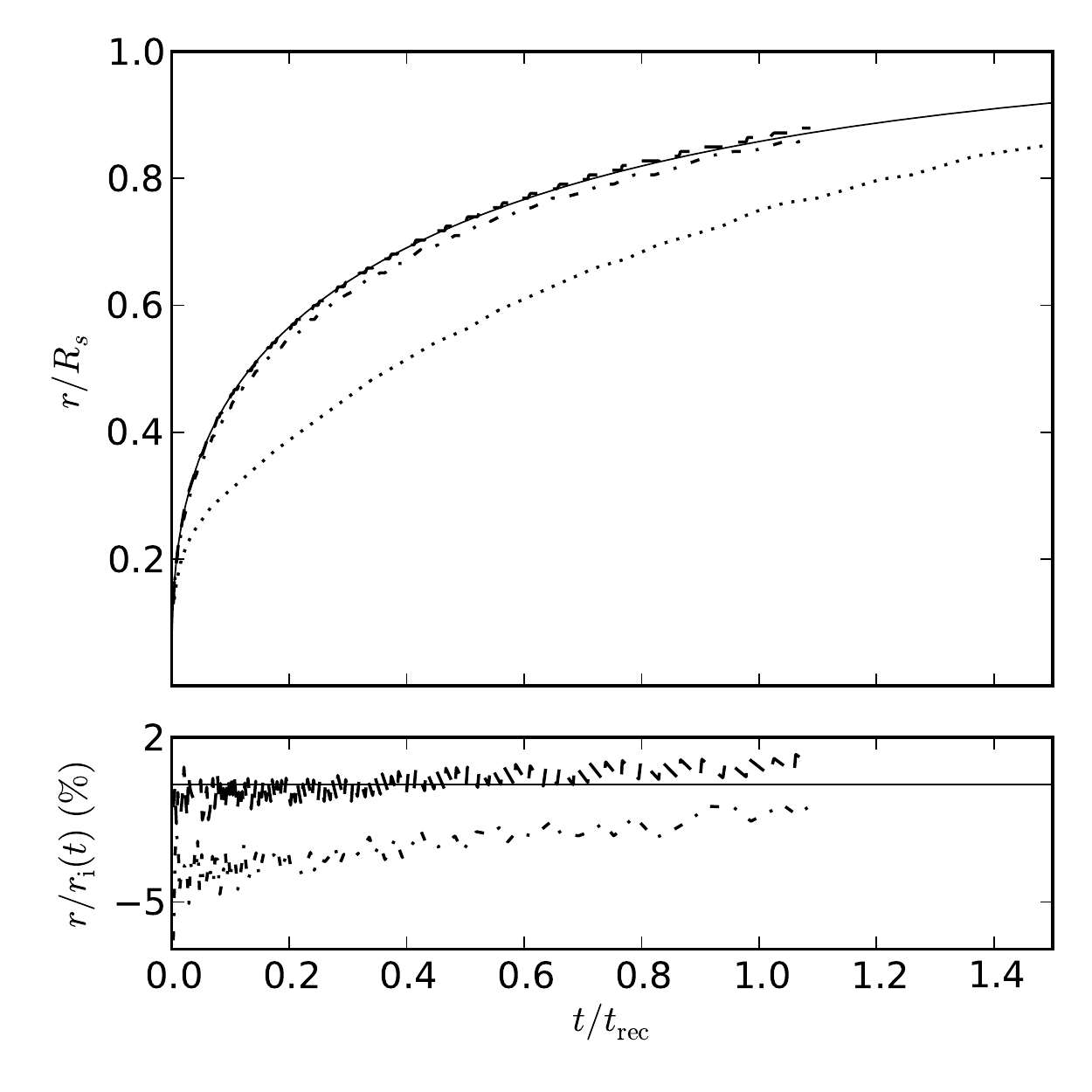}
\caption{{\it Top}: R-type ionization front expansion in a uniform medium. The solid line shows the analytical solution, the dotted line is without a timestep limiter, the dash-dotted line shows the expansion with the change per timestep in the atomic hydrogen fraction limited to 10\% (i.e.\ $f_{\rm H} = 0.1$), and the dashed line shows the case where the change is limited to 1\%. {\it Bottom}: Relative error in our solution compared to the analytic solution. The line styles are the same as in the top panel, but we omit the case without the timestep limiter. \label{fig:rtype_timestep}}
\end{figure}

\begin{figure}
\centering
\includegraphics[scale=0.7]{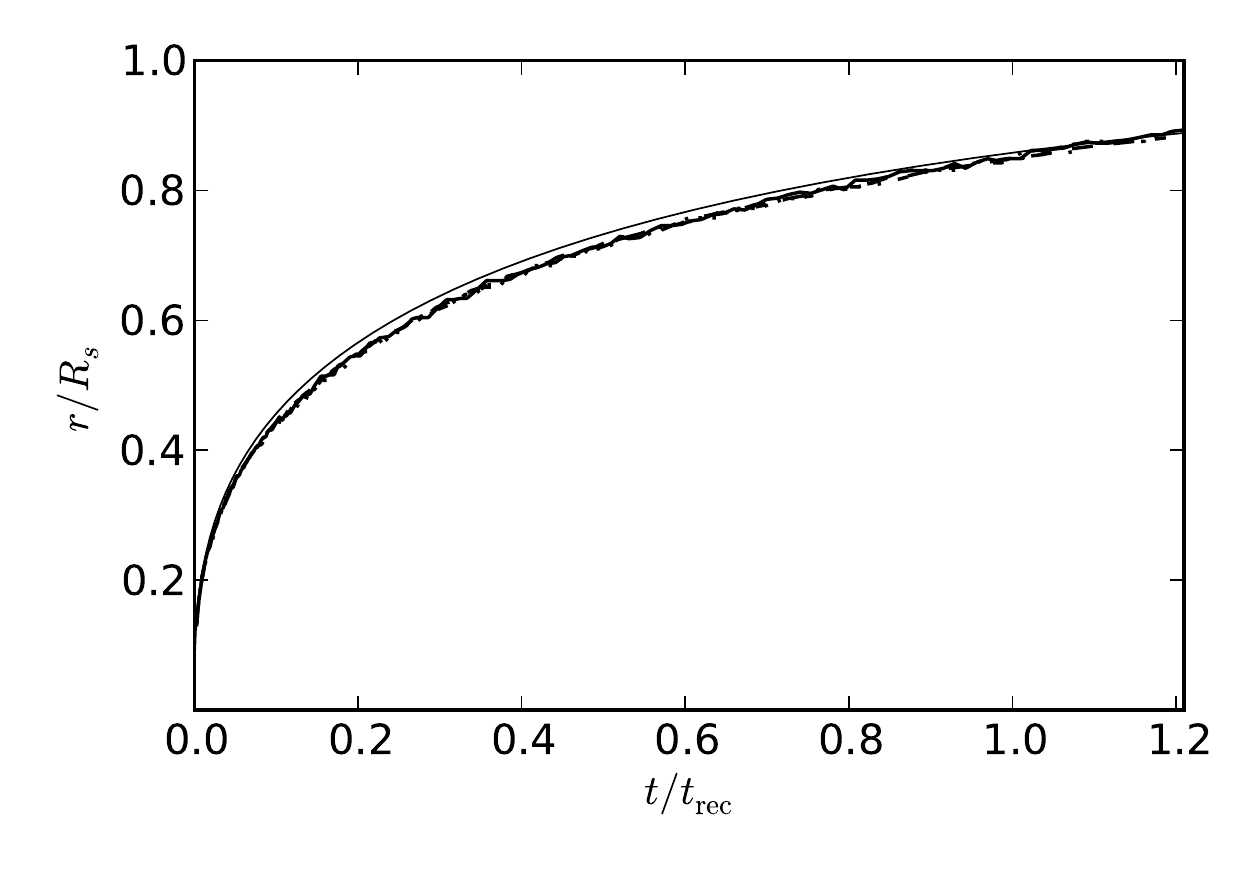}
\caption{Impact of spatial resolution on the expansion of the I-front during its R-type phase. The thin solid line shows the analytical solution, while the four thick lines show simulations with $64^3 $ (dotted), $128^3$ (dash-dotted), $256^3$ (dashed) and $512^3$ (solid) cells. In all four simulations, we set $f_{\mathrm{H}} = 0.1$.
\label{fig:rtype_resolution}} 
\end{figure}

\begin{figure*}
\centering
\includegraphics[scale=0.6]{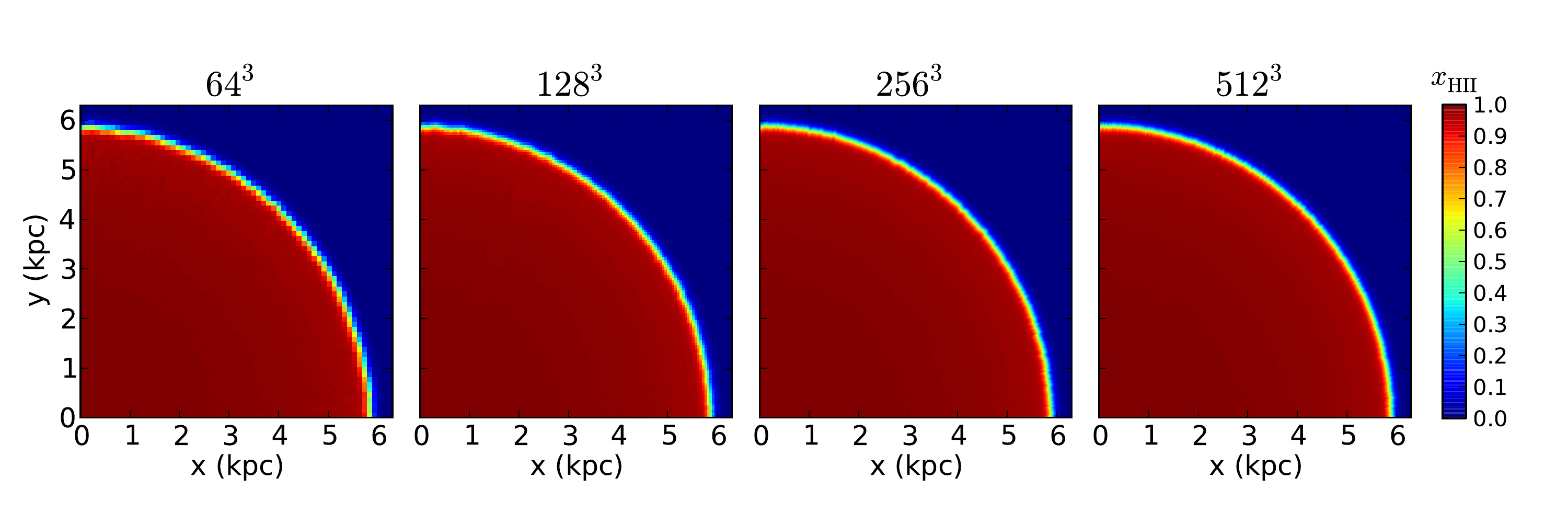}
\caption{Slices through the simulation domain of the R-type test for runs with $64$, $128$, $256$ and $512$ cells per side taken at $t \approx 130 \ \mathrm{Myr}$. 
\label{fig:rtype_slices}} 
\end{figure*}

Figures \ref{fig:rtype_timestep}--\ref{fig:rtype_slices} show the impact of temporal and spatial resolution on the I-front expansion.
In these figures, we define the position of the ionization front as the point where the neutral hydrogen fraction drops below $50 \%$. 

Figure~\ref{fig:rtype_timestep} demonstrates the necessity for a timestep limiter, as without it, we systematically underestimate the I-front radius, in this case by as much as 50\% at $t \sim 0.3 \, t_{\rm rec}$. This occurs because in our calculation of ${\rm d}\tau$ for each grid cell, we implicitly assume that the hydrogen number density does not vary significantly during the timestep. If we do not employ a limiter, then this assumption is not always justified and can lead us to overestimate the number of photons that are absorbed in each cell. However, we also see that if we limit the change of neutral hydrogen to $10 \%$ with $f_{\mathrm{H}} = 0.1 $, the expansion matches the analytical calculation to within an error of only a few percent. Decreasing $f_{\rm H}$ further, to $f_{\rm H} = 0.01$, improves the solution even more, but the improvement is relatively small in view of the large computational cost.

The difference between our numerical solution and the analytical solution is largest at early times, since at this point the neutral hydrogen fraction is changing rapidly  in the intense radiation field close to the source. Our timestep limiter can detect this and reduce the size of the timestep to compensate, but only after we have already taken one step with a too-large timestep. It therefore reduces the error only on subsequent timesteps, which for large $f_{\mathrm{H}}$  is not quite enough to recover the proper I-front expansion.

As far as sensitivity to the spatial resolution is concerned, we see from Figures~\ref{fig:rtype_resolution} and  \ref{fig:rtype_slices} that if we set $f_{\mathrm{H}} = 0.1 $, we detect virtually no change in the expansion behavior as we increase the spatial resolution from $64^{3}$ to $512^{3}$. The main effect of the change in the resolution is to alter the thickness of the I-front, since it has a minimum width of three resolution elements. 
In addition, the initial expansion is slightly influenced by the fact that at high resolution, the cells have smaller volumes and hence will be ionized more quickly compared to cells in more coarsely resolved simulations. This means that high resolution simulations are more sensitive to our choice of $f_{\rm H}$ than low resolution simulations, since the initial change in $n_{\rm H}$ in the central cells occurs more rapidly as we decrease the size of the cells. Note, however, that the error introduced by this only really affects the evolution of the I-front during the first few simulation steps.

Finally, Figure \ref{fig:rtype_slices} also demonstrates that our ray-tracing scheme properly captures the spherical nature of the I-front without introducing any artifacts due to the underlying Cartesian mesh.
%
%
\subsection{Ionization front expansion in an $r^{-2}$ density profile}
H{\sc ii} regions in simulations of star formation usually do not expand into a uniform medium at rest. Instead, the ionization front usually expands along a density gradient from a dense core. 
We test the applicability of our radiative transfer scheme in this case by assuming a density profile of the form
\begin{equation}\label{densprof}
\begin{aligned}
n(r) =
\begin{cases}
n_c & \mathrm{if } \ r \leq r_c \\
n_c (r/r_c)^{-\omega} & \mathrm{if } \ r > r_c,
\end{cases}
\end{aligned}
\end{equation}
with radius $r$ from the core center, and where $n_{c}$ is the hydrogen number density in the core and $r_{c}$ is the size of the constant density center of the core. In such a density profile, equation~\eqref{eq:ionization} becomes
\begin{equation}
\label{eq:ionr2}
\frac{\mathrm{d} N_\mathrm{ion}}{\mathrm{d}t} = 4 \pi r_\mathrm{i}^2 n(r_{i}) \frac{\mathrm{d} r_\mathrm{i}}{\mathrm{d}t} + 4\pi \alpha_{\rm B} \int_{0}^{r_{i}} n(r)^{2}
r^{2} \: {\rm d}r.
\end{equation}
We are interested in the case where the I-front has already left the central dense clump, i.e.\ $r_{\rm i} > r_c$. If we fix the exponent of the density profile to be $\omega = 2$, then it is easy to show that
\begin{equation}
\begin{aligned}
\label{eq:vir2}
\frac{\mathrm{d} r_\mathrm{i}}{\mathrm{d}t} =  \frac{Q}{4 \pi n_c r^2_c } - \frac{4}{3} n_c r_c \alpha_\mathrm{B} + \frac{ n_c r^2_c \alpha_\mathrm{B}}{r_\mathrm{i}},
\end{aligned}
\end{equation}
where $Q = {\rm d}N_{\rm ion} / {\rm d}t$ and $v_{\rm i} = {\rm d}r_{\rm i} / {\rm d}t$. If we now choose $Q$ so that the first two terms on the right hand side of this equation sum to zero, then this equation has the simple analytical solution \citep{franco1990,mellema06,whalen2006}
\begin{equation}\label{eq:profileSolution}
\begin{aligned}
r_\mathrm{i}(t) = r_c (1+ 2 \ t  \, n_c \, \alpha_\mathrm{B})^{1/2}, 
\end{aligned}
\end{equation}
where we have taken $t = 0$ to be the time at which the I-front escapes from the constant density portion of the profile, i.e.\ when $r_{\rm i} = r_{c}$. 

To test whether our code can reproduce this analytical solution, we make the same approximations as before. We set $\gamma = 1.0001$, consider monochromatic photons with $E = 13.6$~eV, so that $\sigma_\mathrm{H} = 6.3 \times 10^{-18} \ \mathrm{cm^2}$ and there is no photo-heating, and fix $\alpha_{\rm B}$ to the constant value
$\alpha_\mathrm{B} = 2.59 \times 10^{-13} \ \mathrm{cm^3 \ s^{-1}}$. We set $Q = 10^{49} \: {\rm photons \: s^{-1}}$ as before, and consider a central density $n_c = 100 \:\mathrm{cm^{-3}}$. With these values, we then need to set $r_c \simeq 1.99 \ \mathrm{pc}$ in order to ensure that the first two terms in equation~\eqref{eq:vir2} vanish.

We consider a simulation domain that extends from 0--6.5~pc in each dimension, and center our spherical density profile within this domain. The source is placed at the center of the density profile. As we want to test the behavior of the I-front in the regime where $r_{\rm i} > r_{c}$, we assume that all of the gas at $r \leq r_{c}$ is already ionized.

\begin{figure}
\includegraphics[scale=0.7]{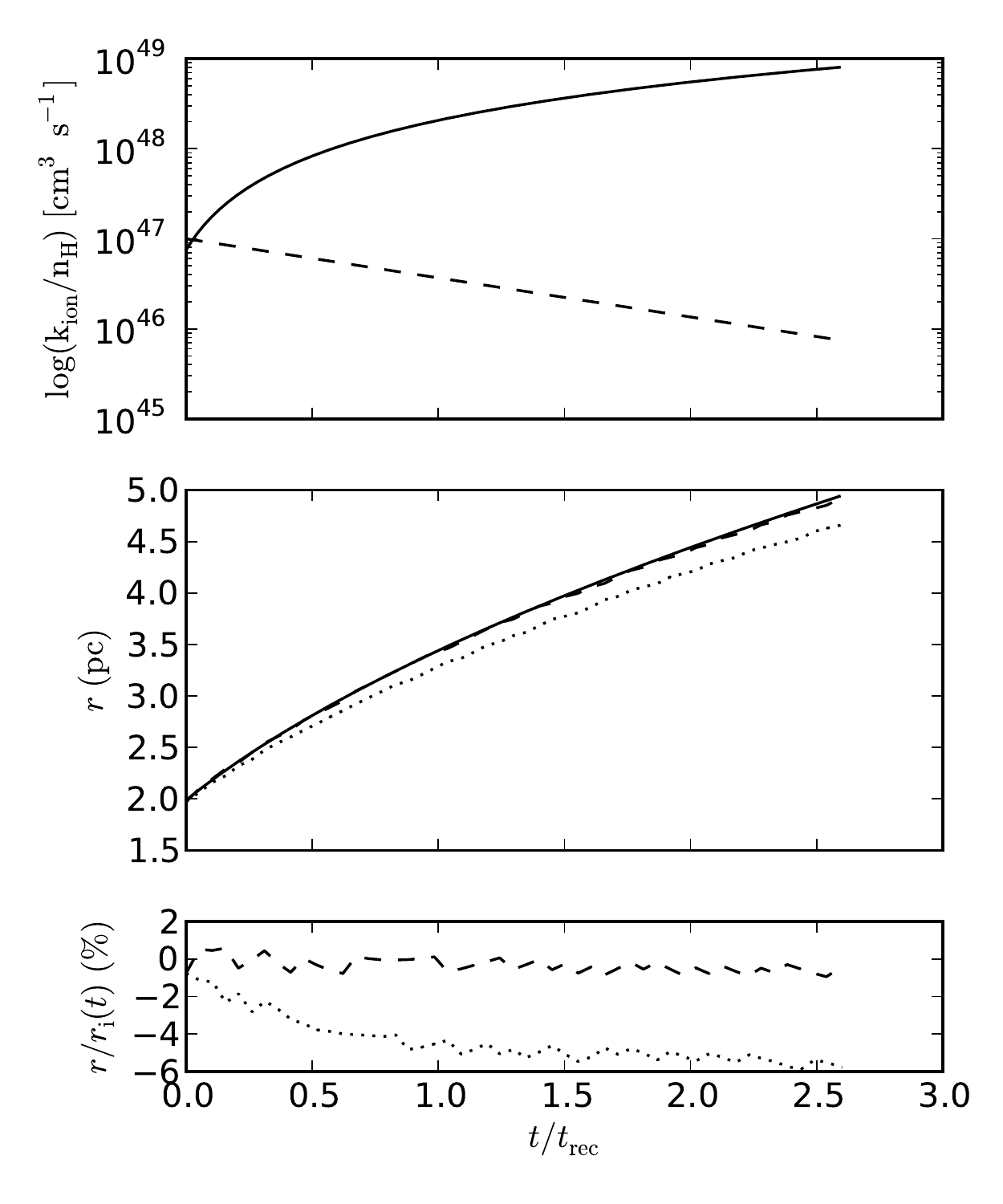}
\caption{{\it Top}: Ionizations per second normalized to the atomic hydrogen number density at the I-front position. The dashed curve corresponds to the case of uniform density ($n_\mathrm{H} = \mathrm{const.}$), while the solid line corresponds to the density profile given by equation~\eqref{densprof}. {\it Middle}: comparison of the numerical to the analytical solution for the I-front expansion with different timestep limiters: The dotted line shows the expansion with $f_\mathrm{H}$ limited to $0.1$, while the dashed line shows the result for $f_{\rm H} = 0.01$. {\it Bottom}: The fractional deviation of the numerical result from the analytical solution.
\label{fig:r2profile}} 
\end{figure}

The main issue tested with this setup is whether our timestep limiter is able to cope with the sustained rapid change of the neutral hydrogen fraction during the expansion of an I-front down a steep density gradient. The top panel of Figure~\ref{fig:r2profile} shows the ratio of the ionization rate to the number density of H atoms at the I-front for both the uniform density and the power-law profile cases. It is clear that this quantity varies much more rapidly with the power-law density profile $n = $ const., particularly when $t < t_{\rm rec}$.

The middle panel in Figure \ref{fig:r2profile} shows the position of the I-front as a function of time in simulations with a spatial resolution of $256^{3}$ grid cells and with $f_{\rm H} = 0.1$ (dotted line) and $f_{\rm H} = 0.01$ (dashed line). The analytical solution is shown as the solid line. 
We find that limiting the change in the atomic hydrogen fraction in each cell to $10 \%$ per timestep allows us to reproduce the analytical solution to within around 5\% (see the bottom panel of Figure~\ref{fig:r2profile}). The error becomes larger as the medium rarefies and the change in ionization fraction in a single cell increases in each timestep. Using a stricter limit with $f_{\rm H} = 0.01$ allows us to reduce the error to less than a percent, but significantly increases the computational cost.
%
%
\subsection{D-type ionization front expansion}
Once the expansion speed of the ionization front drops below the speed of sound in the hot, over-pressured ionized gas, the shock-front traveling on top of the I-front detaches and propagates supersonically into the cold surrounding ambient medium. The shock starts to sweep up a dense shell, and the I-front enters its D-type phase. If the neutral gas is very cold, then we can neglect its thermal pressure. In this case, the rate of change in the momentum of the dense shell located ahead of the I-front is equal to the total force exerted on the shell by the pressure of the hot gas, i.e.\
\begin{equation}\label{eq:shell}
\begin{aligned}
\frac{\mathrm{d} }{\mathrm{d}t} \left( m_\mathrm{i} v_\mathrm{i} \right) = 4 \pi r_\mathrm{i}^2 p_\mathrm{i},
\end{aligned}
\end{equation}
where $p_{\rm i}$ is the pressure of the ionized gas. Note that we have also assumed here that gravity is unimportant and that the radiation pressure of the photons can be ignored.

Let us now consider an ionization front expanding into a constant density medium with $\rho = \rho_{0}$. If we assume that the gas that was initially in the region occupied by the H{\sc ii} region has all been swept up into the shell, then the shell mass is simply
\begin{equation}
m_\mathrm{i} = \frac{4\pi}{3} r^3_\mathrm{i} \rho_0.
\end{equation}
If the shell is thin, then 
the pressure $p_\mathrm{i}$ of the ionized gas is balanced by the ram pressure exerted by the neutral medium, $p_\mathrm{i} = \rho_\mathrm{i} c^2_\mathrm{i} = \rho_0 v^2_\mathrm{s}$, where $v_\mathrm{s}$ is the shock expansion velocity, $c_\mathrm{i}$ is the sound speed in the ionized gas and  $\rho_\mathrm{i}$ the density of the ionized gas.

If we also assume that the H{\sc ii} region is in ionization equilibrium, so that $Q = (4\pi/3) n^2_{i} \alpha_\mathrm{B} r^3_\mathrm{i}$, i.e.\ all of the ionizing photons are used up counteracting the effects of recombination, then we can express the density in the ionized region as 
\begin{equation}
\rho_\mathrm{i} = \rho_0 \left(\frac{r_\mathrm{i}}{R_{\rm s}}\right)^{3/2},
\end{equation}
where $R_{\rm s}$ is the Str\"omgren radius.
With these assumptions, equation~\eqref{eq:shell} reads
\begin{equation}
\begin{aligned}
\frac{\mathrm{d}}{\mathrm{d}t} \left(\frac{4}{3} \pi r^3_\mathrm{i} \rho_0 \dot{r}_\mathrm{i} \right) = 4 \pi r_\mathrm{i}^2 \rho_0 c_\mathrm{i}^2 \left( \frac{r_\mathrm{i} }{R_{\rm s}} \right)^{3/2}.
\end{aligned}
\end{equation}
\cite{hosokawa2006} solve this equation and arrive at the expansion law for a D-type ionization front
\begin{equation}\label{dtypeexpansioneq}
\begin{aligned}
r_\mathrm{i}(t) = R_{\rm s} \left(1  + \frac{7}{4} \sqrt{\frac{4}{3}} \frac{c_\mathrm{i} t}{R_{\rm s} } \right)^{4/7}.
\end{aligned}
\end{equation}
(Note that the more commonly known solution by \cite{spitzerbook} omits the factor of $\sqrt{4/3}$ from the second term inside the parentheses).  
Eventually, the expansion of the I-front should come to rest once the pressure exerted by the ionized region is balanced by the pressure of the ambient medium \citep{raga2012}.

We use the D-type phase of the ionization front expansion as a test for the hydrodynamical response to the over-pressured H{\sc ii} region and the continuous driving of the shell by ionizing radiation. The simulation setup consists of a source at the origin that emits $Q = 10^{49}$ ionizing photons per second, with $2 \ \mathrm{eV}$ per photon in deposited heating energy. The hydrogen ionization cross-section is again fixed at the threshold value of $\sigma_\mathrm{H} = 6.3 \times 10^{-18} \ \mathrm{cm^2}$. 

We set the atomic hydrogen number density to $n_{\rm H} = 1000 \ \mathrm{cm^{-3}}$, and also set $\gamma = 1.66667$ and $\alpha_\mathrm{B} = 2.59 \times 10^{-13} \ \mathrm{cm^3 \ s^{-1}} $. The ambient temperature of the neutral gas is set to $100 \ \mathrm{K}$. The extent of the simulation domain is chosen to be $\pm 64 \ \mathrm{pc}$. 
To balance the photoionization heating, we consider only Lyman-$\alpha$ cooling; the rates of the other cooling and heating terms in our model are set to zero. We perform simulations with spatial resolutions of $64^{3}$, $128^{3}$, $256^{3}$ and $512^{3}$. 

\begin{figure*}
\centering
\includegraphics[scale=0.6]{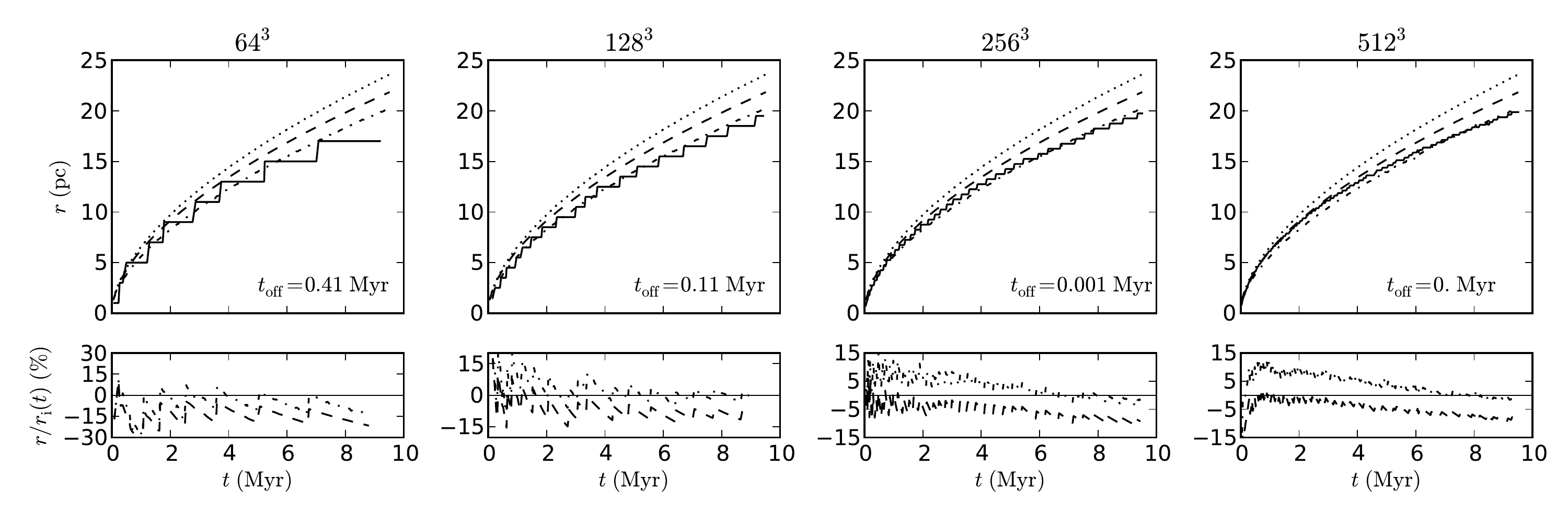}
\caption{Expansion of the D-type ionization front in our test setup with different spatial resolutions. The dotted, dashed and dash-dotted lines show analytical solutions with the sound speed $c_\mathrm{i}$ calculated from the maximum ($T_{\mathrm{max}}$), average ($T_{\mathrm{avg}}$) and minimum ($T_{\mathrm{min}}$) temperature of gas with an ionized hydrogen fraction of $x_\mathrm{H^+} \ge 0.99$. The solid lines plot the expansion obtained from the numerical simulation after correcting for the offset time $t_\mathrm{off}$ needed to generate the initial ionized and over-pressured H{\sc ii} region. The bottom panels show the relative errors calculated from the analytical solutions where we use $T_{\mathrm{avg}}$ (dashed) and $T_{\mathrm{min}}$ (dash-dotted). 
\label{fig:dtype_expansion}} 
\end{figure*}

With this problem setup, the initial Str\"omgren radius is $R_{\rm s}  = 0.68 \ \mathrm{pc}$. This is resolved with multiple grid cells only in our $512^3$ simulation.
Figure \ref{fig:dtype_expansion} shows the expansion behavior of the I-front for different spatial resolutions.
When the initial Str\"omgren sphere is unresolved, the shock does not form until a fully ionized and over-pressured region has been created. 
This is a purely numerical effect and is due to the fact that for low resolution, the ionizing photons are diluted over a large volume, only slowly heating and changing the ionization fraction. Once a minimally sized H{\sc ii} region of one or two cells is created, the D-type expansion progresses as described by the analytical expansion law. Only if the Str\"omgren sphere is at least marginally resolved do we see no delay or a negligible delay (see Figure~\ref{fig:dtype_expansion}). We correct for this time offset by replacing the evolution time $t$ in equation \eqref{dtypeexpansioneq} with $t' = t - t_\mathrm{off}$, where $t_\mathrm{off}$ is the time it takes to ionize the minimum volume of one cell. 

\begin{figure*}
\centering
\includegraphics[scale=0.38]{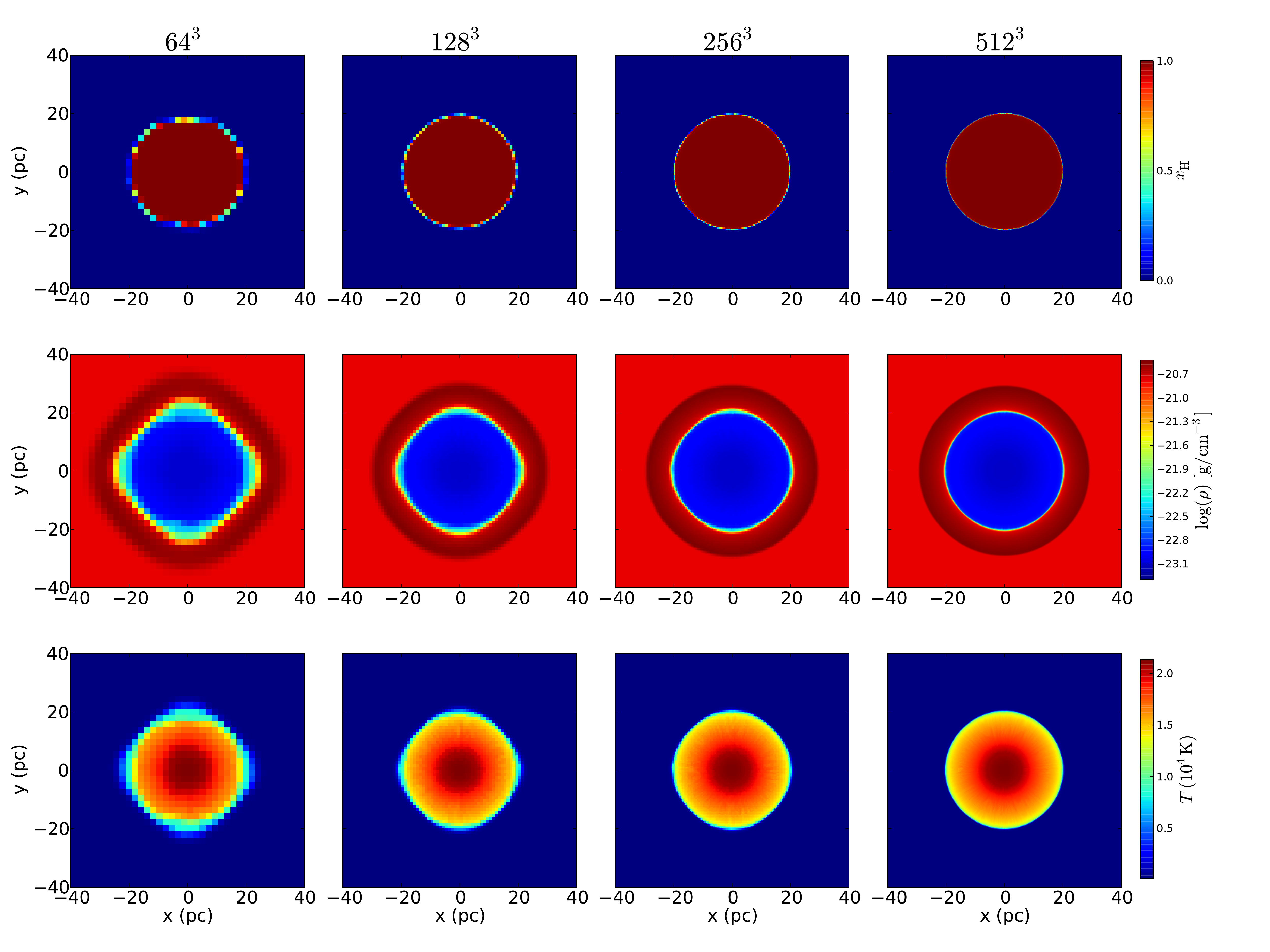}
\caption{Slices of the D-type expansion test through the origin showing different hydrodynamical quantities after $t_\mathrm{rec} \approx  81000$ recombination times or $9.5 \ \mathrm{Myr}$. Spatial resolutions of $2 \ \mathrm{pc} \ (64^3)$, $1 \ \mathrm{pc} \ (128^3)$, $0.5 \ \mathrm{pc} \ (256^3)$ and $0.25 \ \mathrm{pc} \ (512^3)$ are shown.
\label{fig:dtype_slices}}
\end{figure*}

We use the speed of sound calculated from the average temperature in the ionized gas $c_\mathrm{i} = \gamma k_B  T_\mathrm{avg} / \mu $, with the Boltzmann constant $k_B$ and mean molecular weight $\mu$. At later times deviations from the analytical solution emerge that are a consequence of the fact that the ionized gas is not isothermal. We find that in practice, the lowest temperature in the ionized gas, $T_{\rm min}$, is located close to the I-front (see Figure~\ref{fig:dtype_slices}), and that at late times, the expansion of the ionization from follows the analytical solution that we obtain if we set $T = T_{\rm min}$ rather than $T = T_{\rm avg}$. In general, the difference between our numerical solution and the analytical solution is around 5\%, apart from the initial rapid expansion phase, where for an unresolved Str\"omgren sphere the error reaches up to $30 \%$.

Figure \ref{fig:dtype_slices} shows the impact of the spatial resolution on the shape of the shell, as well as the temperature and density structure of the H{\sc ii} region. We see that the
temperature varies by around a factor of two as we move from the central region around the star to the edge of the dense shell. This temperature variation is a consequence of spatial variations in the heating rate and the gas density (which directly affects the cooling rate). We note, however, that the effect is large here because we are not including the contribution of metals to the total cooling rate. If we include the additional metal-line cooling, we find instead that the temperature of the ionized gas evens out at around $8$--$9 \times 10^3 \ \mathrm{K}$ with a negligible temperature gradient. 

Comparing the ionized fraction with the density slice reveals that the shock has a direction-dependent expansion velocity: it propagates slightly faster in directions aligned with the Cartesian mesh and slightly slower in the other directions. This is a well known effect for spherical shock waves on Cartesian grids  and can largely be eliminated by increasing the resolution of the simulation, as Figure~\ref{fig:dtype_slices} demonstrates.

%
%
\subsection{Photon dominated region test}\label{PDRtest}
We follow the test setup outlined in the code comparison project in \cite{roellig2007}. The test was designed to compare dedicated PDR codes that obtain the chemical and thermal structure of a molecular cloud illuminated by the interstellar radiation field. The final, equilibrated state is calculated and compared, where any time evolution is omitted. 
The chemical network used in their test case consists of 31 different chemical species and hence is far more extensive than the simplified network used in \texttt{Fervent}. Despite this, it is interesting to compare how well our code can reproduce the chemical structure of the two main molecular species, H$_{2}$ and CO, and the thermal structure of the PDR.

Table \ref{tab:pdr} shows the elemental composition of the molecular cloud, as well as the test parameters. In our simulations, the chemical abundance of atomic oxygen is tracked using a conservation law: we know the total elemental abundance of oxygen relative to hydrogen ($A_{\rm O}$) and also how much oxygen is incorporated into CO. The fractional abundance of atomic oxygen therefore follows as $x_{\rm O} = A_{\rm O} - x_{\rm CO}$. Our chemical model does not include neutral carbon, and so we assume that any carbon not incorporated into CO is present as C$^{+}$.
\begin{table}
\caption{Parameters of the PDR test \label{tab:pdr}}
\centering
\begin{tabularx}{\columnwidth}{lll}
Symbol & Value & Quantity \\
\hline
$A_\mathrm{He}$   &  $   0.1$ & He abundance \\
$A_\mathrm{O}$    & $3 \times 10^{-4}$ & O  abundance \\
$A_\mathrm{C}$    & $1 \times 10^{-4}$ & C  abundance \\
$k_{\mathrm{cr}}$ & $5 \times 10^{-17} \  \mathrm{s^{-1}}$& cosmic ray ionization rate  \\
$A_\mathrm{V}$    & $6.289 \times 10^{-22} N_\mathrm{H, total} $& visual extinction  \\
$\tau_d $         & $3.02 A_\mathrm{V}$ & dust attenuation  \\
$k_\mathrm{UV}$   & $5 \times 10^{-10} (\chi /10) \ \mathrm{s^{-1}}$ & $\mathrm{H_2}$ dissociation rate \\
$R_\mathrm{H_2} $ & $3 \times 10^{-18} T^{1/2} \ \mathrm{cm^3 s^{-1}}$ & $\mathrm{H_2}$ formation rate  \\
$T_{\mathrm{gas}} $ & $50 \ \mathrm{K}$ & isothermal gas temperature  \\
$T_{\mathrm{dust}} $ & $20 \ \mathrm{K}$ & isothermal dust temperature \\
$f_\mathrm{d/g}$ & $1.0$ & dust to gas ratio\\
\hline
\end{tabularx}
\end{table}

The tests in \cite{roellig2007} are carried out using a UV radiation field that has the spectral shape of the \citet{draine1978} field, and a strength parameterized by $\chi$, where $\chi = 1$ corresponds to the Draine field. This UV radiation field produces an H$_{2}$ photodissociation rate in unshielded gas that is  $k_\mathrm{UV} = 5 \times 10^{-10} (\chi /10) \ \mathrm{s^{-1}}$, as listed in the Table. However, in \texttt{Fervent} we do not directly specify $k_{\rm UV}$, and so we must instead calculate the photon flux in the 11.2--13.6~eV energy bin that we require in order to produce the same dissociation rate. We can write the energy density of the adopted UV radiation field as \citep{draine1978,DBT1996}
\begin{equation}
\begin{aligned}
\lambda u_\lambda = 6.8 \times 10^{-14} \chi  (31.016 \lambda_{3}^{-3}
&-49.913 \lambda_{3}^{-4} \\ &+ 19.897  \lambda_{3}^{-5}),
\end{aligned}
\end{equation}
where $\lambda$ is the wavelength in \AA, $\lambda_{3} = \lambda / 1000$~\AA, and $u_\lambda$ is the energy density per unit wavelength, measured in units of $\mathrm{erg \: cm^{-3}}$~\AA$^{-1}$. Three quantities are needed as input for our scheme: the incident photon flux $N_{\mathrm{11.2}}(r=0)$ in the $E_\mathrm{11.2}$ energy bin
\begin{equation}
\begin{aligned}
N_{\mathrm{11.2}}(r=0) = \frac{A}{2} \int^{1110}_{912} \frac{\lambda u_\lambda}{h} \mathrm{d} \lambda  = 1.06 \times 10^7 \  \chi A,
\end{aligned}
\end{equation}
where $A$ is the area of the illuminated face of our simulation domain and the factor of one half enters because we are considering a semi-infinite plane parallel slab which is exposed to radiation from a total solid angle of only $2\pi$ steradians; the average energy per UV photon in $\mathrm{erg}$, 
\begin{equation}
\begin{aligned}
\left \langle E_\mathrm{11.2} \right \rangle &= \int^{1110}_{912} c u_{\lambda} \mathrm{d} \lambda  \Big/
    \int^{1110}_{912} \frac{\lambda u_\lambda}{h} \mathrm{d} \lambda \\ 
    &= 1.93 \times 10^{-11},
\end{aligned}
\end{equation}
where we have to express the speed of light in $\mathrm{\AA \ s^{-1}}$; and finally, the total energy flux in the $E_\mathrm{5.6}$  energy bin at the cloud surface, measured in units of $\mathrm{erg \ s^{-1} cm^{-2}}$
\begin{equation}
\begin{aligned}
F_{\mathrm{5.6}}(r=0) = \frac{1}{2} \int^{2200}_{1110} c u_\lambda \mathrm{d} \lambda = 1.22 \times 10^{-3} \ \chi.
\end{aligned}
\end{equation}
This last quantity is needed in order for us to be able to account for the effects of photoelectric heating when determining the thermal structure of the PDR.

The simulation domain is an elongated box with a ratio of $(x,y):z$ of $1:10$, with the physical extent set to give a range in visual extinction depending on the gas density used in each test. The ray-tracing is simplified to cast rays without the \texttt{HEALPix} based formalism. Instead, the initial rays are created at the centers of the faces at the minimum $x$ boundary, which we assume to be illuminated by the ISRF. They propagate in the positive $x$ direction, where they are split into four child rays if they encounter a more highly resolved region, one for each child cell (see Appendix \ref{AppendixA} for a brief description of the mesh structure). 

The PDR comparison project described in \citet{roellig2007} consists of eight tests. Four of these are isothermal with fixed gas and dust temperatures (denoted F1 to F4). The other four are carried out with the same density and ISRF strengths as the isothermal tests, but the gas and dust temperatures are allowed to evolve self-consistently with the chemical state. These non-isothermal tests are denoted as V1-V4. The tests explore two different densities and two different radiation field strengths: a cloud with a hydrogen number density of  $n_{\mathrm{H}} = 10^3 \ \mathrm{cm^{-3}}$ illuminated by an ISRF with $\chi = 10$ or $\chi = 10^5$ (tests F1-2, V1-2) and a cloud with a hydrogen number density of  $n_{\mathrm{H}} = 10^{5.5} \ \mathrm{cm^{-3}}$ with the same variation in the ISRF field strengths (tests F3-4, V3-4). 

In the isothermal test cases, we use an adiabatic index of $\gamma = 1.0001$, while in the non-isothermal tests we adopt the standard $\gamma = 1.6667$. The extent of the simulation domain is chosen so as to provide a sufficiently high spatial resolution to sample small values of visual extinction $A_{\rm V}$. For tests F1-2 and V1-2, the smallest cell size is set to $2.44 \times 10^{14} \ \mathrm{cm}$ and the simulation domain extends to $\pm 2 \times 10^{18} \ \mathrm{cm}$ in the $x$ and $y$ directions and $4 \times 10^{19} \ \mathrm{cm}$ in the $z$ direction. For tests F3 and V3, the bounding box size is $\pm 0.5 \times 10^{13} \ \mathrm{cm}$ in $x$ and $y$ and $ 10^{14} \ \mathrm{cm}$ in the $z$ direction, meaning that the most resolved cell has a size of  $6.1 \times 10^{8} \ \mathrm{cm}$. For tests F4 and V4 the box extends to $\pm 0.5 \times 10^{16} \ \mathrm{cm}$ in $x$ and $y$ and $ 10^{17} \ \mathrm{cm}$ in the $z$ direction, with the smallest cell having a size of $2.44 \times 10^{12} \ \mathrm{cm}$.

In contrast to many dedicated PDR codes, our radiative transfer scheme has to be evolved explicitly in time until the chemical network reaches its equilibrium. During this evolution, we do not follow the hydrodynamical response of the gas to any changes in the thermal pressure, i.e.\ the density field remains fixed in space and constant.

\begin{figure*}
\centering
\includegraphics[scale=0.5]{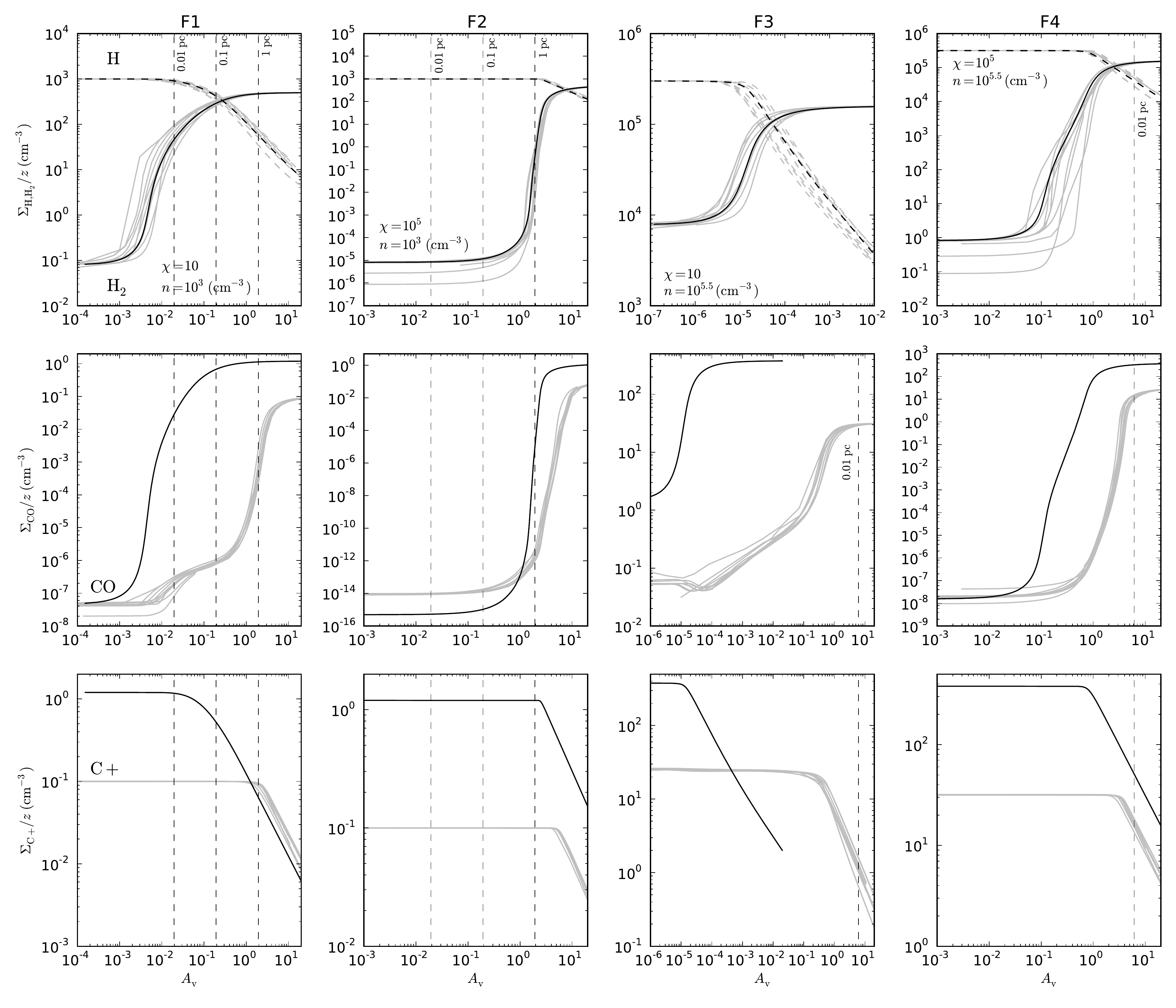}
\caption{Profiles of the isothermal PDR tests F1-F4 after they converged in time. The light gray lines are the results from the eight PDR codes compared in \citet{roellig2007} and the solid lines show the results obtained using \texttt{Fervent}. The surface densities are divided by the physical distance from the cloud edge to yield number densities. The  density of the molecular cloud and the intensity of the incoming radiation field are annotated in the first row of the plot. The dashed lines show physical distances from the cloud edge. Our simplified chemistry is able to reproduce the results for the hydrogen species but only roughly approximates the carbon chemistry to an order of magnitude.
\label{fig:pdr_chemistry_isothermal}} 
\end{figure*}

Figure~\ref{fig:pdr_chemistry_isothermal} shows the results that we obtain for the four isothermal tests. For comparison, we also plot the results from the eight PDR codes compared in \cite{roellig2007},  focusing on the surface densities of H, H$_{2}$, C$^{+}$ and CO.
We see from the Figure that \texttt{Fervent} does a good job of reproducing the transition from atomic hydrogen to molecular hydrogen at the edge of the cloud. The PDR codes compared in \citet{roellig2007} disagree somewhat on the details of this transition, and our results lie well within this spread.  

On the other hand, it is clear from the lower panels of Figure~\ref{fig:pdr_chemistry_isothermal} that we do not reproduce the behavior of the CO or C$^{+}$ surface densities. The main reason for this is our assumption that we can obtain the CO dissociation rate from the H$_{2}$ dissociation rate simply by applying a conversion factor. In reality, this is accurate only in optically thin gas. Once the gas starts to become optically thick, H$_{2}$ self-shielding plays a major role in decreasing the H$_{2}$ photodissociation rate \citep{DBT1996}. However, the analogous process for CO is much less effective, on the grounds of the low abundance of CO relative to H$_{2}$, and in any event does not have the same functional dependence on column density as the H$_{2}$ self-shielding correction \citep[see e.g.][]{lee96}. By directly coupling the CO and H$_{2}$ photodissociation rates, we therefore overestimate the rate at which the CO photodissociation rate falls off as we move into the cloud, and hence overestimate the CO surface density. In addition, our extremely simplified treatment of CO formation is also known to overestimate the CO formation rate relative to more accurate models \citep{glover2012}. We therefore see that we cannot use \texttt{Fervent} in its current form for predicting accurate CO abundances. However, we stress that our goal when adding the current simplified treatment of CO chemistry was simply to be able to approximately distinguish between regions of the ISM that are cold and CO-bright and ones which are warm and CO-faint when performing large-scale simulations \citep[see e.g.][]{walch14}, and for this particular purpose, our current treatment is adequate. In the future, we intend to improve on \texttt{Fervent}'s treatment of CO by adding an additional energy bin specifically to treat CO photodissociation, and by using a more sophisticated treatment of CO formation, such as the model of \citet{nl99}, but this is outside the scope of the present work.


Finally, we also see from Figure~\ref{fig:pdr_chemistry_isothermal} that \texttt{Fervent} over-estimates the amount of ionized carbon in the PDR by roughly an order of magnitude. This is another consequence of our reduced chemical model: since we do not include neutral atomic carbon, we mis-classify as C$^{+}$ some carbon that should in fact  be present as C. Again, we plan to address this in the future work by improving the model of the carbon chemistry.

\begin{figure*}
\centering
\includegraphics[scale=0.5]{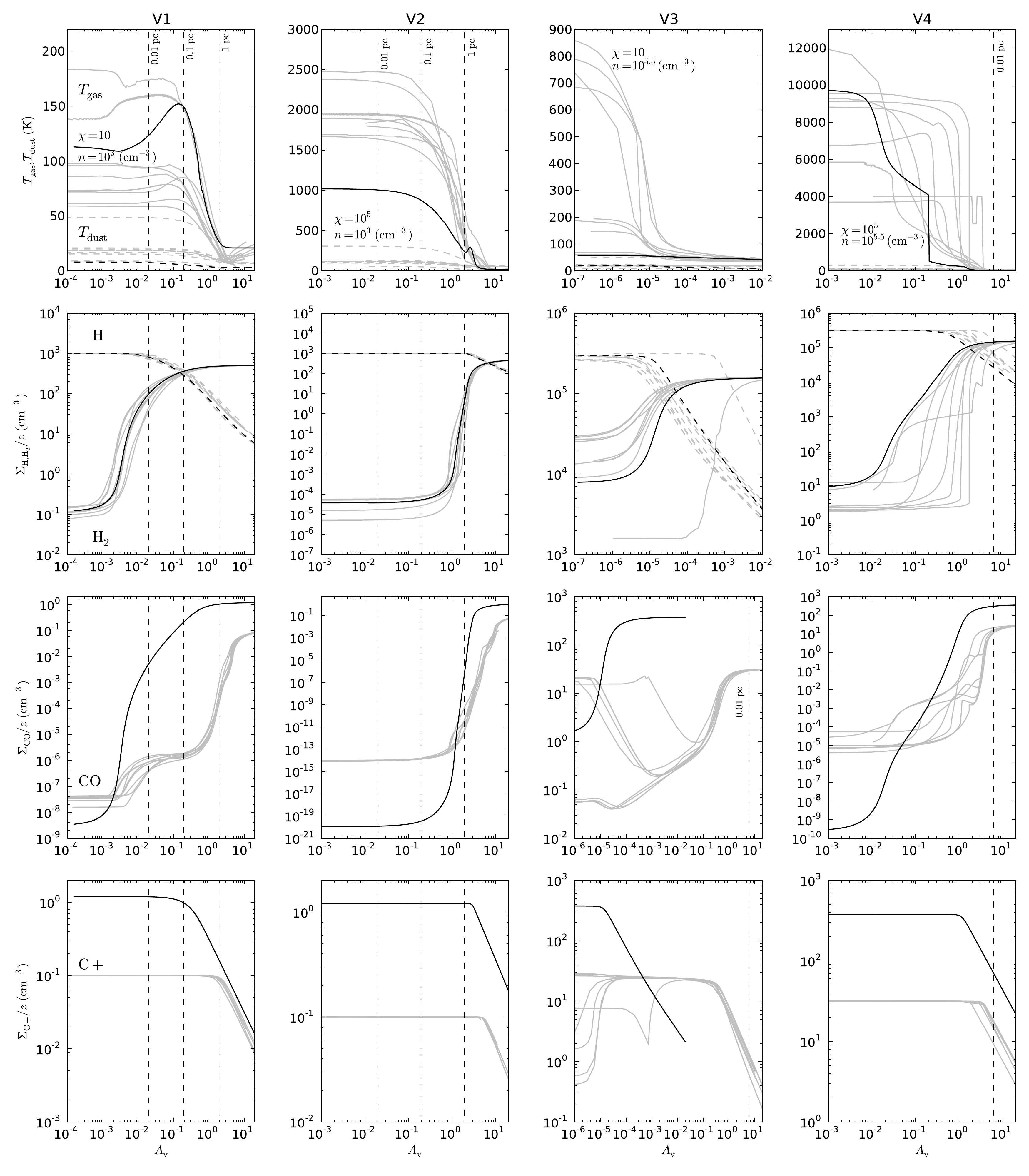}
\caption{Profiles of the non-isothermal PDR tests V1-V4 after the gas has reached thermal and chemical equilibrium. The light gray lines are the results presented in \citet{roellig2007} and the solid lines show the results we obtain using \texttt{Fervent}. The surface densities are divided by the physical distance from the cloud edge to yield number densities. The  density of the molecular cloud and the intensity of the incoming radiation field are annotated in the first row of the plot. The dashed lines show physical distances from the cloud edge. Our simplified chemistry is able to reproduce the results for the hydrogen species but only roughly approximates the carbon chemistry to within about an order of magnitude. The temperature we calculate lies within the spread of the other models. \label{fig:pdr_chemistry_temperature}} 
\end{figure*}

In Figure~\ref{fig:pdr_chemistry_temperature}, we show the results from the non-isothermal tests. 
We again find a good agreement in most cases between our results for the H$_{2}$ and H surface densities and those computed by the PDR codes compared in \citet{roellig2007}, particularly for the lower density PDRs. The gas and dust temperature structure itself lies within the very large spread obtained from the dedicated codes, except for test V2 where we 
find a temperature that is around a factor of two lower than in \citet{roellig2007}. We note that one factor that may contribute to this is the fact that in our treatment of fine structure cooling from atomic oxygen, we use collisional excitation rate coefficients for O-H collisions taken from \citet{akd07}. These are a factor of two to three higher than the older \citet{lr77} values available at the time that the \citeauthor{roellig2007} code comparison was carried out, and so in conditions where atomic oxygen cooling dominates (warm, dense, atomic gas), we would naturally expect to recover lower temperatures than in these calculations.
Finally, we again see that we do not do a good job of reproducing the behavior of the CO and C$^{+}$ surface densities, for the same reasons as explained above.
%
\subsection{Isolated source in a dense molecular medium}
In this test, we again examine the expansion of an H{\sc ii} region around a point source embedded in a uniform density gas. The difference is that in this case, we take the gas to be fully molecular initially, and we account for the PDR that forms ahead of the I-front. This is a complex multi-physics problem, where numerical simulations are actually the best tool for exploring the multiple combined heating and dissociation processes. No simple analytical solutions or published test results are available for comparison to our results. Instead, we set up a test that checks whether the
behavior we recover seems physically reasonable, and which allows us to look for any apparent numerical artifacts. Importantly, we want to check that the behavior that we see during the transition from an R-type to a D-type I-front is reasonable. 
Is the PDR in the proper position in relation to the ionization front, and does it evolves over time with the general expansion of the H{\sc ii} region in the way that we would expect? 
While the ionization front remains R-type, the physical separation between the I-front and the edge of the PDR is very small and so we do not expect to resolve the PDR layer. However, once the I-front transitions to the D-type, we expect the PDR to broaden and so we should start to be able to resolve it. 

We use the parameters given in Table \ref{tab:pdr} for the composition of the medium and as input for the chemistry module, except for the initial gas and dust temperatures which we adjust to $10 \ \mathrm{K}$ and $100 \ \mathrm{K}$, respectively, and $k_\mathrm{UV}$, which is set from the properties of the radiation source. 
The simulation domain extends $\pm 6 \ \mathrm{pc}$ in each direction and we choose an initial number density of $1000 \ \mathrm{cm^{-3}}$. Our setup corresponds roughly to the test case F2 in the previous section if we were to omit the radiation beyond the Lyman limit from the present test.

We resolve the Str\"omgren sphere radius of $ 1 \ \mathrm{pc}$, which is only strictly defined for a purely atomic hydrogen medium, with a minimum cell size of $0.094 \ \mathrm{pc}$ for the coarsest simulation with $128^3$ cells. This guarantees that the initial ionized region is captured so that the I-front expands properly. In addition, this high spatial resolution is needed to resolve the thin transition layer from molecular to atomic hydrogen. 

The radiation source is located at the center of the domain and is fully characterized by its blackbody spectrum of $T_\mathrm{eff} = 4 \times 10^4 \ \mathrm{K}$ and a stellar radius of $5 \times 10^{11} \ \mathrm{cm}$. The derived values for $N_{11.2}$, $N_{13.6}$ etc.\ that are used in our radiative transfer scheme are shown in Table \ref{tab:source}.
\begin{table}
\caption{Radiation source parameters}
\centering
\begin{tabularx}{\columnwidth}{l@{\hskip 1in}l@{\hskip 1in}}\label{tab:source}
Symbol & Derived value  \\
\hline
$N_\mathrm{11.2}$   & $3.23 \times 10^{48} \ \mathrm{s^{-1}}$ \\
$N_\mathrm{13.6}$   & $1.61 \times 10^{48} \ \mathrm{s^{-1}}$ \\
$N_\mathrm{15.2+}$  & $4.70 \times 10^{48} \ \mathrm{s^{-1}}$ \\
$F_\mathrm{pe}$     & $1.51 \times 10^{38}  \ \mathrm{erg \ s^{-1}}$  \\
$\left<E_\mathrm{11.2}\right>$ & $12.33 \  \mathrm{eV}$  \\
$e_\mathrm{13.6}$    & $0.72 \  \mathrm{eV}$ \footnote{Deposited energy per photon in the subscripted energy range\label{footnoteA}} \\
$e_\mathrm{15.2+}$   & $2.94 \  \mathrm{eV}$ \\
$e_\mathrm{H_{2}}$   & $4.44 \  \mathrm{eV}$  \\
$\sigma_\mathrm{13.6} $& $5.38 \times 10^{-18} \ \mathrm{cm^2}$ \\
$\sigma_\mathrm{15.2+}$ & $2.43 \times 10^{-18} \ \mathrm{cm^2}$ \\
$\sigma_\mathrm{H_{2}}$ & $6.01 \times 10^{-18} \ \mathrm{cm^2}$ \\
\hline
\end{tabularx}
\end{table}
Finally, all of the physics described in Section \ref{model} is used in this test, including the effective cross-sections based on the effective blackbody temperature of the source, metal line cooling and the coupling of ionizing radiation to molecular hydrogen. 

In the remainder of this section, we look at the evolution of the combined ionization and photodissociation fronts at four times: during the R-type phase, at the transition time between R-type and D-type, during the early D-type phase, when the density contrast between the shell and the cavity is a factor of a few, and during the late D-type phase, when the density contrast between the shell and the cavity is large. 

\begin{figure}
\centering
\includegraphics[scale=0.65]{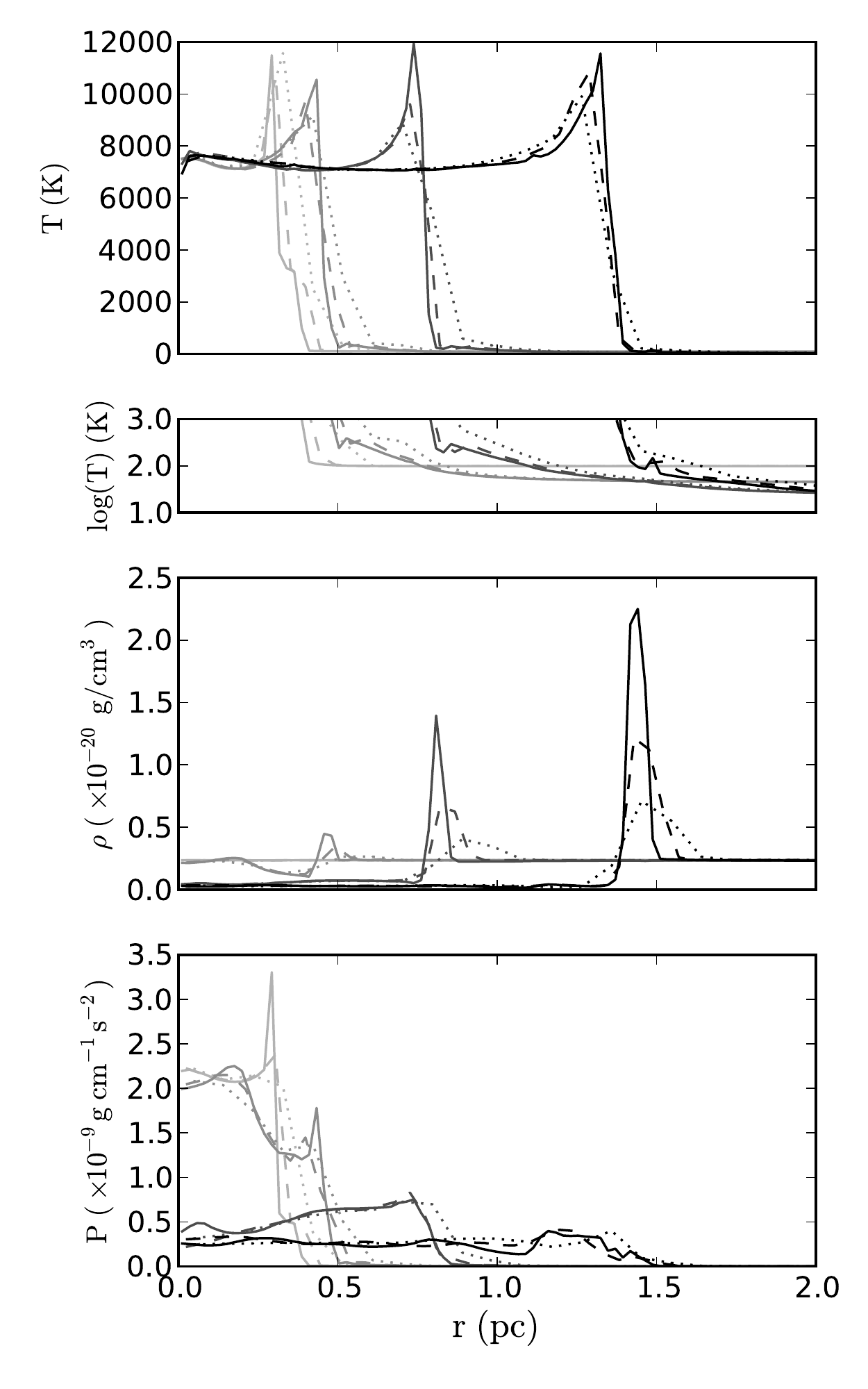}
\caption{Expansion of a combined PDR and ionization front into a fully molecular uniform medium with a number density of $1000 \ \mathrm{cm^{-3}}$. The top panel shows the gas temperature and the small panel below it shows logarithmically scaled temperatures in the range of $10$ to $1000 \ \mathrm{K}$ to highlight the pre-heating due to the PDR. The third panel from the top shows density and the bottom pressure profiles from the source up to the position of the I-front for four different times, $t = 1, 100, 500, 1500 \ t_\mathrm{rec}$, which are shaded from light gray to black. The dotted, dashed and solid lines show results for spatial resolutions with $128^3$, $256^3$ and $512^3$, respectively. 
\label{fig:oneSource_thermodynamics}} 
\end{figure}

\begin{figure*}
\centering
\includegraphics[scale=0.8]{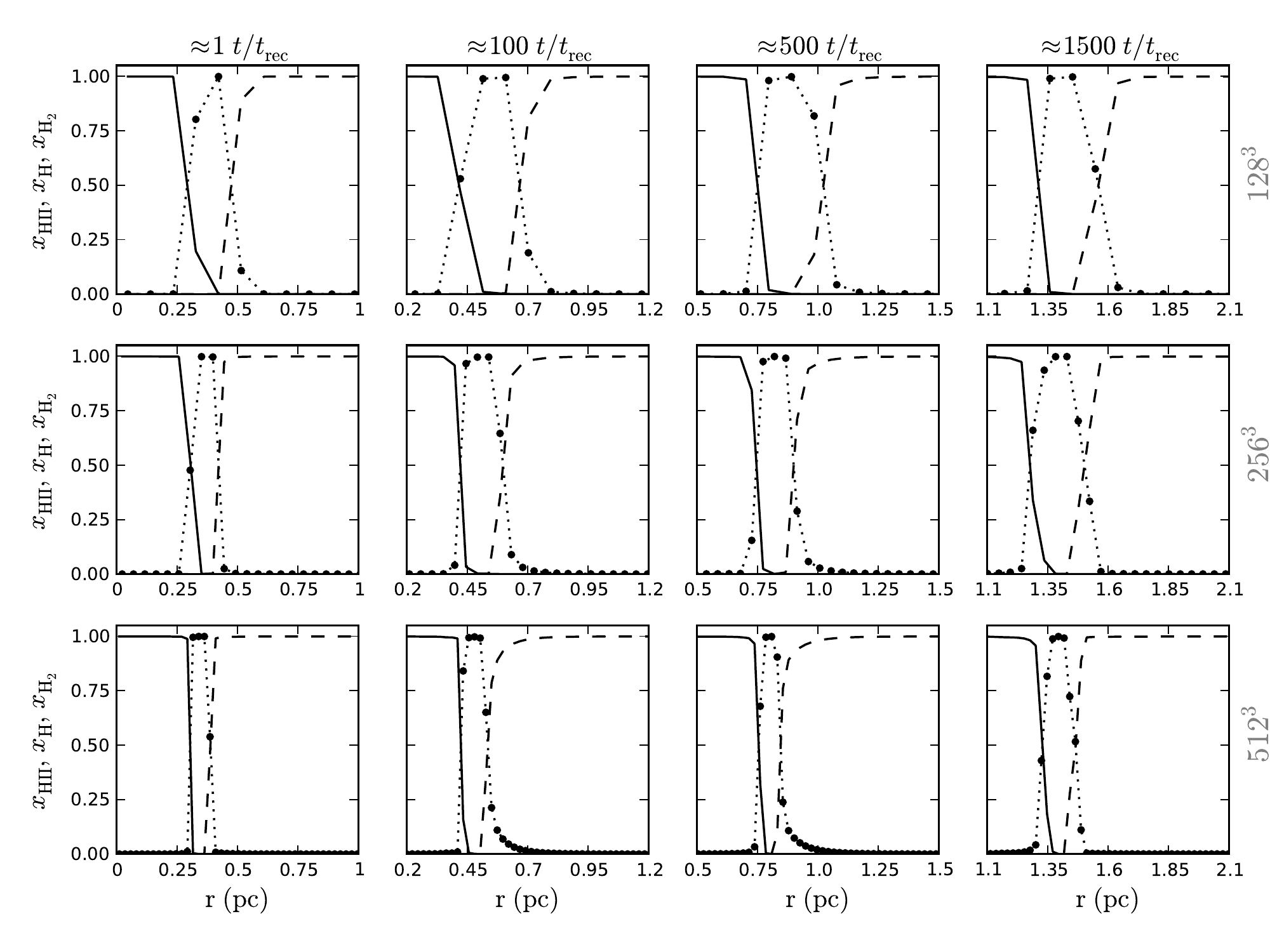}
\caption{Evolution of the thickness and chemical structure of the ionization and PDR fronts expanding in a uniform medium. We show the fronts at times of $t = 1, 100, 500, 1500 \ t_\mathrm{rec}$ for spatial resolutions of $128^3, 256^3$ and $512^3$ cells from the top to bottom row. The black dots mark the position of the cell centers. The solid, dashed and dotted lines denote ionized, atomic and molecular hydrogen fractions of the gas. (Note that here we define $x_{\rm H_{2}}$ such that a value of one corresponds to fully molecular gas).
\label{fig:oneSource_chemical_evol}} 
\end{figure*}

Figure \ref{fig:oneSource_thermodynamics} shows profiles of the temperature, density and pressure at these times.
We find that the evolution of the combined fronts is independent of the spatial resolution. The temperature of the H{\sc ii} region is at around $7-8 \times 10^3 \ \mathrm{K}$, far below the value of $10 \times 10^3 \ \mathrm{K}$ that is commonly assumed. At all times, the gas density contrast and thickness of the shell is limited by the spatial resolution. This leads to an ionized region that is slightly smaller for low resolution, as the shell always has a width of two to three cells, and the cells are larger in the low resolution simulations. 
In Figure \ref{fig:oneSource_chemical_evol}, we quantify the chemical evolution of the PDR and I-front structure by examining how the fractional abundances of the different chemical species change as we move across the structure.
During the R-type phase, the ionized, atomic and molecular hydrogen layers are as thin as numerically possible, i.e.~only two to three cells wide. Gradually, the PDR spreads out which can be seen best in the thickness of the atomic hydrogen layer. At late times the dissociation region becomes several cells thick for all resolutions. However, due to the finite minimum cell sizes, they differ in total physical size, i.e.~the PDR structure is not yet converged. The transition of $\mathrm{C^+}$ to $\mathrm{CO}$ follows closely the transition of $\mathrm{H}$ to $\mathrm{H_2}$, owing to our assumption that the CO photodissociation rate scales directly with the H$_{2}$ photodissociation rate.

\begin{figure*}
\centering
\includegraphics[scale=0.55]{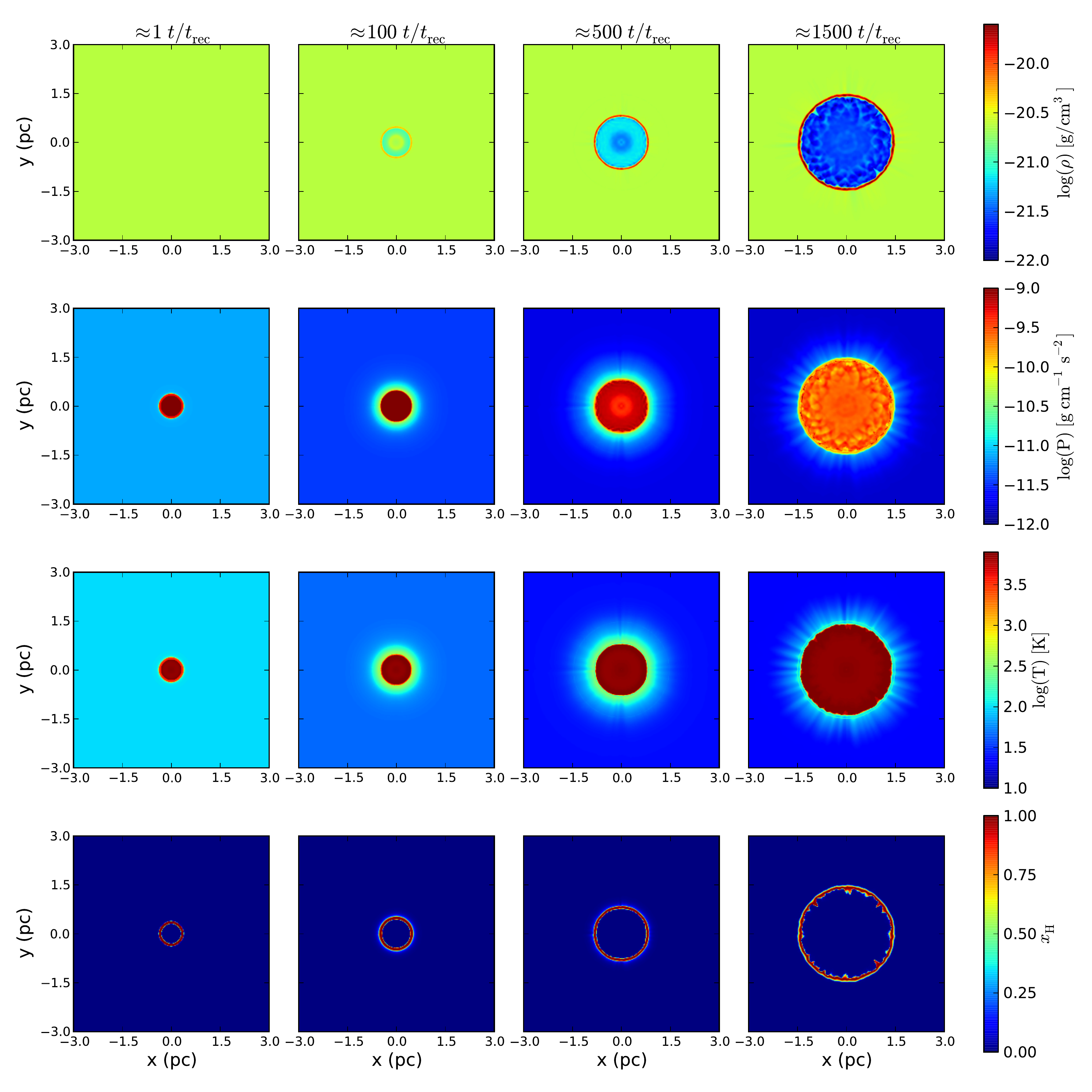}
\caption{Slices of the combined PDR and ionization front expansion into a uniform medium. We show the density, pressure, temperature and atomic hydrogen fraction for four different output times. 
\label{fig:oneSource_time_evol}} 
\end{figure*}

Finally, we also show slices through the simulation domain in Figure \ref{fig:oneSource_time_evol} for the highest resolution simulation with $512^3$ cells. 
We note that the gas in the PDR layer is heated to several hundred Kelvin. This leads to a reduced pressure contrast between the H{\sc II} region and the ambient medium. The simulation was not set up to be in equilibrium, which results in a change in the temperature and pressure of the undisturbed gas over time. This gas cools from its initial temperature of 100~K to a temperature of a few tens of K at late times, close to its equilibrium value.
%
%
\subsection{Photo-evaporation of a dense clump by two sources}
The environment in a typical star-forming region is not uniform. It consists of large temperature and density contrasts with corresponding differences in the chemical structure. After a massive star forms in one of these regions inside a dense molecular core, it clears out its own vicinity of any remaining gas not used up in its creation. An initially isolated H{\sc ii} region inside this core is formed which later breakes out and merges with others and affects the cloud as a whole. At this point, any remaining dense structures not host to the initial star formation are impacted by the radiative feedback of several stars. 

We investigate the capability of our code to treat multiple sources illuminating a dense cloud in an idealized setup. The test consists of a uniform medium in a $\pm 16 \ \rm{pc}$ box with two identical sources at positions $p_1 (x,y,z) = (-14,0,0) \ \mathrm{pc}$  and $p_2(x,y,z) = (0,-14,0) \ \mathrm{pc}$ illuminating a homogeneous spherical over-density positioned at the origin with a radius of $r_{\mathrm{clump}} = 4 \ \mathrm{pc}$. The parameters of the radiation sources are listed in Table~\ref{tab:source}. We model two scenarios distinguished by the contrast between the clump density and the density of the ambient medium. In the first scenario, the hydrogen nuclei number density in the ambient medium is set to $n = 1 \: {\rm cm^{-3}}$ and the temperature of the gas is set to $1000 \ \mathrm{K}$. The dense clump is initialized with  a density a thousand times higher and a temperature of $10 \ \mathrm{K}$.

In these conditions, the initial recombination time in the ambient medium is $\sim 0.12 \ \mathrm{Myr}$ and each source creates an H{\sc ii} region with a Str\"omgren radius of $R_{\rm s} = 58.3 \ \mathrm{pc} $ which encompasses the whole simulation domain. The ionization fronts from both stars therefore reach the dense clump during their R-type phase, leaving no time for the ambient medium or the clump to react hydrodynamically. Accordingly, we only solve for the chemical and thermal structure within the first recombination time. This setup allows us to study whether there are any artifacts arising from our adaptive ray-tracing scheme and rate calculation as well as the overlap and interaction of both ionization fronts.

In the second scenario, we change the density contrast between the ambient medium and the clump to a factor of ten by increasing the density of the ambient medium density to $n = 10 \ \mathrm{cm^{-3}}$ and decreasing the clump density to $n = 100 \ \mathrm{cm^{-3}}$. This results in a recombination time of $0.012 \ \mathrm{Myr}$ and the Str\"omgren radius decreases to $R_{\rm s} = 12.6 \ \mathrm{pc}$. These parameters allow us to study the hydrodynamical response of the ambient medium and the clump to heating by stellar radiation and the pressure from the surrounding ionized gas. 

Both setups have metal abundances and dust-to-gas ratios that are the same as in the PDR test (see Table \ref{tab:pdr}) and adiabatic indexes of $\gamma = 1.6667$. We choose to initialize all gas to be fully molecular to highlight the effects of the PDR although realistically the ambient gas should be nearly atomic for both scenarios. 

\begin{figure*}
\centering
\includegraphics[scale=0.80]{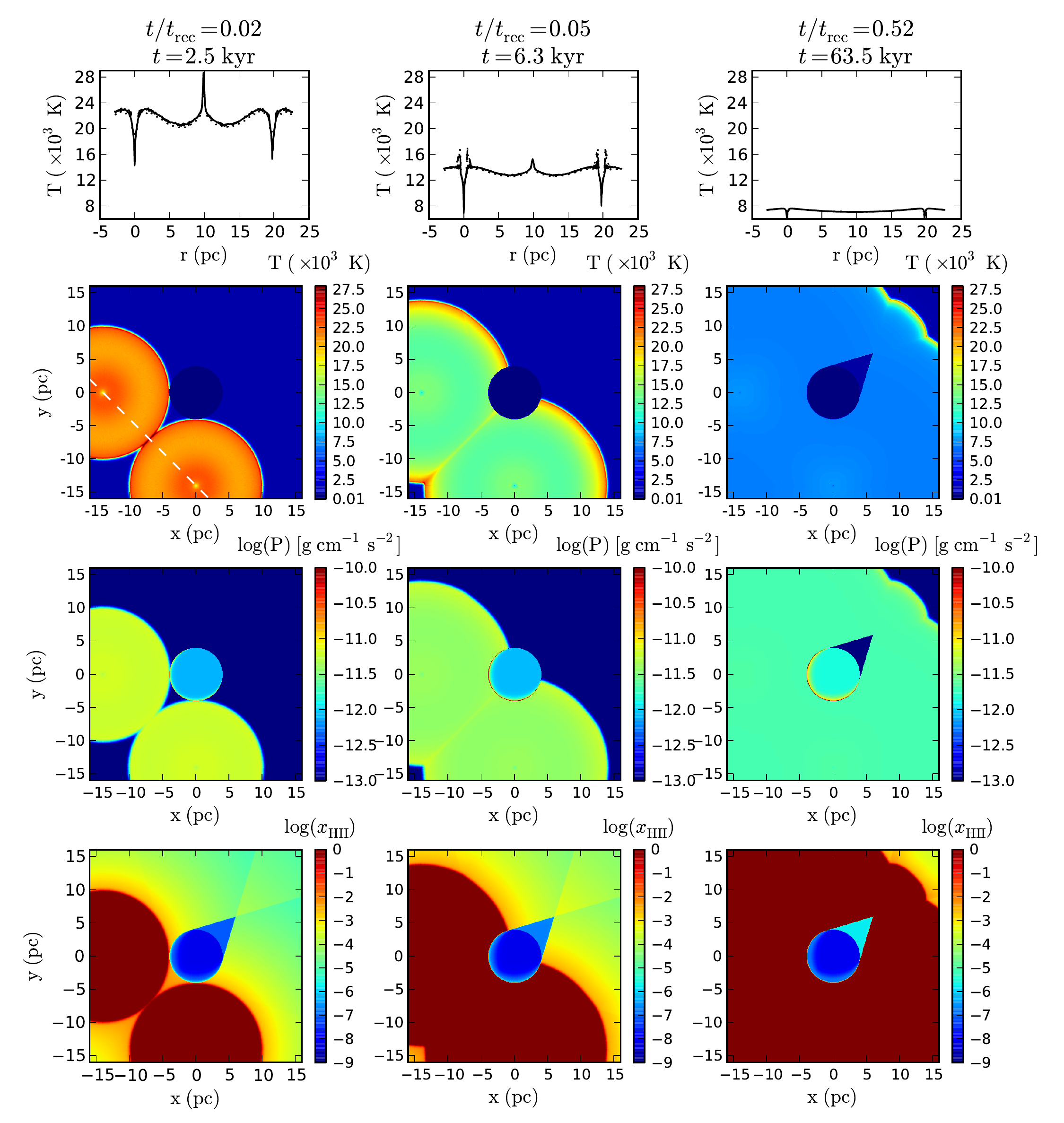}
\caption{High density contrast test of a dense clump irradiated by two sources. Shown is the evolution of the combined I- and PDR- front with no hydrodynamic evolution. The temperature profiles in the top row are taken along the dashed white line, the connecting line between both stellar sources. The profiles are plotted for resolutions of $128^3$ (dotted) $256^3$ (dashed) and $512^3$ (solid). The slices shown in the figure are taken from the highest resolution simulation. The profiles and slices are taken at three different times: when the I-fronts have just met, at an intermediate time when the ionization fronts have passed the clump, and at equilibrium. The spikes in the temperature profile at intermediate times are a transient feature introduced by the adaptively split rays, which introduce a small error in the sampling of the radiation field close to the source positions.  
\label{fig:twoSource_time}} 
\end{figure*}

Figure \ref{fig:twoSource_time} shows the evolution in the chemical composition and the temperature of the gas for the first scenario for a series of different output times. 
The temperatures and chemical structure equilibrate after just $0.5$ recombination times, with no expected change over the next few recombination times until the I-front reaches its D-type phase. During that phase, the temperature of the H{\sc ii} region drops dramatically from around $28 \times 10^3 \ \mathrm{K}$ in the ambient medium when the two ionization fronts meet to $7.5 \times 10^3 \ \mathrm{K}$ at the quasi-steady state of the setup. 

The spike in the temperature at the midway point connecting the two sources is the result of the added up heating and ionizing rates from both stars. Once both H{\sc ii} regions fully merge and a large optically thin bubble is formed, the heating rate in this region becomes more similar to that near the sources and the temperature spike disappears. At all times the evolution of both I-fronts is symmetric, i.e.\ there is no biasing based on source position or the order in which the rays traverse the domain.

At the surface of the dense clump, a PDR forms that is dominated by atomic hydrogen. Because of the high clump density, this PDR layer is narrow: the transition from fully ionized to fully molecular gas takes place in a distance of only around 1~pc.

Very close to the radiation sources, the temperature drops by two thousand Kelvin at equilibrium compared to the temperature at intermediate distances. This artifact is a consequence of the fact that when solving the chemical rate equations, we require our ODE solver to produce accurate results only for those chemical species whose abundances exceed some specified absolute tolerance. We do this on the grounds of computational efficiency -- it makes little sense spending a large amount of computational time to accurately compute the H{\sc i} abundance if this is e.g.\ only $10^{-15}$ -- but it means that when the ionization rate is very large, the atomic hydrogen abundance can become so small that it falls below this tolerance. If it does so, then the solver is at liberty to set it to zero. From the point of view of the chemical evolution, this introduces negligible error, as the gas is dominated by H$^{+}$. However, it does affect the thermal evolution, since if $x_{\rm H}$ is zero, there is no photoionization heating, whereas if $x_{\rm H}$ is merely very small, the photoionization heating rate can be significant if the ionization rate is high. 
This artifact can be eliminated by reducing the absolute tolerance used in the solver, although this has the effect of increasing the computational cost of the entire simulation. Alternatively, it can also be eliminated by putting a floor on the atomic hydrogen abundance, preventing it from ever dropping completely to zero, although again this has a measurable computational cost. However, in practice we only see this effect in extremely close proximity to very strong ionizing sources, and we do not find it to have any significant dynamical effects, suggesting that in more realistic simulations, this effect is likely to be harmless.

\begin{figure}
\centering
\includegraphics[scale=0.45]{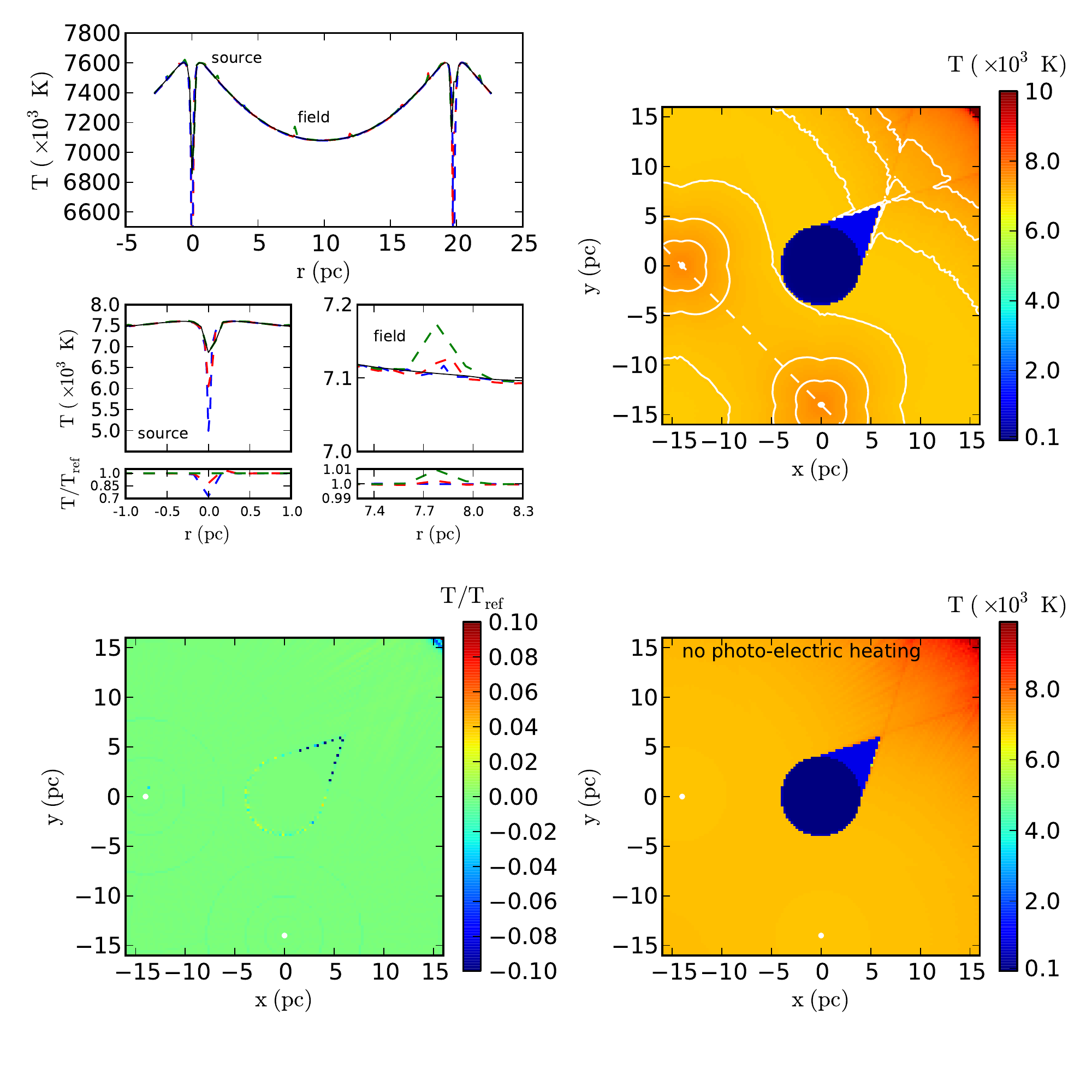}
\caption{Errors and temperature structure after the high density contrast setup reached equilibrium. The upper left plot quantifies the error from ray-splitting and the effect of spatial resolution. The plotted temperature profiles are taken along the white dashed line for simulations with $128^3$ (dashed green), $256^3$ (dashed red) and $512^3$ (dashed blue) cells. The relative error is computed from a control simulation with no adaptive ray-splitting at a resolution of $128^3$, denoted $T_\mathrm{ref}$ (solid black line). Iso-temperature contours of $7000$, $7200$ and $7400\ \mathrm{K}$ are over-plotted on the control run temperature slice in the top right panel. The bottom left panel shows the error incurred from adaptive ray-splitting in the temperature map and the panel on the bottom right shows a control run with no photo-electric heating. Note the different temperature scalings in the temperature slices. 
\label{fig:twoSource_split}} 
\end{figure}

Figure \ref{fig:twoSource_split} illustrates other numerical artifacts of our radiative transfer scheme. 
Some error is introduced by the adaptive splitting of rays that leads to a change in the sampling of the radiation field at the splitting radius. This results in slight offsets in the accumulated gas columns and total intersection length of all crossing rays which changes the calculated rates in comparison to an un-split case. We compare a control simulation with no adaptive splitting, where the initial number of rays is large enough to sample all cells sufficiently at all distances to one with ray-splitting. The errors shown in the upper and lower left panel in Figure \ref{fig:twoSource_split} are generally of the order of $1\%$ at most. Another, transient feature stemming from the adaptively split rays are the spikes observed in the temperature profiles at intermediate times in Figure \ref{fig:twoSource_time}. The amplitude of the spikes can be reduced by increasing the initial \texttt{HEALPix} level at additional computational cost. For the default parameters the error around the source position is of 15 \%. We tested the impact of the initial \texttt{HEALPix} level in control simulations utilizing the low density setup and found that it had no impact on the overall evolution.

The fluctuations around the shadow cast by the dense clump and in the PDR stem from the varying number of rays that enter the cells. This depends on the current rotation of the \texttt{HEALPix} sphere, which leads to the calculated rates fluctuating slightly. As the edge of the shadow and PDR react very sensitively to changes in these rates, the error grows accordingly. 

Spatial resolution has an impact on the temperature at the source positions, since the smaller the volume of the cell with the source, the stronger the radiation field within the cell.
Effectively, the grid is unable to resolve the point-like source which leads to a resolution dependence. 

The iso-temperature contours show that the heated regions close to the stars are not perfectly spherical. The origin of the lobe-like structures lies in the approximation used in modeling the photo-electric heating. It is a geometric artifact introduced in equation~\eqref{pefinal} where the cell face area is used to recover a flux per area. To avoid it, one would have to calculate the area seen by the ray by using a projection onto the ray. We perform a control run with no photo-electric heating to make sure that there are no additional artifacts from other processes. As can be seen from the bottom right plot in Figure~\ref{fig:twoSource_split} the resulting temperature field is nearly flat with no discernible lobes or other features. 

\begin{figure*}
\centering
\includegraphics[scale=0.7]{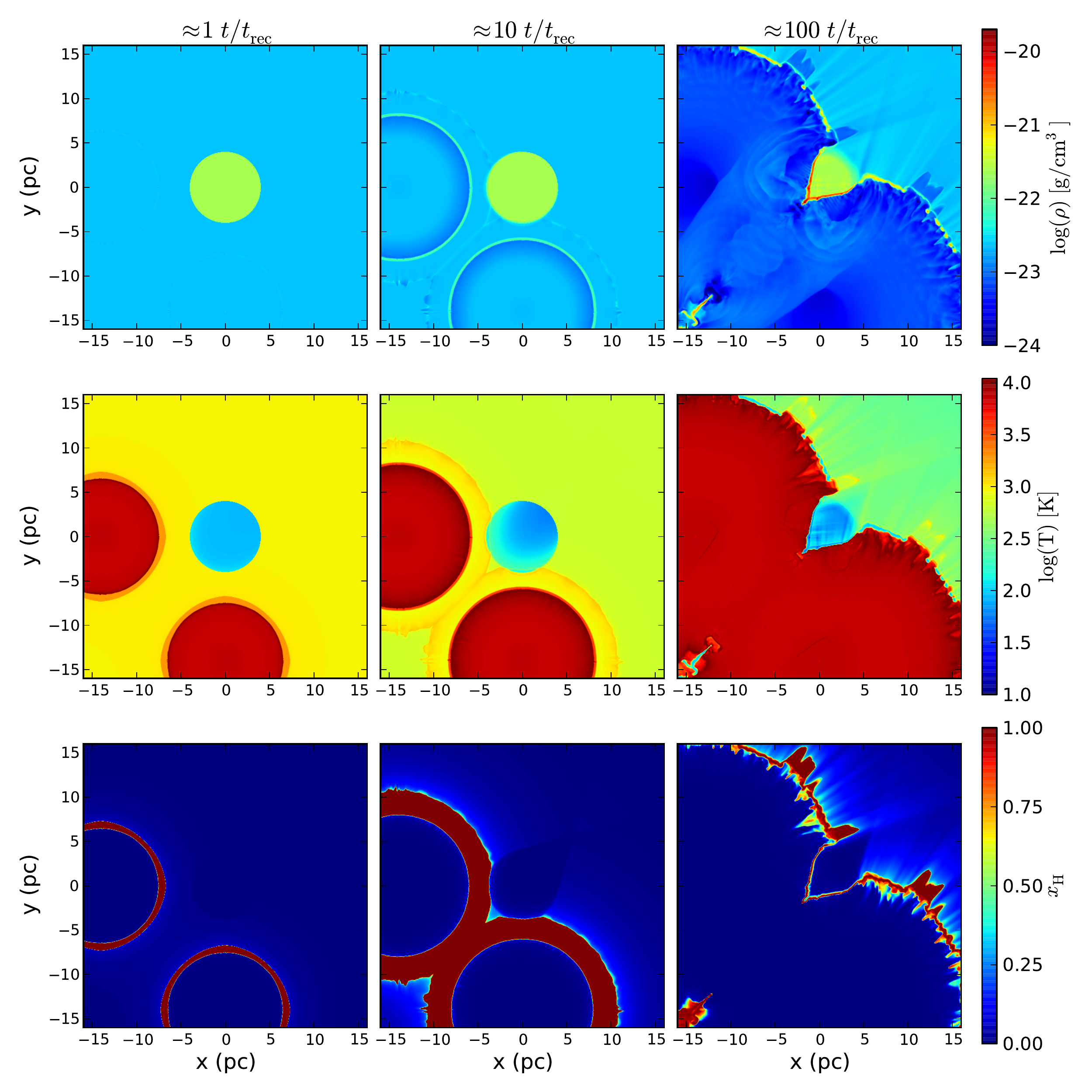}
\caption{The low density contrast test case of the expansion of two H{\sc ii} regions impinging on a dense clump with hydrodynamic response. Slices in density, temperature and atomic hydrogen fraction of the highest resolution runs with $512^3$ cells are shown. The evolution is plotted at three snapshots in time which correspond to recombination times of $1$, $10$ and $100$ in the ambient medium.   
\label{fig:twoSource_hydro}} 
\end{figure*}

The evolution of the second scenario is shown in Figure~\ref{fig:twoSource_hydro}.
The ionization fronts meet after approximately ten recombination times, calculated for the ambient medium. At around the same time they transition to their D-type phase. Two shells around the H{\sc ii} region can be made out in the density slice, one generated directly by the over-pressured ionized gas, the other from the heat deposited in the PDR, upstream of the I-front. The edge of the shell is unstable as can be seen along the cardinal directions where the outer, thinner shell breaks up. 

The shock front is reflected off the dense clump and after one hundred recombination times an interference pattern emerges from the incoming and the reflected shocks. The clump itself is symmetrically compressed from each side over the whole evolution, with a corresponding increase in temperature of a few thousand Kelvin on the surface, but no significant heating on its inside. 

In the low density ambient medium, a thick PDR forms which pre-heats the gas before the I-front reaches the same position. It can be easily made out in the atomic hydrogen fraction slice shown in the bottom row of Figure \ref{fig:twoSource_hydro}. It becomes thinner as it moves into the over-dense region. Similarly, at later times the PDR preceding the I-front becomes narrower as the shell becomes denser and the visual extinction increases over a shorter distance.

\begin{figure*}
\centering
\includegraphics[scale=0.7]{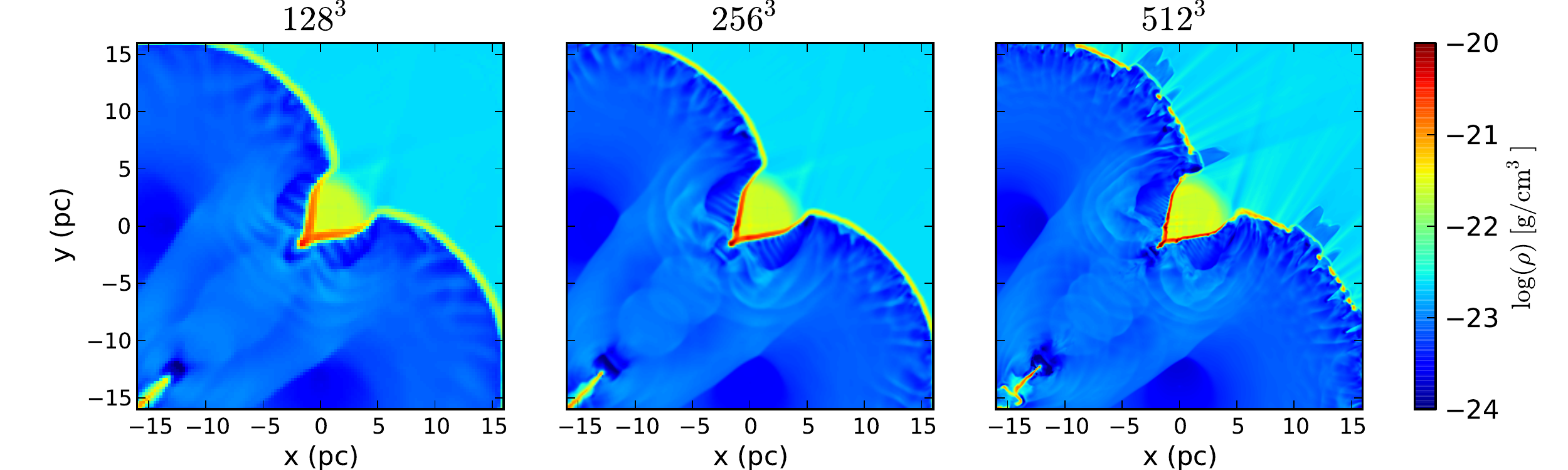}
\caption{Resolution study for the low density contrast case. The late stage evolution after one hundred recombination times is shown for simulations with resolutions of $128^3$, $256^3$ and $512^3$. 
\label{fig:twoSource_res}} 
\end{figure*}

Finally, Figure \ref{fig:twoSource_res} shows a spatial resolution study of the second scenario. 
As described in the previous section, the thickness and density of the shell depends on the spatial resolution. Additionally, this test shows that the stability of the shell is influenced as well: the smaller the cell size, the larger the density fluctuations. For all simulation runs, we find convergence in the position of the shells and fronts as well as in the temperature structure and the geometry of the compressed clump.
\section{summary}\label{summary}
In this paper, we present \texttt{Fervent}, a radiative transfer code module for the magnetohydrodynamical adaptive mesh refinement code \texttt{FLASH 4}. \texttt{Fervent} is designed to model the effects of radiation from massive stars without assuming either thermal or chemical equilibrium, as is otherwise common in simulations of stellar feedback. It allows us to self-consistently evolve the chemical, thermal and density structure of the interstellar medium surrounding a massive star over time.

We are able capture the combined effect of the I-front and PDR front expansion necessary to accurately simulate H{\sc ii} regions. 
This is in contrast to most dedicated photon-dominated region (PDR) codes that only calculate the final equilibrated state with no explicit time evolution and in a static density field.  

\texttt{Fervent} is based upon the ray-tracing scheme outlined in \cite{moray}, which likewise utilizes the \texttt{HEALPix} library \citep{HEALPix} to adaptively split rays to sample the radiation field. This approach is well-suited for combining with the adaptive mesh structure in \texttt{FLASH 4}, as described in detail in the Appendix. Our chemical and thermal treatment is designed for modelling the impact of stellar feedback on the present day ISM. 
We account for all of the main radiative heating processes: photoionization (of H and H$_{2}$), H$_{2}$ photodissociation, the vibrational pumping of H$_{2}$ by FUV photons, and photoelectric heating by dust grains. We show that to properly capture these effects requires us to split up the photons into at least four frequency bins.

We describe how to calculate photoionization and photodissociation rates in a fashion that is independent of spatial resolution, and show how to couple them to a 
fast bare-bones chemical network used to model the hydrogen chemistry in the ISM. We also show how to treat a single chemical reaction that overlaps several energy bins and photons in a single energy bin that couple to multiple species in a photon conservative way. 

We test the code extensively from simplified to full setups where all physics modeling is included. \texttt{Fervent} reproduces the well-known analytic results of I-front expansions in an atomic medium. In addition, we find very good agreement with results obtained from dedicated PDR codes for the structure of the hydrogen dissociation front as well as the temperature
structure of this region. Disagreements arise for $\mathrm{CO}$, which at present we can only approximate to within an order of magnitude owing to the simplifications made in our chemical network in order to decrease its computational cost. 

We explore under what conditions the hydrodynamical response to the thermal feedback is captured. We show that we have to limit the change in the hydrogen fraction to $10\%$ over a single simulation time step and resolve the initial Str\"omgren-sphere  by at least one resolution element for a proper expansion of the I-front. 
The effect of changing the spatial resolution is also thoroughly examined, and we show that apart from a few well-known issues to do with numerical diffusivity (shell thickness and stability, deviations from perfect spherical symmetry), all of our results converge in simulations with both low and high resolutions.  

In its current state \texttt{Fervent} is an extremely capable radiative transfer module with many potential applications in the field of star formation and the dynamics of the interstellar medium, which we plan to improve further in the future (e.g.\ proper treatment of the carbon chemistry).

\section*{Acknowledgments}
We thank Stefanie Walch, Richard W\"unsch and the other members of the SILCC collaboration (\url{http://hera.ph1.uni-koeln.de/~silcc/}) for allowing us to use their modified version of \texttt{FLASH 4}. We also thank Jan-Pieter Paardekooper for his feedback on an earlier draft of the paper. RSK, SCOG and CB thank the DFG for funding via the SFB 881 ``The Milky Way System" (subprojects B1, B2, and B8). CB thanks Mordecai-Mark Mac Low for his hospitality at the AMNH and the Kade foundation for its support. SCOG and RSK also acknowledge support from the DFG via SPP 1573, ``Physics of the Interstellar Medium'' (grant number GL 668/2-1). RSK furthermore acknowledges support from the European Research Council under the European Community's Seventh Framework Programme (FP7/2007-2013) via the ERC Advanced Grant STARLIGHT (project number 339177).
\appendix{
\section{Ray-tracing implementation details}\label{AppendixA}
In the following, we describe the general structure of the adaptively split ray-tracing implemented in  \texttt{FLASH 4}. 
\subsection{Data structures}
We distinguish two major data structures, the adaptive mesh used to decompose the  computational domain and a two-dimensional array exclusive to the ray-tracing. The mesh structure is implemented as a fully-threaded tree that holds all of the hydrodynamical quantities for each resolution element on each refinement level in the simulation. The ray array saves all information used in box-ray intersections, radiation source properties and some of the data gathered during sampling of the radiation field. Table \ref{table:a1} shows the quantities carried in the mesh data structure that are accessed and changed. Other quantities not relevant to the routine, e.g.~pressure etc., are omitted. The data saved in the ray array is shown in Table~\ref{table:a2}.
\begin{table}
\caption{quantities accessed and saved in mesh data structure \label{table:a1}}
\centering
\begin{tabularx}{\columnwidth}{l@{\hskip 0.5in}l@{\hskip 0.5in}}
name & symbol \\
\hline
gas density \footnote{names in normal typeface are only accessed} & $\rho$   \\
$\mathrm{H}$ abundance    & $x_\mathrm{H}$   \\
$\mathrm{H_2}$ abundance  & $x_\mathrm{H_2}$ \\
$\mathrm{H^+}$ abundance  & $x_\mathrm{H^+}$ \\
cell size & $d$ \\
\textbf{dissociation rate} \footnote{names in bold are changed by the ray-tracing}  & $k_\mathrm{UV}$ \\
\textbf{ionization rate}   & $k_\mathrm{ion}$ \\
\textbf{direct dissociation rate}   & $k_\mathrm{dis}$ \\
\textbf{ionization heating rate}   & $\Gamma_\mathrm{ion}$ \\
\textbf{direct dissociation heating rate}   & $\Gamma_\mathrm{dis}$ \\
\textbf{photo-electric energy flux} & $F_\mathrm{pe}$ \\
\hline
\end{tabularx}
\end{table}
Besides these two main structures, auxiliary arrays are used in the parallel communications routines, as described in Section~\ref{parallel}. During the ray-tracing step, heating and reaction rates are saved into the tree structure for later use in the chemistry module. These quantities are shown in bold in Table \ref{table:a1}. 

The ray array is allocated at the start of the simulation  and persists throughout the run. This has the advantage that we do not have to allocate and deallocate each array in each ray-tracing step, but has the disadvantage that we have to specify a maximum number of rays. 
\begin{table}
\caption{quantities saved in ray data structure}
\centering
\begin{tabularx}{\columnwidth}{l@{\hskip 0.5in}l@{\hskip 0.5in}}\label{table:a2}
name & symbol \\
\hline
direction, x component & $x_{\mathrm{dir}}$  \\
direction, y comp.& $y_{\mathrm{dir}}$  \\
direction, z comp.& $z_{\mathrm{dir}}$  \\
source position, x comp.& $x_{\mathrm{pos}}$  \\
source position, y comp.& $y_{\mathrm{pos}}$  \\
source position, z comp.& $z_{\mathrm{pos}}$  \\
\it{number of ionizing photons in $E_\mathrm{13.6}$}\footnote{names in italics are communicated by message passing interface}\footnote{see section \ref{model} for the extents of the energy ranges} & $N_{\mathrm{13.6}}$  \\
\it{number of ionizing photons in $E_\mathrm{15.2+}$} & $N_{\mathrm{15.2+}}$  \\
$\mathrm{H}$  ionizing photon energy in $ E_\mathrm{13.6}$ & $ \left<E_{\mathrm{13.6}}\right>$  \\
$\mathrm{H}$  ionizing photon energy in $E_\mathrm{15.2+}$ & $ \left<E_{\mathrm{15.2+}}\right>$  \\
$\mathrm{H_2}$  ionizing photon energy & $\left<E_{\mathrm{dis}}\right>$  \\
average UV photon energy & $\left<E_{\mathrm{11.2}}\right>$ \\
number of UV photons at source & $N_{\mathrm{UV}}$  \\
$5.6$ to $11.2 \ \mathrm{eV}$ energy flux at source  & $F_{\mathrm{pe}}$  \\
$\mathrm{H}$ ionization cross section for photons in $E_\mathrm{13.6}$   & $\sigma^\mathrm{13.6}_{\mathrm{H}}$  \\
$\mathrm{H}$ ionization cross section for photons in $E_\mathrm{15.2+}$& $\sigma^\mathrm{15.2+}_{\mathrm{H}}$  \\
$\mathrm{H_2}$ ionization cross section & $\sigma_{\mathrm{H_2}}$  \\
\it{total $\mathrm{H}$ column}  & $\mathrm{N_H}$  \\
\it{$\mathrm{H_2}$ column} & $\mathrm{N_{H_2}}$  \\
\it{\texttt{HEALpix} level} & $\mathrm{l_h}$  \\
\it{\texttt{HEALpix} number} & $\mathrm{n_h}$  \\
\it{total distance from source} & $r_s$  \\
\it{block identifier }& $\mathrm{b_{id}}$  \\
\it{source identifier} & $\mathrm{s_{id}}$  \\
domain identifier & $\mathrm{d_{id}}$  \\
\hline
\end{tabularx}
\end{table}
\subsection{Algorithmic overview}
The ray tracing module is divided into several routines motivated by the organization of resolution elements (cells) in uniform mesh blocks with extents $N_\mathrm{cells}\times N_\mathrm{cells}\times N_\mathrm{cells}$ in the \texttt{PARAMESH} \citep{paramesh} tree data structure, the default AMR module in \texttt{FLASH} 4. Blocks, numbered by a locally unique identifier (ID), are the nodes of the tree and each time a block is marked to be refined, $2^m$ child blocks are created on the next higher level of the tree, where $m$ is the spatial dimension of the simulation. For blocks with $2\times 2\times 2$ cells this tree reduces to the commonly-used oct-tree. \\
\begin{figure*}
\centering
\includegraphics[scale=0.7]{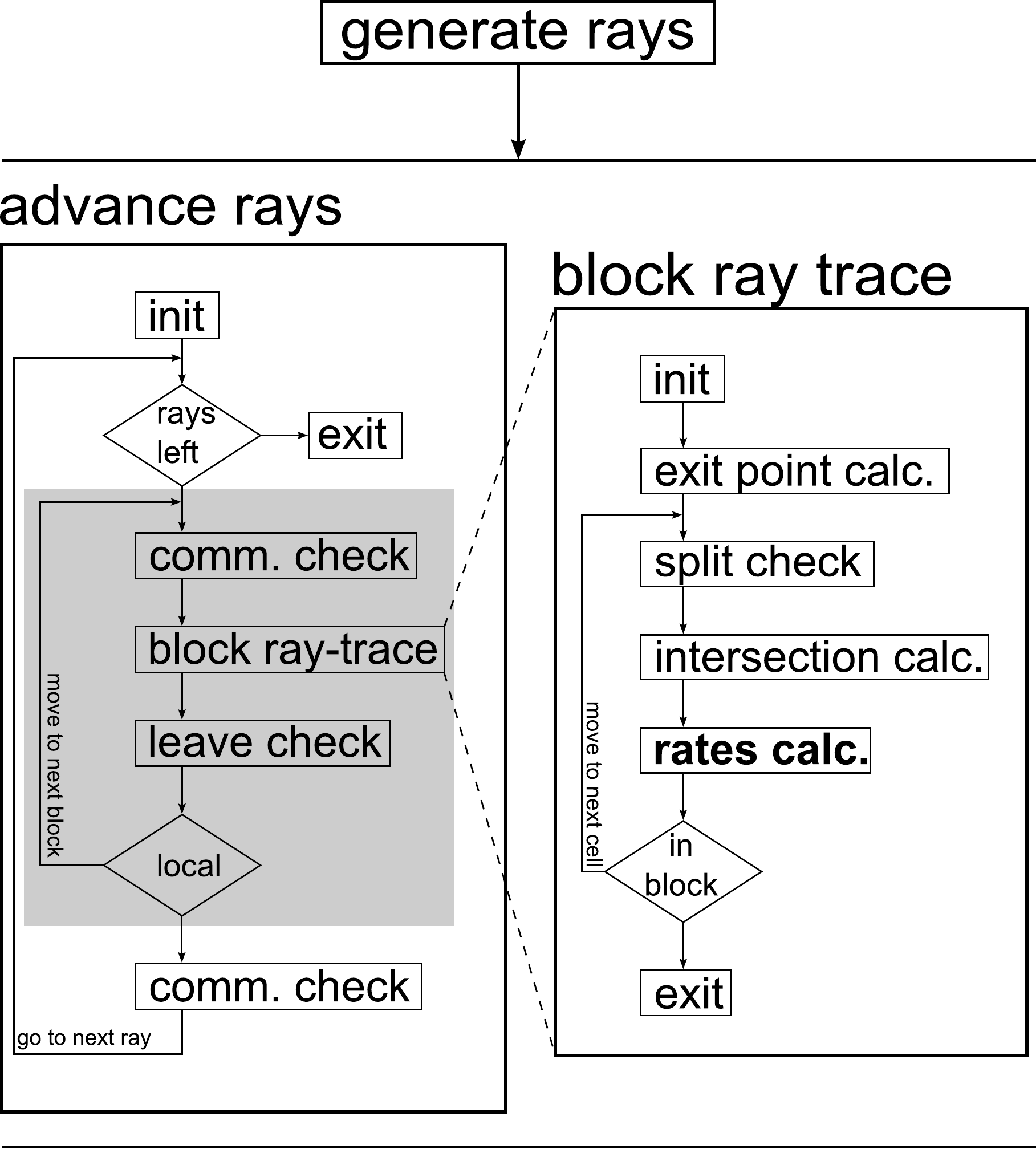}
\caption{Structure of the ray-tracing module. The highlighted gray area marks the code for the traversal of a single ray. The routine name in bold is the only point in the module where outside data and data structures are requested and used. 
\label{fig:progStructure}} 
\end{figure*}
The main routine of the ray-tracing module is called from inside the chemistry module before the chemical network is solved, to make sure that the hydrodynamical data used in the chemistry and ray-tracing modules are the same. In principle, it could be called from any point during one timestep in the simulation. The main routine consists of two steps (see Figure \ref{fig:progStructure}): the generation of rays with properties taken from a global source list, and a ray-tracing step, where the rays are advanced through the computational domains. 

For our ray-tracing module we assume a global list of all source positions and properties is available on all computational domains. The source data structure carries the effective temperatures and luminosities of the stars, from which the properties that describe the emitted radiation are calculated (see Table \ref{table:a2}). Before any ray-tracing is performed, initial rays distributed on a \texttt{HEALPix} sphere are generated around each source location. The spheres are rotated by a randomly chosen angle every timestep to wash out any alignment artifacts with the Cartesian mesh structure. All generated rays are appended in the ray array without distinguishing their origin. This has the consequence that we are not able to explicitly track individual contributions to the chemical state of a given cell from a given source. 

All domains containing a source begin ray-tracing the first ray in the ray data structure, while all others wait to receive rays from neighboring domains. A ray is followed along through the blocks of the local domain it intersects until one of the following criteria is fulfilled: the ray leaves the global computational domain, it is stopped because the visual extinction $A_{\rm V}$ becomes large enough that there is no longer a need to follow the ray, or it reaches a block that is part of a neighboring domain and is queued for transport. Rays are moved between blocks by a tree lookup, where the proper neighbor is chosen during the intersection calculation. If the found neighbor block ID is on a different computational domain, the corresponding ray is sent via message passing interface (MPI) communication to its target. Afterwards, the advancement routine moves to the next available ray, or if none are left locally, it stays in communication until all domains are done. \\
\begin{table}
\caption{Ray-tracing parameters}
\centering
\begin{tabularx}{\columnwidth}{l@{\hskip 0.5in}l@{\hskip 0.5in}}\label{table:a3}
Parameter name & Default value \\
\hline
pixel to cell face ratio ($f_\Omega$) & $4.0$  \\
initial \texttt{HEALpix} level & $4$  \\
maximum number of rays per comp. domain& $30 \times 10^4$  \\
rotate rays & True  \\
split rays & True  \\
stopping $A_{\rm V}$ & 10.0  \\
bundle rays & 5 \\
comm.\ interval & 20  \\
periodic in x & False  \\
periodic in y & False  \\
periodic in z & False  \\
periodic box length & 1.00  \\
\hline
\end{tabularx}
\end{table}
The actual ray-box intersection calculation is done on a block level, where we take advantage of the uniform mesh structure to optimize the ray-cell intersection (see next section and Figure \ref{fig:intersection} for details). During the ray traversal of a single block, each time a new cell is entered the ray  splitting criterion $f_\Omega$ is evaluated. If the ray has to be split, the current slot in the ray array is overwritten by one of the child rays, and the additional three child rays are appended to the end of the ray data structure. The splitting routine also checks whether any of the child rays move off the global domain, which might be the case if rays are split in the vicinity of the simulation domain boundaries. After splitting, the current cell iteration inside the block is terminated and a new iteration is called recursively with the updated child ray data instead of iterating to the next ray in the ray array. In this way, we always try to follow one ray or one of its child rays directly to the edge of the local domain, where the ray is sent without waiting for all local rays to traverse the local domain. 

The ray splitting, rotation of the \texttt{HEALPix} sphere and parallelization approach combine to randomize the order in which rays intersect   cells in each timestep. Which and how many sources illuminate a cell, as well as the total number of rays that enter it, is therefore unpredictable during the ray-tracing step itself. For multiple sources, to assure a properly converged result, one would have to take into account all contributions from all sources per cell simultaneously, call the chemical update during the ray-tracing, and re-distribute the attenuated photon fluxes onto the intersecting rays, which are then moved to the next cell. 

Instead, as rays are indistinguishable between sources, we trust that the employed timestep criterion holds for multiple sources.  This can be seen if a cell illuminated by two identical sources on each side is considered. The rays enter and see the same chemical composition and calculate the same rates, which are added up to twice the value. Thus, the composition changes twice as fast and the timestep should adjust accordingly. A caveat is that we employ a corrector timestep, i.e.~the timestep is reduced after a fast change was detected, and the first time two or more radiation fronts interact (or a radiation source turns on), the rate of change of the chemical composition will be underestimated, and we may therefore employ a timestep that it larger than we would desire. 
As the validation study described in the main body of the text demonstrates, the errors arising from this are usually very small. Our ray-tracing approach does not suffer from any biasing based on the order that radiation sources are treated, simply because we calculate and sum all rates from all sources in one step. 

The ray-tracing module interfaces with the mesh data during the iteration over all intersected cells in a block  via one single routine. It concentrates the physics modeling in the form of rate calculations in one place, and deposits all of the information required for the subsequent chemistry step in the appropriate fields in the tree data structure. This makes it easy to include additional physics, such as coupling to a more extensive chemical network, or adding additional energy bins. 

Finally, we give an overview of all parameters and their default settings that determine the behavior of the ray-tracing module (see Table \ref{table:a3}). The parameters can be roughly divided into two sets, one that describes how the radiation field is sampled and the other describing the communication pattern. The two most important parameters are $f_\Omega$, whose value approximately corresponds to the number of rays per cell and secondly the boolean flag that indicates whether rays are to be split at all. In some cases, especially for known geometries, it might be computationally more efficient to fix the number of rays by specifying the initial \texttt{HEALPix} level, and not splitting the rays. In that case, one has to make sure that all cells are intersected by at least one ray. 

The parameters that control the behavior of the parallel communication are described in Section \ref{parallel}. Each simulation boundary can be set to be periodic, in which case a ray is allowed to leave a border and enter the opposite one. The ray is terminated if it traverses a total length given by the periodic box length parameter.
\subsection{Box-ray intersection}
There are two stages to the intersection calculation of an infinite line (ray) with an axis-aligned box (cell) in our ray-tracing module.
First, once the ray enters a new block, the point the ray exits is computed by checking each block boundary individually,
\begin{equation}\label{eq:intersection}
\begin{aligned}
\mathbf{t}_\mathrm{u} = \left( \frac{b^u_x-s_x}{n_x} ,\frac{b^u_y-s_y}{ n_y},\frac{b^u_z-s_z}{n_z} \right) \\
\mathbf{t}_\mathrm{l} = \left( \frac{b^l_x-s_x}{n_x} ,\frac{b^l_y-s_y}{ n_y},\frac{b^l_z-s_z}{n_z }\right),
\end{aligned}
\end{equation}
picking the set of boundaries that point away from the source, $\mathbf{t}_\mathrm{next} = \mathrm{max}( \mathbf{t}_\mathrm{l},  \mathbf{t}_\mathrm{u})$, and finally choosing the shortest distance to one of the boundaries in the set, $t = \mathrm{min}(\mathbf{t}_\mathrm{next})$.
\begin{figure}
\centering
\includegraphics{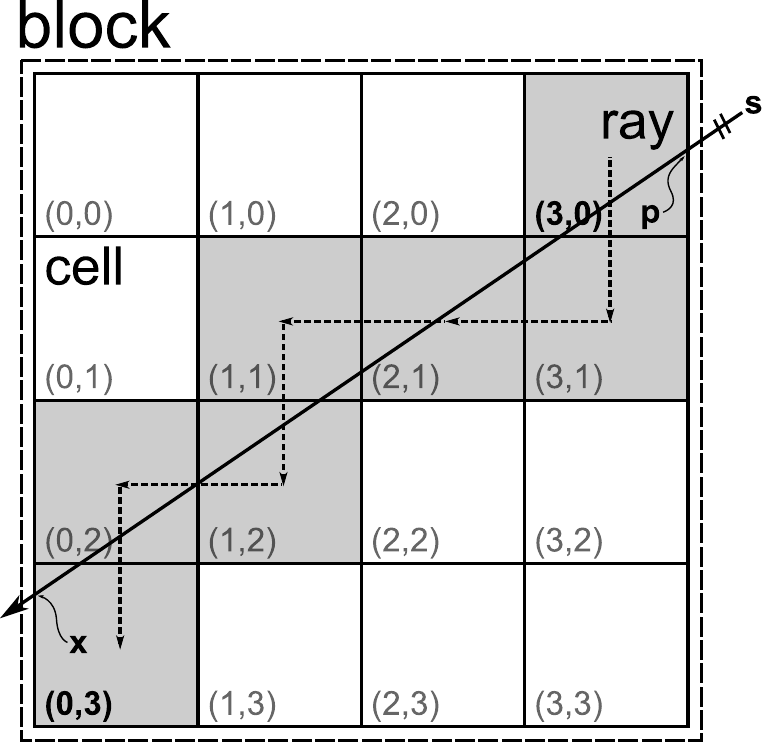}
\caption{An example of a ray-block intersection. In this case the entry index is $\mathbf{(3,0)}$ and the exit index is $\mathbf{(0,3)}$. The short-dashed arrows indicate the order in which the cells are traversed.
\label{fig:intersection}}
\end{figure}
Here, we have denoted the source position by $\mathbf{s}$, the upper and lower block boundary coordinates by $\mathbf{b}^u$ and $\mathbf{b}^l$ and the direction vector by $\mathbf{n}$. The exit point $\mathbf{x}$ is obtained by adding the scaled unit vector to the source position, $\mathbf{x} = t \cdot \mathbf{n} + \mathbf{s}$. Figure \ref{fig:intersection} illustrates the block-ray intersection. 

Initially all rays start at the source with $t = 0$. Each time the advancement routine iterates to a new ray, the current position $\mathbf{p}$ of the ray at the beginning of the intersection calculation is computed from the radius and source position saved in the ray data structure. 

Instead of using the source position in the intersection calculation, one could use an intermediate point by adding, e.g.~the last intersection point in the previous block. However, due to limits in numerical precision the distance calculated from $\mathbf{s}$, $r_s = t \cdot \mathbf{n}$, typically does not match the one from adding up all previous block intersection lengths $\mathrm{d}r_i$ to the current block $k$,
\begin{equation}\label{eq:radii}
\begin{aligned}
r_s \ne \sum^k_{i=0} \mathrm{d}r_i.
\end{aligned}
\end{equation} 
The small numerical drift is due to the square-root operation necessary to obtain $\mathrm{d}r$ and can lead to the ray-tracing routine failing if both approaches are mixed. For example, depending on how close a ray comes to a vertex or edge, calculation of the ray position from $r_s$ might intersect the block while calculation from the summed intersection lengths might not. 

The computation of the exit point also allows us to look up the direction of the next block the ray enters during its traversal. The position of the smallest element in the $\mathbf{t}_\mathrm{u}$ or $\mathbf{t}_\mathrm{l}$ set gives the direction in the positive or negative  $x$-, $y$-, or $z$-direction, respectively, and can be directly used in the tree lookup. 

The operation with the highest  computational expense is the intersection length calculation for each cell inside one block. We optimize the operation by interpolating $\mathbf{x}$ and $\mathbf{p}$ onto the uniform mesh inside the block with cell size $\Delta x$. This gives integer entry and exit indices, and each time a cell is crossed the index is incremented or decremented according to the cell face the ray leaves. The intersection length itself is computed from equation \eqref{eq:intersection}, where $\mathbf{b}$ is replaced with the cell boundary position $\mathbf{c}$. We only consider upper or lower cell boundaries in each intersection calculation, depending on the direction of the ray. This halves the number of computations necessary per cell. Additionally, we update and recompute $t$ only for the boundary updated in the direction the ray left the current cell. We find the smallest of the new set $\mathbf{t}_\mathrm{next} = (t_x, t_y, t_z)$ and from this the intersection length $\mathrm{d}r = \lvert \mathrm{min}(\mathbf{t}_\mathrm{next}) \cdot \mathbf{n} \rvert$. We call the rates calculation routine each time $\mathrm{d}r$ is computed for a given cell. Once the ray enters the last cell of the block it intersects, the entry and exit indices become equal, and we stop the cell iteration. The integer stepping is highlighted in gray in Figure \ref{fig:intersection}. 

Every time the ray's total distance property $r_s$ is updated, the splitting criterion is evaluated. If the ray is split, one of the child rays is selected, and a new exit point is calculated. The block ray-trace is then called recursively and the ray-tracing continues.
\subsection{Parallelization}\label{parallel}
Arguably, the most performance relevant part of the ray-tracing module is its parallelization for use on distributed memory machines. Such machines consist of nodes, each containing several computing units, interconnected by a high bandwidth network. Each node has a fixed amount of memory that holds part of the global simulation data, i.e.~the local domain. The simulation data is usually decomposed by space, where each node is assigned some part of the total volume. 

Radiation from point sources is a difficult problem to parallelize, owing to the fact that sources are not evenly distributed in space. A single local domain could potentially end up with most sources. This unbalances the work load, and in the worst case leads to a large fraction of the computing units idling while they wait for the ones holding sources to finish their ray-tracing step.

For our ray-tracing module, two parallelization schemes are possible. The first option is a synchronous communication pattern, where we wait until all rays are traversed locally before doing any communication. Rays that leave the local domain are saved in a buffer, and once all computing units finish their ray-tracing, a global communication step is performed. The buffered rays are sent and then another ray-tracing step is performed. This process is repeated until no rays are left in the global domain. For a single source, this can be seen as a domain-by-domain approach. First, the domain holding the source ray-traces, while all others wait. In the second step, the neighboring domains ray-trace, while again all others, including the source domain, wait. With this communication pattern, the work load moves out in shells around the source domain. From an implementation perspective, the module would be cleanly split into a ray-tracing step and a communication step, where all network sends are posted and completed. 

The second parallelization option is an asynchronous communication pattern. In this pattern, rays are bundled and sent off in intervals, without postponing communication until all rays are traversed locally. The network sends are completed when the target computing unit calls the communication routine, usually after it has traversed a number of local rays itself, or no local rays are left. This is set by the communication interval parameter (see Table \ref{table:a3}). In this communication pattern, the work imbalance is reduced by involving neighboring domains earlier in the ray-tracing. This leads to an overall decrease in computational time, as time is not wasted idling while waiting for the synchronous communication step. 

One additional important aspect is the amount of data sent during MPI communication. This sets the size of the communication buffers and in turn determines the number of messages that can be sent at the same time. We try to optimize this by only communicating the bare minimum amount of data necessary. Specifically, the ray properties that characterize the ray direction, \texttt{HEALpix} level and unique pixel identifier, from which the direction vector can be looked up, the attenuated ionizing photon flux, the total and molecular hydrogen columns, the block and source identifier and the traversed distance of the ray from the source. The communicated data is shown in italics in Table \ref{table:a2}. The other ray properties are locally reconstructed from the global source list. In total, we reduce the amount of information sent per ray to a third of what it otherwise would be.
}
\bibliography{references}{}

\end{document}